%% file: Cov-CodeV20-Quantumv3.tex
\documentclass[a4paper,noarxiv,twocolumn,accepted=2020-02-22]{quantumarticle}
\usepackage{amsmath}
\usepackage{amssymb}
\usepackage{epsfig}
\usepackage{color}
\usepackage{amsthm}
\usepackage[top=50pt,bottom=50pt,left=60pt,right=60pt]{geometry}
\usepackage{mathtools}
\usepackage[makeroom]{cancel}
\usepackage{hyperref}
\usepackage{appendix}
\usepackage{cancel}
\usepackage{stmaryrd}
\usepackage{soul}
\usepackage{qcircuit}
\usepackage{verbatim}
\usepackage[numbers,sort&compress]{natbib}

\newtheorem{theorem}{Theorem}

\newtheorem{corollary}{Corollary}

\newtheorem{definition}{Definition}

\input{MischaCommands}
\newcommand{\inpp}{\textup{L}}
\newcommand{\outpp}{\textup{P}}
\newcommand{\env}{\mathfrak{E}}
\newcommand{\code}{\textup{Co}}
\newcommand{\F}{\textup{F}}
\newcommand{\envcode}{\boldsymbol{\xi}}
\newcommand{\M}{{\scriptscriptstyle M}}

\newcommand{\K}{{\scriptscriptstyle K}}
\newcommand{\Nnew}{L}
\newcommand{\wor}{\textup{worst}}

\newcommand{\ketbra}[2]{|#1\rangle\langle #2|}

\newcommand{\tr}[0]{\textup{tr}}

\newcommand{\id}{{\mathbbm{1}}} 

\newcommand{\alv}[1]{{\color{red}Alv: #1}}

\begin{document}
\title{Continuous groups of transversal gates for quantum error correcting codes from finite clock reference frames}

\author{Mischa P. Woods}
\affiliation{Institute for Theoretical Physics, ETH Zurich, Switzerland}
\author{\'Alvaro M. Alhambra}
\affiliation{Perimeter Institute for Theoretical Physics, Waterloo, Canada}

\begin{abstract}
	Following the introduction of the task of \emph{reference frame error correction} \cite{hayden2017error}, we show how, by using reference frame alignment with clocks, one can add a continuous Abelian group of transversal logical gates to \emph{any} error-correcting code. With this we further explore a way of circumventing the no-go theorem of Eastin and Knill, which states that if local errors are correctable, the group of transversal gates must be of finite order. We are able to do this by introducing a small error on the decoding procedure that decreases with the dimension of the frames used. Furthermore, we show that there is a direct relationship between how small this error can be and how accurate quantum clocks can be: the more accurate the clock, the smaller the error; and the no-go theorem would be violated if time could be measured perfectly in quantum mechanics.  The asymptotic scaling of the error is studied under a number of scenarios of reference frames and error models. The scheme is also extended to errors at unknown locations, and we show how to achieve this by simple majority voting related error correction schemes on the reference frames. In the Outlook, we discuss our results in relation to the AdS/CFT correspondence and the Page-Wooters mechanism.
\end{abstract}

\maketitle

\section{Introduction and Overview of Results}\label{Introduction and Overview of Results}

In order to build a functional universal quantum computer, full fault-tolerance must be achieved. The idea behind fault-tolerance is that the errors that occur at particular points during the computation do not propagate or amplify along the whole computation to the point of being uncorrectable. Due to fundamental physical constraints such as no-cloning, achieving this is a notoriously challenging task, with a number of different requirements on how to prepare, manipulate, and protect the quantum states with error-correcting codes. One of the most desirable features of the codes used in fault-tolerant computation is the ability to apply logical gates \emph{transversally}, which one can implement while still being able to correct for local errors.

The framework for error correction is based on considering three spaces | a \textit{logical} $\mathcal{H}_\inpp$, \textit{physical} $\mathcal{H}_\outpp$ and a \textit{code} $\mathcal{H}_\code\subseteq \mathcal{H}_\outpp$ space.\footnote{Some authors use the convention of not considering the code space explicitly, in which case one sets $\mathcal{H}_\code= \mathcal{H}_\outpp$.} Logical states $\rho_\inpp$ containing quantum information are encoded via an encoding map $\mathcal{E}: \mathcal{B}(\mathcal{H}_\inpp)\rightarrow \mathcal{B}(\mathcal{H}_\code)$ onto the code space, which is a subspace of some larger physical space where errors | represented via error maps $\{\env_j\}_j: \mathcal{B}(\mathcal{H}_\code) \rightarrow \mathcal{B}(\mathcal{H}_\outpp)$ | can occur. Decoding maps $\{\mathcal{D}_j\}_j : \mathcal{B}(\mathcal{H}_\outpp)\rightarrow \mathcal{B}(\mathcal{H}_\inpp)$ can then retrieve the information while correcting for errors; outputting the logical state $\rho_\inpp$. That is:
 \begin{equation}\!\!
\Qcircuit @C=2em @R=.5em {\rho_\inpp \quad \quad & \gate{\mathcal{E}}& \gate{\env_j} & \gate{\mathcal{D}_j} & \qw \quad \quad \quad \quad \quad \rho_\inpp, \quad \quad \quad}\label{eq:perferct encoding deconding}
\end{equation}
for all $j$ and for all states $\rho_\inpp\in\mathcal{S}\left(\mathcal{H}_\inpp \right)$. Depending on the error model, the index $j$ indicating which error occurred may or may not be known. If it is unknown, the decoding map $\mathcal{D}_j$ cannot depend on $j$. We say that a logical gate $V_{\inpp}$ can be applied transversally if for any state $\rho$, the encoder $\mathcal{E}$ 
 is such that
\begin{equation} \label{eq:cov}
\mathcal{E}(V_\inpp \rho V_\inpp^\dagger)= V_\code^{\otimes \K} \mathcal{E}( \rho )V_\code^{\dagger \otimes \K},
\end{equation}
where the tensor product structure ``$\otimes \scriptstyle{K}$" represents the division of the code into different subsystems or ``blocks" in which errors can be independently corrected. This condition means that the action of the encoding map commutes with the action of the logical gate $V_\inpp$, which is represented by $V_{\code}^{\otimes \K}$ in the physical space. An interesting case to consider is that of codes $\mathcal{E}$ and groups $G$ for which all group elements (indexed by $g$) can be applied transversally using a unitary group representation $U_\inpp(g)$,
\begin{equation} \label{eq:cov 2}
\mathcal{E}(U_\inpp(g) \rho U_\inpp(g)^\dagger)= U_\code(g)^{\otimes \K} \mathcal{E}( \rho )U_\code(g)^{\dagger \otimes \K} \quad \forall g\in G.
\end{equation}
We will refer to codes whose encoding has this property as \emph{covariant} codes. 
There are a number of results that restrict the existence of such codes, in particular for stabilizer codes \cite{zeng2011transversality,bravyi2013classification,pastawski2015fault,jochym2018disjointness,brown2016quantum}. Most notably, the no-go theorem of Eastin and Knill \cite{eastin2009restrictions} states that in any finite-dimensional code (not necessarily stabilizer) in which local errors can be corrected, the groups of logical gates that can be applied transversally must be finite. This thus excludes dense sets of logical gates, as well as any continuous Lie subgroup of $\text{U}(d)$.


Since this no-go result imposes a fundamental limitation on the possible transversal gates, it is interesting to find ways of circumventing it, via schemes that do not satisfy all of the assumptions. A number of alternatives have been thoroughly explored, including the protocols of magic state distillation \cite{bravyi2005universal}, and other more specific schemes  (see for instance \cite{knill1996threshold,bombin2007topological,paetznick2013universal,jochym2014using,bombin2015gauge,yoder2016universal}), many of which propose a relaxation of the transversality condition in some fault-tolerant way.

Recently, a new kind of circumvention was put forward in \cite{hayden2017error}, where they show examples of codes with physical spaces of \emph{infinite dimension} that allow for the transversal implementation of Lie groups. The need for infinite dimensions (and seemingly infinite energy too, as we will soon discuss) limits their practical relevance, but the idea motivates the following question: do there exist large (but finite) covariant codes in which \emph{approximate} error correction can be performed? In other words, can we circumvent the no-go results by allowing for small errors that decrease with the size of the code? 

Here we explore this question using the notion of reference frames and clocks. We construct imperfect codes in which Abelian $\text{U}(1)$ groups can be implemented transversally. To that aim, we use a simple finite dimensional version of the encoding map from \cite{hayden2017error}, which shows that perfect covariant codes are possible provided one has access to a \emph{perfect} reference frame. A perfect reference frame \cite{bartlett2007reference} is defined as a quantum system that encodes, without error, information about a particular group element. 
That is, given a group representation $U(g)$, such that $U(0)=\id$, the state $\ket{\psi}$ is a perfect reference frame iff $\forall \,g$
\begin{equation}\label{eq:dela g} 
\bra{\psi} U(g) \ket{\psi}= \delta_{0,g},
\end{equation}
where $\delta_{0,g}$ is a Kronecker delta for finite groups and Dirac delta in the case of Lie groups.
Hence, each point of the orbit $\ket{\psi(g)}=U(g) \ket{\psi}$ is orthogonal to all the others (and thus perfectly distinguishable). 

This connects with the work by Pauli on quantum clocks in the case of an Abelian U$(1)$ groups. If the group $\{U(g)\}_g$ is a one-parameter compact Lie group generated by solving the Schr\"odinger equation, namely $U(g)=\me^{-\mi g \hat H}$ for some Hamiltonian $\hat H$, where $g=t$ is the time transcribed, then the constraint Eq. \eqref{eq:dela g} is analogous to requiring the existence of a \emph{perfect} time operator $\hat t$. Indeed, defining
\be \label{time op idealised}
\hat t= \int_G dg\, g\, \ketbra{\psi(g)}{\psi(g)},
\ee 
where the integral is over the Haar measure, one finds $\hat t \ket{\psi(g)}= g \ket{\psi(g)}$ for all $g\in G$.
Pauli \cite{pauli1,pauli2} already concluded that such time operators require quantum systems that | while mathematically well defined | cannot exist as they require infinite energy.\footnote{The infinite energy manifests itself either with unbounded from below Hamiltonians (when $t$ belongs to an unbounded interval) or infinitely strong potentials producing strict confinement of the wave function (for when $t$ belongs to an bounded interval)~\cite{Garrison1970}.} These are known as \textit{Idealised} clocks. However, this does not rule out the existence of \emph{approximate} time operators $\hat t$ and initial clock states $\ket{\psi}$ which serve as a reference frame for the observable $t$,  \cite{salecker1958quantum,peres1980measurement,woods2016autonomous} either in the form of a stopwatch \cite{PhysRevLett.82.2207} or ticking clock \cite{Pauletal2017,RaLiRe15,woods2018quantum}

Here we explore how certain finite-sized reference frames (which we call ``clocks") can be used to build imperfect covariant codes, and we give upper bounds to the errors induced by their finite size. We use the construction of \textit{Quasi-Ideal} clocks \cite{woods2016autonomous} and \textit{Salecker-Wigner-Peres} (SWP) clocks \cite{salecker1958quantum,peres1980measurement}, to provide simple encoding and decoding protocols based on the task of reference frame alignment. We show using Quasi-Ideal clock states entangled over $L$ subsystems, that the worst-case entanglement fidelity between the input $\rho_\inpp$ and decoded output  
$\mathcal{D}_j(\env_j\left(\mathcal{E}_{\textup{cov}}(\rho_\inpp)\right))$ denoted $f_\wor$, and defined in Section \ref{sec:entfid}, in our protocol satisfies (up to log factors) the lower bound $1-f_\wor\leq \bo( 1/(L {d_\clo})^2)$ where $L$ is the number of entangled Quasi-Ideal clocks and $d_\clo$ is their dimension. In \cite{faist2018prep} the generic upper bound $1-f_\wor \geq \bo(1/(L {d_\clo})^2)$ is derived for all covariant encoding maps generated by isometries. 
A direct consequence of the combination of our results with those of \cite{faist2018prep} adapted to our setting, is that \emph{all} error correcting codes $\{\map,\{\env_j,\mathcal{D}_j\}_j\}$, can be made covariant w.r.t. the $\text{U}(1)$ groups considered here with an optimal fidelity $f_\wor$ (largest possible value) satisfying 
\be 
\bo\left(\frac{1}{(L{d_\clo})^2}\right) \leq 1-f_\wor  \leq \bo\left( \frac{\ln^{6}(L{d_\clo})}{(L{d_\clo})^2} \right),
\ee 
for a large $L d_\clo$. 

In order to study the role of different resources in our protocols, we define $t$-incoherent clock states $\rho_\clo$ as those for which there exits $g\in G$ such that their group evolved initial state, $U_\clo(g)\rho_\clo U_\clo^\dag(g)$, commutes with the projective measurements used in our protocol to measure them.\footnote{See text around Eq. \eqref{eq:t-incoherent clock def} for full definition.}
As we later explain, these states require minimal coherent resources to be created in comparison with other clock states. 
 We find that all codes which use $t$-incoherent clock states satisfy the upper bound $1-f_\wor\geq \bo( 1/(L {d_\clo} ))$. We then consider the more commonly used SWP clocks (see for instance \cite{salecker1958quantum,peres1980measurement,LUNARDI2011415,PhysRevA.96.022120,PhysRevA.79.012110,Teeny2016} and references therein), which belong to this class, and show that they yield an error of order $1-f_\wor= \bo( 1/(L {d_\clo}))$ hence saturating the bound for $t$-incoherent clocks. However, we also show that while coherent clock states are necessary to achieve $1-f_\wor\leq \bo( 1/(L {d_\clo} ))$, they are not sufficient, since finite dimensional analogues of coherent states can only achieve the same scaling as the SWP clocks. 

The results discussed so far consider codes in which the error model on the clock is just erasure at known locations. However, what if one cannot discern the location at which the error occurred?  Our final result is to prove that one can also correct local unknown phase errors which occur at an unknown location in the clocks. For this case, we are able to achieve $1-f_\wor\leq \bo( 1/(L {d_\clo}))$ up to log factors.




\section{Covariant codes based on reference frames}
\label{sec:perfectcode}

We now outline the generic error correction scheme we consider, which is based on a generalisation of the reference frame-based scheme in \cite{hayden2017error}. 
Let $\left\{\map,\{\env_j, \mathcal{D}_j\}_j \right\}$ be any perfect error correcting code satisfying Eq. \eqref{eq:perferct encoding deconding} (not necessarily covariant). 

Now, let $\rho_\F^{(\M)}$ be a reference frame for a group isomorphic to $\{U_\inpp(g)\}_g$ and $\{U_\code(g)\}_g$, with unitary group representation $\{U_\F^{\otimes \M}(g)\}_g$. 
 We define the following for all $g\in G$
\begin{equation}
\mathcal{E}_g(\cdot) := U_\code(g)^{\otimes \K}\mathcal{E}(U_\inpp^\dagger (g)(\cdot) U_\inpp (g))U_\code(g)^{\dagger \otimes \K}.
\end{equation}
The covariant code is then:
\begin{equation}\label{eq:code}
\mathcal{E}_{\text{cov}}(\cdot) := \int_G \text{d} g \,  \mathcal{E}_g (\cdot) \otimes U_\F(g)^{\otimes \M}\,\rho_{\F }^{(\M)}\,U_\F(g)^{\dagger\otimes \M},
\end{equation}
where $\text{d} g $ is the uniform or Haar measure over the group.\footnote{It is noteworthy that all our results hold for all $\K$ | the number of gates being applied transversally. 
In fact, one can choose $\K=1$ when considering a physical space in which the gates are not applied locally.} Note that $\mathcal{E}_{\text{cov}}$ is from $\mathcal{H}_\inpp$ to $ \mathcal{H}_\outpp\otimes\mathcal{H}_\F$ while usually the encoding map takes logical states to states in the physical space. As such, it is convenient to think of the reference frames as an extension of the code space to $\mathcal{H}_{\widetilde{\code}}:= \mathcal{H}_\code\otimes\mathcal{H}_\F$. Moreover, the channel $\mathcal{E}_{\text{cov}}$ is now covariant in the sense of Eq. \eqref{eq:cov 2} if the symmetry group on the r.h.s. is defined in the extended physical space, namely $U_\code(g)^{\otimes \K}\otimes U_\F(g)^{\otimes \M}$, so that $U_\inpp(g)$ can be implemented transversally. 
This way, the notion of covariance defined in Eq. \eqref{eq:cov 2} is now over $\mathcal{H}_{\widetilde\code}$  rather than $\mathcal{H}_{\code}$ alone. We note, however, that unlike the code subspace, none of the logical information is encoded directly into frames $\F$, and they merely serve as a reference for $g$, for which both the group $\{U_\F^{\otimes \M}\}_g$ and initial frame state $\rho_\F^{(\M)}$ can be \emph{chosen} at our convenience to optimise our protocol. The purpose of the error maps $\env_j$ is that they take information from the code space and ``mix" it with the rest of physical space resulting in errors. On the other hand, the reference frames play the following role in the encoding map $\map_\textup{cov}\;\!$: the logical information is still encoded into the code space, but now the information about \textit{which} transversal gate is applied to it is encoded in the reference frames, hence without knowledge from the reference frames during the decoding, it would not be possible to discern which gates had been applied. If the reference frames are damaged via an error (which is a reasonable assumption since they are now part of the extended code space), this will inhibit the ability to decode correctly resulting in a worse decoding of the logical information. 

The protection of the information to be encoded relies on the channel $\mathcal{E}$, which is arbitrary. On the other hand, the protection for the reference frame in the enconding of Eq. \eqref{eq:code} will come merely from having a few copies of them which may or may not be entangled. In the case of no entanglement, $\rho_{\F }^{(\M)}=\rho_{\F }^{\otimes \M}$, the states are only classically correlated due to the twirling over $G$ in Eq. \eqref{Thm:3 clock phase error}, and thus bears strong analogies with a classical repetition code. 

The encoding-error-decoding procedure consists of the following steps:
\begin{itemize}
	\item [1)] An arbitrary state $\rho_\inpp$ is encoded as $\mathcal{E}_{\textup{cov}}(\rho_\inpp)$.
	\item [2)] Errors occur: The error maps considered here are now of the form $\env_{j,q}=\env_j\otimes \envcode_{q}$, with $\envcode_{q}$ acting on the reference frames and the index $q$ | as with $j$ | may or may not be known depending on the error model. This may include the loss of up to $M-1$ frames, or in the case of Theorem \ref{Thm:3 clock phase error}, an unknown phase error at an unknown location.
	\item [3)] Frames are measured: In the case of erasure errors at known locations of the reference frames, only the $N$ non erased clocks are measured. The remaining $N$ frames are measured projectively in some basis, and after tracing out the reference frames, we write
	\begin{align}\label{eq:stage1E}
	\quad	\tr_\F[\env_{j,q}\left(\mathcal{E}_{\text{cov}}(\rho_\inpp)\right)]=	\env_j \left( \mathcal{E}_{g'} (\rho_\inpp) \right)+\hat E_{g'}'',
	\end{align}
	where $\hat E''_{g'}$ represents the error term and $g'$ is chosen based on the measurement outcomes, the initial clock states and the error maps $\envcode_q$. If the reference frames are good, the error $\hat E_{g'}''$ will be small since the measurements will be able to distinguish approximately the group elements. Indeed, in the case of perfect reference frames Eq. \eqref{eq:dela g} or equivalently the Idealised clock Eq. \eqref{time op idealised}, the optimal choice of $g'$ leads to no error, $\hat E_{g'}''=0$.

	\item [4)] Finally, we apply the decoding map $\bar{\mathcal{D}}_{j,{g'}}$ defined as
	\begin{align}
		\qquad\bar {\mathcal{D}}_{j,g}(\cdot)&=\\
		\qquad U_\inpp(g) &\map^{-1}\left( U_\code^{\dag\otimes\K} (g) \map\left( \mathcal{D}_j(\cdot) \right) U_\code^{\otimes\K}(g) \right) U_\inpp^\dag(g),\nonumber
	\end{align}
	where $\mathcal{D}_j\left(\env_j(\cdot)\right)= \map^{-1}(\cdot)$, with $\map^{-1}$ being the inverse channel for the encoder $\map$ (that is, the channel that undoes the encoding when no errors occur).
	
	This achieves
	\be 
	\bar{\mathcal{D}}_{j,g'}\big(\tr_\F[\env_{j,q}\left(\mathcal{E}_{\text{cov}}(\rho_\inpp)\right)]\big)=\rho_\inpp+\hat E_{g'},
	\ee 
	where $\hat E_{g'}$ is the final error term. 
\end{itemize}

The circuit diagram is as follows.

\begin{equation}\scalebox{0.8}{\quad \quad
\Qcircuit @C=0.8em @R=1em {
& & \mbox{encoding} & & \mbox{error} & & \mbox{Decoding} & & 
\\ 
\rho_\inpp & & \multigate{2}{\mathcal{E}}  & \multigate{5}{\mathcal{E}_{\text{cov}}} & \multigate{2}{\env_j} & \multigate{2}{U_\code(g')^{\otimes \K}} & \multigate{2}{\bar{\mathcal{D}}_{j,g}} & & 
\\  
\ket{0}^{\otimes \K-1 \,\,\,} & & \ghost{\mathcal{E}}  & \ghost{\mathcal{E}_{\text{cov}}} & \ghost{\env_j} & \ghost{U_\code(g')^{\otimes \K}} &\ghost{\bar{\mathcal{D}}_{j,g}} &  \qw &\,\quad\quad\rho_\inpp\!+\!\hat E_{g'}
\\
& & & & & & & & 
\\
& \ket{\includegraphics[width=0.022\textwidth]{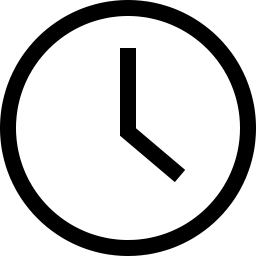}} &  & \ghost{\mathcal{E}_{\text{cov}}} &  \meter   & \cctrlo{-1} & \cctrlo{-1} &  & 
\\
& \ket{\includegraphics[width=0.022\textwidth]{clockimage3.png}} &  & \ghost{\mathcal{E}_{\text{cov}}} & \measureD{\includegraphics[width=0.022\textwidth]{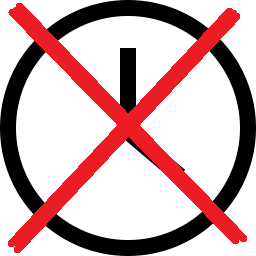} }  & & & & 
\\
& \ket{\includegraphics[width=0.022\textwidth]{clockimage3.png}} &  & \ghost{\mathcal{E}_{\text{cov}}} & \measureD{\includegraphics[width=0.025\textwidth]{clockimage3no.png} } &  & & & 
}} \nonumber
\end{equation}

 This scheme is closely related to the task of \emph{reference frame alignment}, in which the group of ``transversal gates" corresponds to the unknown rotation that happens between two parties, here the encoder and decoder. This is such that the channel connecting them is effectively a decoherence channel. The unknown rotation is inferred by measuring the frame (see \cite{bartlett2007reference,bartlett2009quantum} for further details). Reference frame alignment can in fact be thought of as the particular case where the encoding, error and decoding are identity channels. We discuss this connection further in \App~ \ref{app:quantumcomm}, where we show that the error probability using the Quasi-Ideal clock is $(\ln d_\clo)^3/d_\clo^2$ |  a quadratic improvement over the SWP clock, which only achieves $\propto 1/d_\clo$ as shown in \cite{bartlett2009quantum}.
 
Another question of relevance is what happens if the original encoding $\mathcal E$  already had a group of transversal gates $\{V_\inpp,V_\code\}$ satisfying Eq. \eqref{eq:cov}. If $[V_\inpp,U_\inpp(g)]=0$ and $[V_\code, U_\code(g)]=0$ for all $g\in G$ then these gates are also transversal for $\map_\textup{cov}$. Moreover, while these gates are not transversal for $\map_\textup{cov}$ in general, the gates $\{U_\inpp(g') V_\inpp U_\inpp^\dag(g'),U_\code(g') V_\code U_\code^\dag(g')\}$, which can be viewed as merely a change of basis on the original gates, are | up to a small error $E''_{g'}$ | pairs of transversal gates on $\map_\textup{cov}$ after the reference frames have been measured and a value $g'$ has been discerned in step 3); which is performed before decoding. See \App~ \ref{app:tranversal} for details.

\subsection{Finite-sized clocks}\label{sec:finite}

We now detail the unitary group representations to which our results apply to.
For the logical and physical spaces, we consider all compact representations of the Abelian U$(1)$ group. These can be written in the form $U_\inp(t)=\me^{-\mi t \hat H_\inp}$, $U_\code^{\otimes \K}(t)=e^{-i t \hat H_\code}$ respectively, where using the shorthand $\textup{S} \in \{\inpp, \code\}$, the generators are $\hat H_\textup{S}=\sum_{n=0}^{d_\textup{S}-1} \omega h_{\textup{S},n}  \ketbra{n}{n}_\textup{S}$ for a $d_\textup{S}$ dimensional space, for some frequency $\omega>0$. They have fixed gaps but arbitrary degeneracy, so that  $h_{\textup{S},n} \in \{0,1,2,\ldots,\Delta h_\textup{S}\}$ with $\nn\ni\Delta h_\textup{S}\le d_\textup{S}$. As we will see, the performance improves as the range $\Delta h_\code$ decreases, so in fact the best choice is a trivial generator $\hat H_{\code}\propto \id$, for which we can set $\Delta h_\code=0$.
 As motivated by our discussion of Pauli's findings, in this context, one can think of the group elements $U_\F(g)$ as providing the dynamics of quantum clocks $\rho_\F$ where $g=t$ is time. We will therefore refer to the reference frame as a ``clock" and use the labels $\F$ and $\clo$ interchangeably in this context. The clock is chosen as the compact non-degenerate representation, namely $U_\clo(t)=\me^{-\mi t \hat H_\clo}$ with  
$\hat H_{\clo}=\sum_{r=0}^{d_{\clo}-1} \omega r \ketbra{r}{r}_\clo$ where $\{\ket{r}_\clo\}$ is an arbitrary orthonormal basis. Physically, energy gaps which are all integer multiples of $\omega$ are required to create degeneracies between systems $\inpp,\code,\clo$. Technically, it means that the group representations on $\inpp$, $\code$, $\clo$ are compact, and that the irreps of $U_C(g)$ contain those of $U_\inpp (g)$ and $U_\code (g)$. The compactness allows replacement of the integration range in Eq. \eqref{eq:code} with $[0,T_0]$ where $T_0=2\pi/\omega$ is the recurrence time and the measure becomes $\text{d}g=dt/T_0$.

The main clocks we use in Eq. \eqref{eq:code} are based upon the Quasi-Ideal clocks  \cite{woods2016autonomous} and the SWP clocks \cite{peres1980measurement}. They both share the same Hamiltonian $\hat {H}_\clo$. The definition of SWP clock states are simple, they are any one of the pure states of the Fourier-transformed basis of eigenbasis of $\hat H_\clo$, namely $\rho_{\textup{SWP}}= \ketbra{\theta_k}{\theta_k}$ with
\be \label{eq:SWP clcok states}
\ket{\theta_k}=\frac{1}{\sqrt{{d_\clo}}}\sum_{j=0}^{{d_\clo}-1} \me^{-\mi 2\pi n j/{d_\clo}}\ket{j}_\clo,
\ee 
where $\ket{\theta_k}=\ket{\theta_{k \mod d.}}$, $k\in\zz$; and is referred to as the \emph{time basis}. On the other hand, the Quasi-Ideal clock states $\rho_{\textup{QI}}= \ketbra{\Psi(k_1^0)}{\Psi(k_1^0)}$ are defined as a coherent superposition of SWP clocks,
\begin{align}\label{eq:clockdef}
\ket{\Psi(k_1^0)}=\sum_{k \in \mathcal{S}_{{d_\clo}}(k_1^0)}A \me^{-\frac{\pi}{\sigma^2}(k-k_1^0)^2 }\me^{\mi 2\pi n_0 (k-k_1^0)/{d_\clo}} \ket{\theta_k},\nonumber
\end{align} 
where $A$ is a normalization constant, $\mathcal{S}_{{d_\clo}}(k_1^0)$ is the set of ${d_\clo}$ consecutive integers centred about $k_1^0\in\rr$ and $\omega n_0\in (0,\omega {d_\clo})$ is approximately the mean energy of the clock. $k_1^0$ can be thought of as an approximate initial time marked by the clock. The parameters $n_0$, $k_1^0$ and $\sigma\in(0,{d_\clo})$ can be tuned to our convenience.

Both the Quasi-Ideal and SWP clocks are usually associated with the same projective operators in the ``time" basis, defined as
\be \label{eq:time operator}
\hat {t}_\clo=\sum_{k=0}^{d_\clo-1} k\frac{T_0}{d_\clo} \ketbra{\theta_k}{\theta_k}.
\ee 
When measuring individual clocks, we will do so in this time basis. When the clocks are entangled, we will use a time operator of this form but with the projectors $\ketbra{\theta_k}{\theta_k}$ replaced with non-local projectors over the multiple entangled clocks, as explained later. For a detailed description of the measurement protocols, see \App~ \ref{app:gen}.

The Quasi-Ideal clock states have been shown to yield a good performance in the context of quantum control \cite{woods2016autonomous} and measurement of time \cite{woods2018quantum}. On the other hand, in \cite{woods2016autonomous,woods2018quantum} the SWP clocks appeared to be suboptimal. For the task at hand, we will prove that this difference of performance still occurs. 

\subsection{Entanglement fidelity}\label{sec:entfid}
The figure of merit we will use for quantifying the performance of codes which cannot decode perfectly is the worst-case entanglement fidelity. Given a sequence of encoding, error and decoding which we label as the channel $\mathcal{K}: \mathcal{H}_{{\inpp}} \rightarrow \mathcal{H}_{\inpp}$, the entanglement fidelity is defined as
\begin{equation}
f_{\wor}(\mathcal{K})=\min_{\phi} \bra{\phi} \mathcal{K} \otimes \mathcal{I} (\ket{\phi}\bra{\phi}) \ket{\phi}, 
\end{equation}
where $\mathcal{I}$ is an identity channel acting on a copy of $\mathcal{H}_{\inpp}$, and the optimization is over all bipartite states on which the map $\mathcal{K} \otimes \mathcal{I}$ acts.\footnote{Note that it is sufficient to take an ancillary space of size $d_\inpp$.} Having a perfect code as those satisfying Eq. \eqref{eq:perferct encoding deconding}, is equivalent to $f_{\wor}(\mathcal{K})=1$, which is achieved with perfect reference frames or Idealised clocks.
\noindent For a justification of why this is a good measure of approximate codes, see for instance \cite{knill1997theory,nielsen1996entanglement}. 

The channel $\mathcal{K}$ that we consider here is for the error correction codes $\{\map_\textup{cov},\{\env_{j,q},\tilde{\mathcal{D}}_{j,q}\}_{j,q}\}$ as described in section \ref{sec:perfectcode}, in which we use clocks that are not idealised.

\section{Results}\label{Results}
For simplicity, we will only state our bounds to leading order in $d_\clo,d_\outpp,d_\inpp,L$. Here $d_\outpp\geq d_\code$ is the dimension of the physical space and $L$ is introduced later. 
Except for Theorem \ref{Thm:3 clock phase error}, explicit bounds | not just the leading order terms | can be found  in the~\app. Furthermore, in all theorems and corollaries in which the Quasi-Ideal clock is involved, the width $\sigma$ scales logarithmically with $d_\clo$. The exact value can be found in the proofs.

Our first result covers the case in which at least one of the clocks is left untouched by erasure errors. 

\begin{theorem}\label{re:1clock}
	Consider the covariant code $\map_{\textup{cov}}$ in Eq. \eqref{eq:code}, with $M$ Quasi-Ideal clocks, namely
	\be
	\mathcal{E}_{\textup{cov}}(\cdot) 
	= \frac{1}{T_0}\int_0^{T_0} dt \,  \mathcal{E}_t (\cdot) \otimes U_\clo(t)^{\otimes \M}\,\rho_{\textup{QI}}^{\otimes\M}\,U_\clo(t)^{\dagger\otimes \M},\label{eq:cov code for N non entangled}
	\ee
	with error channels $\{\env_{j,q}=\env_j\otimes \envcode_{q}\}_{j,q}$, where $\{\envcode_q\}_q$ are erasure at known locations of at most $M-1$ clocks and $\{\env_j\}_j$ are the error channels for the code $\map$. Then for all $M\in\nn^+$, and for all error correcting codes $\left\{\map,\{\env_j, \mathcal{D}_j\}_j \right\}$, there exists decoding channels $\tilde{\mathcal{D}}_{j,q}$ for $\map_{\textup{cov}}$ such that the worst-case entanglement fidelity of the covariant encoding satisfies
	\begin{align}\label{eq:f worst result 1}
	f_{\wor} \ge 1&- \frac{3 \pi \sqrt{d_{\inpp}}d_{\outpp}}{4}\left(\frac{\ln^{3}({d_\clo})}{{d_\clo}}\right)^2 (\Delta h_\inpp+\Delta h_\code)^2 \nonumber \\&+\mathcal{O}\left(\sqrt{d_{\inpp}}d_{\outpp} \frac{\ln^{3/2}({d_\clo})}{{d_\clo}^3}\right).
	\end{align}
\end{theorem}
	For the proof, we need to write the encoding-error-decoding scheme in the form described in \app~\ref{app:gen} when we have a single clock unaffected by errors. After working out the decoding protocol explicitly, we find a bound that, together with the results on the entanglement fidelity of \app~\ref{app:entfid} allows us to derive the Theorem in \app~ \ref{app:1clockproof}, after choosing $\sigma=\ln^{3}({d_\clo})$ for the single clock we measure.


In order to study the role that the different resources play in the current scheme, consider the following definitions. We call a clock state $\rho_\textup{inc}\in\mathcal{S}(\mathcal{H}_\inpp)$, with time operator $\hat t_\clo$ and unitary group representation $\{U_\clo(t)\}_{t\in\rr}$,   $t$-\emph{incoherent} if there exists a state in its periodic orbit which commutes with the time operator, namely if there exists $t_0\in\rr$ such that 
\begin{align}\label{eq:t-incoherent clock def}
[U_\clo(t_0)\rho_\textup{inc}U_\clo^\dag(t_0),\hat t_\clo]=0,
\end{align} where $\hat t_\clo$ is the operator used in step 3) to measure the clock.\footnote{Note that in the case of time operators which are not projective POVMs, $t$-incoherent clocks can still be defined by replacing $\hat t_\clo$ by the 1st moment operator of the POVM. We do not need to consider such generalisations here however.} Conversely, clock states which are not $t$-incoherent are called $t$-\emph{coherent}.

Observe that if one can construct the code $\map$ and decoders $\mathcal{D}_j$, then for a given initial clock state $\rho_\clo^{(\M)}$, in order to construct the covariant code $\map_\textup{cov}$ in Eq. \eqref{eq:code}, one needs to be able to apply the transversal gates to the clock and code $\map$, and create classical correlations (a separable state) between them. For the decoders $\tilde{\mathcal{D}}_j$, the only additional resource required to apply them is the ability to measure the clocks projectively in the time basis. The ability to perform such operations is always required for the construction of any error correction code $\{\map_\textup{cov},\{\env_{j,q},\tilde{\mathcal{D}}_{j,q}\}_{j,q}\}$. In the case of $t$-incoherent clocks, the initial clock state $\rho_\clo^{(\M)}$ can then easily be constructed by simply measuring any state on the clock Hilbert space and applying the appropriate traversal gate. However, in the case that the initial clock state $\rho_\clo^{(\M)}$ is $t$-coherent (such as in the case of Quasi-Ideal clock states in Theorem \ref{re:1clock}), one cannot constructed it with any of the above mentioned operations since they do not allow for the creation of the necessary coherence (that is, with respect to the basis of $\hat t_\clo$). Therefore, from a resource-theoretic standpoint, it is interesting to characterise how good codes can be if they only use $t$-incoherent clocks:

\begin{theorem}\label{re:diag}
	Consider the covariant code $\map_{\textup{cov}}$ in Eq. \eqref{eq:code}, with one $d_\clo$ dimensional $t$-incoherent clock $\rho_\textup{inc}$, with time operator $\hat t_\clo$ defined in Eq. \eqref{eq:time operator}, 
	\be
	\mathcal{E}_{\textup{cov}}(\cdot) 
	= \frac{1}{T_0}\int_0^{T_0} dt \,  \mathcal{E}_t (\cdot) \otimes U_\clo(t)\,\rho_\textup{inc}\,U_\clo^\dag(t).\label{eq:dia clocks thorem}
	\ee
	Allow for no errors on the clock, $\{\env_{j,q}=\env_j\otimes \envcode_{q}=\env_j\otimes \mathcal{I}\}_{j,q}$. Then for all error correcting codes $\left\{\map,\{\env_j, \mathcal{D}_j\}_j \right\}$, the entanglement fidelity that can be achieved is upper bounded by
	\begin{equation}\label{eq:f worst result 2}
	f_{\wor} \le 1- \frac{C}{{d_\clo}},
	\end{equation}
	where $C>0$ is independent of ${d_\clo}$. Moreover, there exists a scheme using the Salecker-Wigner-Peres clock, $\rho_\textup{inc}=\rho_\textup{SWP}$ that achieves
	\begin{equation}\label{eq:f worst result 2 2}
	f_{\wor} = 1- \frac{C^*}{{d_\clo}},
	\end{equation}
	where $C^* \ge C$ is also independent of ${d_\clo}$.
\end{theorem}
The full proof is shown in \app{} \ref{app:diagonalclockproof} and it goes as follows: we write explicitly the state of the code after the clock has been measured, and we apply an arbitrary decoder. Then we write the entanglement fidelity explicitly and show that the error term decays at best linearly with the dimension of the clock. The performance achieved by the SWP clock is calculated by looking at the scaling of the errors in the decoding scheme of Section \ref{sec:perfectcode}.

The setup in the above Theorem is the same as in Theorem \ref{re:1clock} in the special case that $M=1$ and one exchanges the Quasi-Ideal clock for a $t$-incoherent one. So by comparing Eqs. \eqref{eq:f worst result 1} and \eqref{eq:f worst result 2} we see that there is essentially a quadratic improvement in the scaling w.r.t. the clock dimension $d_\clo$. This demonstrates the necessity of $t$-coherent 
clock states to achieve the improved scaling in our protocols. Furthermore, in all our protocols, the final state of the clock after applying a decoder $\tilde{\mathcal{D}}_j$ is $t$-incoherent and thus in protocols in which the initial clock state was $t$-coherent, the coherence was ``used up" in the process.

Interestingly, we have observed the same effective quadratic advantage by using the Quasi-Ideal clock rather than the SWP clock for the related task of reference frame alignment (See Section \ref{sec:perfectcode} and \App~\ref{app:quantumcomm}) and ticking clocks \cite{woods2018quantum}.

While $t$-coherent clock states are necessary to achieve the improved scaling, they are clearly not sufficient. A case in point are clock states which are incoherent in the energy basis. Indeed, by inverting Eq. \eqref{eq:SWP clcok states} one observes that they are $t$-coherent yet since they do not evolve in time are useless for this task. 
	
Changing the width $\sigma$ of the Quasi-Ideal clock state allows one to understand these differences better. The uncertainty in both	the energy and time basis, denoted $\Delta E$ and $\Delta t$ respectively, satisfy $\Delta E \Delta t = 1/2$ to leading order in $d_\clo$ for all Quasi-Ideal clock states. They are thus approximately minimum uncertainty states. By changing the value of $\sigma/\sqrt{d_\clo}$ from small to large one can achieve the limiting cases of a time eigenstate $\Delta E \gg \Delta t$, and an energy eigenstate $\Delta E \ll \Delta t$. Eq. \eqref{eq:f worst result 1} in Theorem \ref{re:1clock} corresponds to the optimal value of $\sigma=\ln^{3}({d_\clo})$, which corresponds to states which are time squeezed ($\Delta E \geq \Delta t$), but not by too much | the states do not become ($t$-incoherent) time eigenstates. On the other hand, Quasi-Ideal clock states which are energy squeezed ($\Delta E \leq \Delta t$), yield an entanglement fidelity $f_{\wor}$ which scales with $d_\clo$ even worse than $t$-incoherent states. In-between, one finds the non squeezed ($\Delta E = \Delta t$) Quasi-Ideal states which have the same $d_\clo^{-1}$ scaling as the $t$-incoherent SWP clock. 

The latter Quasi-Ideal clock states behave analogously to ``classical" coherent states | their expectation values in the time and energy bases oscillate like simple harmonic oscillators, and they minimize the Heisenberg uncertainty with equal uncertainty in each basis. However, the analogy ends here. The time squeezed Quasi-Ideal clock states \textit{remain} time squeezed under the application of $U_\clo(t)$ for all $t\in\rr$, while squeezed coherent states in one quadrature (e.g. position) become squeezed in the complementary quadrature (e.g. momentum) under evolution of its Hamiltonian | broadening in the initial quadrature basis. Intuitively, this is expected to be an important difference between squeezed coherent states and squeezed Quasi-Ideal clock states, at least regarding the present task. This is because good decoding maps would require the states to be squeezed during the entire periodic orbit in the basis in which they are measured in step 3) of the decoding protocol. 

We also find a significant improvement to the entanglement fidelity when one has access to a larger number of  clocks. To achieve it, we embed a large entangled clock within the Hilbert space of $\Nnew\in\nn^+$ smaller ones, effectively creating a clock of dimension $d(L):= L(d_\clo-1)+1$. 

\begin{theorem}\label{re:Nclock} Given a covariant error correction code $\{\map_\textup{cov},\{\env_{j,q},\tilde{\mathcal{D}}_{j,q}\}_{j,q}\}$ as described in section \ref{sec:perfectcode}, with entanglement fidelity $f({d_\clo})\leq f_{\wor} \leq f'({d_\clo})$ in which a single ${d_\clo}\in\mathcal{S}_N \subseteq \nn^+$ dimensional clock is used, there exists another covariant error correction code with the same error channels $\{\env_{j,q}\}_{j,q}$ in which $\Nnew$ clocks of dimension $d_\clo$ are entangled, such that $f(\Nnew({d_\clo}-1)+1)\leq f_{\wor} \leq f'(\Nnew({d_\clo}-1)+1)$, for all $\Nnew,d_\clo\in\nn^+$, s.t. $L(d_\clo-1)+1\in \mathcal{S}_N$.
\end{theorem}	
The embedding needed for this theorem leads naturally to an entangled discrete Fourier transform basis on the Hilbert space of $\Nnew$ clocks
\begin{equation}
\ket{\theta_k(\Nnew)}=\frac{1}{\sqrt{d(\Nnew)}}\sum_{n=0}^{d(\Nnew)} \me^{-\mi 2 \pi n{k}/{d(\Nnew)}}\ket{E_{n,1}},
\end{equation}
where $\ket{E_{n,1}}$ are a non-degenerate set of $\Nnew$ eigenvectors of the generator $\bigoplus_{i=1}^{\Nnew} \hat H_\clo $ \footnote{The symbol $\bigoplus$ denotes the Kronecker sum.} 
 (see \app~\ref{app:Nclock} for details).

When this embedding is applied to the SWP clock and Quasi-Ideal clock, it gives rise to a $\Nnew$-site SWP Entangled clock state, $\rho_{\textup{SWPE},\Nnew}=\ketbra{\theta_k(\Nnew)}{\theta_k(\Nnew)}$ and an $\Nnew$-site Quasi-Ideal Entangled clock, $\rho_{\textup{QIE,} \Nnew}=\ketbra{\Psi^\Nnew(k^0_1)}{\Psi^\Nnew(k^0_1)}$,
\begin{align}
\ket{\Psi^\Nnew(k^0_1)}=\sum_{k \in \mathcal{S}_{d(\Nnew)}(k^0_1)}A e^{-\frac{\pi}{\sigma^2}(k-k^0_1)^2 }e^{i 2\pi n_0 \frac{(k-k^0_1)}{d(\Nnew)}} \ket{\theta_k(\Nnew)},\nonumber
\end{align}
with the new time operator,
\be \label{eq:time operator new}
\hat {t}_\clo(\Nnew)=\sum_{k=0}^{d_\clo-1} k\frac{T_0}{d_\clo} \ketbra{\theta_k(\Nnew)}{\theta_k(\Nnew)}.
\ee

The combination of Theorem \ref{re:Nclock} with Theorems \ref{re:1clock}, \ref{re:diag}, yields an important corollary:
\begin{corollary}\label{re:Nclockoptimal}
	Consider the setup in Theorem \ref{re:1clock}, but with $M$ $\Nnew$-site Quasi-Ideal Entangled clocks $\rho_{\textup{QIE,\Nnew}}^{\otimes \M}$, rather than  $M$ Quasi-Ideal clocks and the erasure now being on at most $M -1$ of the $\Nnew$-site Quasi-Ideal Entangled clocks. The worst-case entanglement fidelity of the covariant encoding now satisfies 
	\begin{align}\label{eq:new coherent with L copies}
f_{\wor} \ge 1&- \frac{3 \pi \sqrt{d_{\inpp}}d_{\outpp}}{4}\left(\frac{\ln^{3}(\Nnew\, {d_\clo})}{\Nnew\, {d_\clo}}\right)^2 (\Delta h_\inpp+\Delta h_\code)^2 \nonumber \\&+  \mathcal{O}\left(\sqrt{d_{\inpp}}d_{\outpp} \frac{\ln^{3/2}(\Nnew\,{d_\clo})}{(\Nnew\,{d_\clo})^3}\right).
\end{align}	
Similarly, consider the setup in Theorem \ref{re:diag}, but with a $d_\clo$ dimensional $t$-incoherent clock state $\rho_{\textup{inc},\Nnew}$, with a time operator $\hat t_{\clo}(\Nnew)$, rather than a SWP clock with time operator $\hat t_\clo$. The worst-case entanglement fidelity of the covariant encoding now satisfies
\begin{equation}\label{eq:f worst result 2 3}
f_{\wor} \le 1- \frac{C}{{\Nnew\,d_\clo}},
\end{equation}
where $C$ is independent of $d_\clo$, $\Nnew$; and equality is obtained for the $\Nnew$-site SWP Entangled clock $\rho_{\textup{SWPE},\Nnew}$.
\end{corollary}

If one compares the scaling with the number of copies $\Nnew$ in Eqs. \eqref{eq:new coherent with L copies},\eqref{eq:f worst result 2 3} one finds effectively a quadratic advantage in the case of the $t$-coherent states. The difference in scaling between the two cases is the same as that found in metrology when comparing the classical shot noise scaling with the quantum Heisenberg scaling~\cite{StephLody}. 

The bound Eq. \eqref{eq:new coherent with L copies} in Corollary \ref{re:Nclockoptimal} effectively saturates the upper bound derived in \cite{faist2018prep}, which is proven to hold for all covariant error correction codes generated by isometries. Applying it to our constructions, it takes the form
\begin{equation}
f_{\wor} \le 1- \frac{\Delta h_\inp^2}{16 (\Delta h_\code +L d_\clo)^2}.\label{eq:up bound general}
\end{equation}
For the details about how this inequality follows from the results of \cite{faist2018prep} see \App~\ref{app:converse}.\footnote{Note that the extra additive factor $\Delta h_P$ in Eq. \eqref{eq:up bound general} is to be expected since the code $\map$ could itself contain a clock.}

If we keep the code parameters $\Delta h_{\code},\Delta h_{\inpp}, d_\outpp, d_\inpp$ fixed and scale up the clock size and the number of clocks, the combination of lower and upper bounds Eqs. \ref{eq:new coherent with L copies}, \ref{eq:up bound general} prove that our construction achieves an optimal scaling with both the dimension of the clock ${d_\clo}$ and the number of them $L$, up to the logarithmic factors. Furthermore, this proves that the bound is tight for all choices of $\left\{\map,\{\env_j, \mathcal{D}_j\}_j \right\}$.  

So far we have only considered erasure errors at known locations on the clock. 
However, what if one cannot detect where the error occurred without damaging the code? This is the case in the most elementary error correcting codes. If one has many clocks and the error occurs with an approximately uniform error probability distribution over the clock locations, one simple approach would be to choose one of the clocks, erase the other clocks, and perform error correction with this clock. If the probability of the error occurring on this clock is low, then this would work well on average. However, this approach is wasteful since it does not use the encoded information in the other clocks and requires a low probability of error on a particular clock.  Now we will investigate how well we can recover in the case of unknown phase errors at unknown locations which also works well for a small number of clocks. Consider the case in which one clock (or an entangled block of $L$ clocks in the sense of Theorem \ref{re:Nclock}) whose location is unknown has a random phase applied to it (i.e. an un-known group element $U_\clo^{\otimes L}(t_\textup{ph})$, $t_\textup{ph}\in\rr$). We call this a 2-unknown phase error and denote it $\envcode_{\textup{ph},q}(t_\textup{ph})$, where $q\in\nn$ and $t_\textup{ph}\in\rr$ are the unknown site location and phase respectively. The following result shows that such errors are correctable up to an arbitrarily small error.
\begin{theorem}\label{Thm:3 clock phase error}
	Consider the covariant code $\map_{\textup{cov}}$ in Eq. \eqref{eq:code}, with 3 blocks of $\Nnew$-site Quasi-Ideal Entangled clocks, namely
	\be
	\mathcal{E}_{\textup{cov}}(\cdot) 
	= \frac{1}{T_0}\int_0^{T_0} dt \,  \mathcal{E}_t (\cdot) \otimes U_\clo(t)^{\otimes 3\Nnew}\,\rho_{\textup{QIE,} \Nnew}^{\otimes 3}\,U_\clo(t)^{\dagger\otimes 3\Nnew},\label{eq:cov code for 3 block clocks}
	\ee
	with error channels $\{\env_{j,q}=\env_j\otimes \envcode_{\textup{ph},q}(t_\textup{ph})\}_{j,q}$, $q\in1,2,3$ where $\{\env_j\}_j$ are the error channels for the code $\map$, and $\envcode_{\textup{ph,}q}$ is a 2-unknown phase error acting on one of the three $\Nnew$-site Quasi-Ideal Entangled clocks, $\rho_{\textup{QIE,} \Nnew}$. Then for all $\Nnew\in\nn^+$, $q\in\{1,2,3\}$, $t_\textup{ph}\in\rr$ and error correcting codes $\left\{\map,\{\env_j, \mathcal{D}_j\}_j \right\}$, there exists a decoding channel $\tilde {\mathcal{D}}_j$ for $\map_\textup{cov}$, which is \text{independent} of the unknown block $q$ and phase $t_\textup{ph}$, such that 
	\ba 
	f_{\wor} \geq 1-& \sqrt{d_\inpp} d_\outpp \frac{10\pi(\Delta h_\inpp+\Delta h_\code)}{\sqrt{3}} \frac{\ln^7(\Nnew\, {d_\clo})}{\Nnew\, {d_\clo}} \nonumber \\&+\bo\left( \sqrt{d_\inpp} d_\outpp \frac{\ln^{11}(\Nnew\, {d_\clo})}{(\Nnew\, {d_\clo})^2} \right).\label{eq:q:fworst lowbound}
	\ea 
\end{theorem}
This result extends trivially to the more general case in which one has $M$ blocks (rather than 3) of $L$-site Quasi-Ideal Entangled clocks in the covariant code and has erasure errors at up to $M-3$ known locations, and the 2-unknown phase error on one of the 3 blocks of the remaining copies. See \app~\ref{3 clocks proof} for proof.

\begin{figure}[]
	\includegraphics[width=0.85\linewidth]{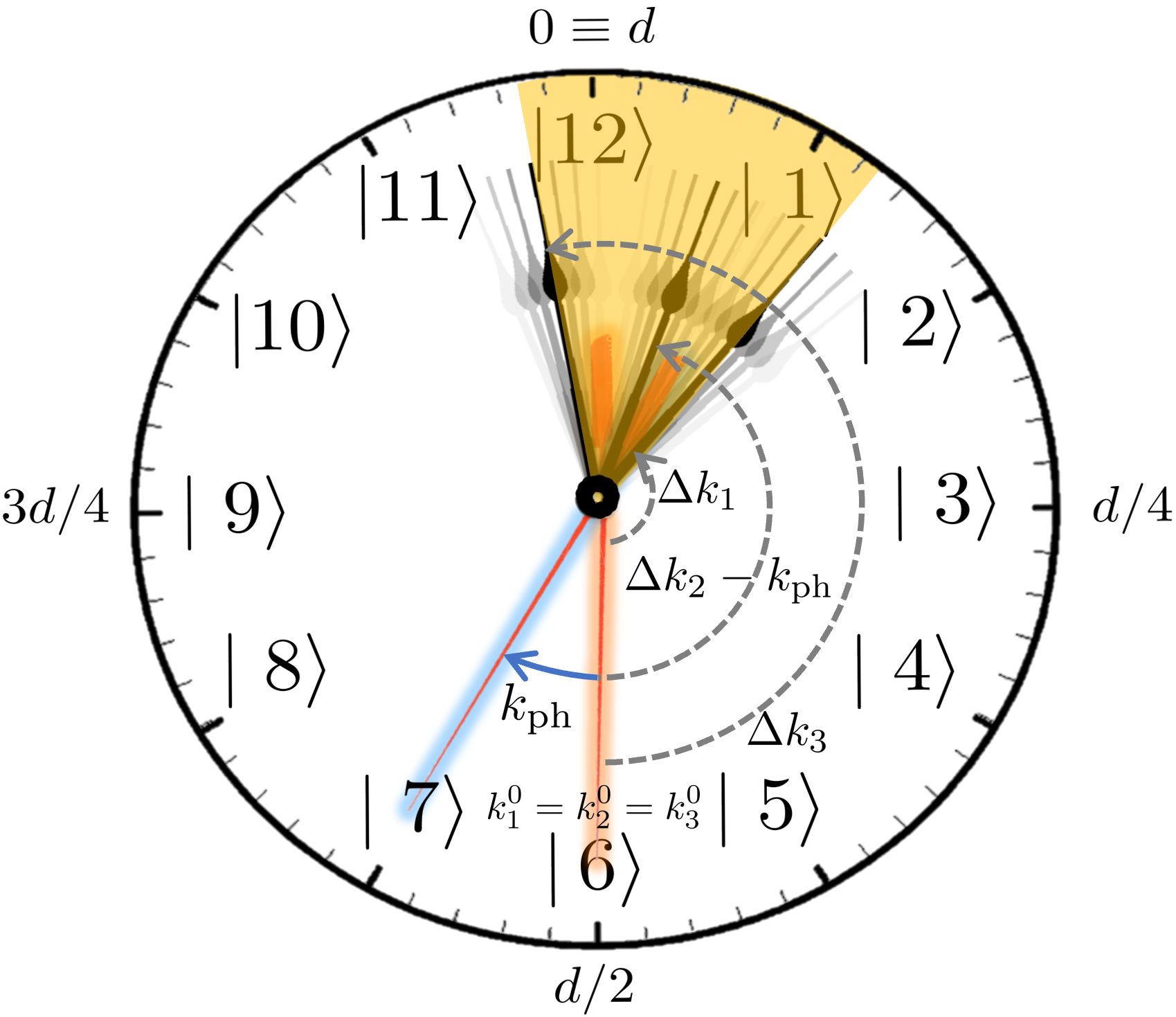}
	\caption{\textbf{Illustration of 3 clocks superimposed with a 2-unknown phase error}. All clocks are set to have the same initial time, $k_1^0=k_2^0=k_3^0$ (red glowing hand). The three measurement outcomes of the three clocks (black hands) attain similar values with high probability. The yellow shaded region represents possible values of $k_\alpha$  for this example, where $k_\alpha=g' (d_\clo-1)/(2\pi T_0)$, and $g'$ is described in step 3) of the decoding protocol in Section \ref{sec:perfectcode}. An unknown phase error occurs on an unknown clock (for the purpose of illustration, this is the 2\textsuperscript{nd} clock). It has the effect of shifting the initial time of the 2\textsuperscript{nd} clock by an unknown amount $k_\textup{ph}$. The apparent elapsed times are $
	\Delta t_1=2\pi (k_1-k_1^0) T_0/({d_\clo}-1)$, $
	\Delta t_2=2\pi (k_2-k_2^0-k_\textup{ph}) T_0/({d_\clo}-1)$, $
	\Delta t_3=2\pi (k_3-k_3^0) T_0/({d_\clo}-1)$, however, due to the 2-unknown phase error, $\Delta t_2$ will give the incorrect prediction. Nevertheless, in this example, the phase error is small and no correction is needed. \label{fig:3clocks}}
\end{figure}

To gain an intuitive picture of how the protocol works, it is best to consider the case $L=1$ (the general $L$ case is then a direct consequence of Theorem \ref{re:Nclock}). For $L=1$ the protocol is similar to our previous ones: we measure the 3 clocks locally in the time basis, and based upon the information from the 3 measurement outcomes, we calculate a time $g'=t'$ [step 3)] and apply a corresponding decoding map $\bar D_{j,t'}$ [step 4)] on the physical space. Due to the classical correlations between the code and clocks, the outcomes of the 3 clocks are correlated. This is such that, if there were no 2-unknown phase errors, the clocks would all indicate approximately the same ``elapsed time" and one could correct the code up to an error of order Eq. \eqref{eq:q:fworst lowbound} with the knowledge of only one of the 3 measurement outcomes. However, when a phase error occurs in the $q$\textsuperscript{th} clock, its initial time $k_q^0$ shifts by an unknown phase, making it an unknown variable and the reported elapsed time by the clock measurement of the $q$\textsuperscript{th} clock, is thus incorrect. Since the value of $q$ is also unknown, one cannot simply ignore the corresponding measurement outcome, so our protocol is to order the three measured elapsed times in ascending order and apply a unitary corresponding to the elapsed time which is neither the smallest nor the largest out of the three. There are two possibilities: 1) there was no phase error on this clock, in which case the marked elapsed time is approximately correct. In this case, the phase error could have been large. 2) the phase error occurred on this clock. In this case, the phase error must have been small, since the (incorrect) measured elapsed time is upper and lower bounded by that of the other two clocks (which must both be correct, since there is at most one 2-unknown phase error). This corresponds to the case of Fig. \ref{fig:3clocks}.

This processing of the measurement outcomes is very closely related to majority voting used in a classical repetition code. 
Here the main difference is that the outcomes of the clocks are not binary and are unlikely to agree exactly.

Note how our protocol does not rely on any assumptions about the probability distributions over blocks $q=1,2,3$ and phases $t_\textup{ph}\in\rr$ over which the 2-unknown phase error occurs. However, if one does assume that the probability of the 2-unknown phase error $\envcode_{\textup{ph},q}(t_\textup{ph})$ is small and allows for an arbitrary number of 2-unknown phase errors to occur in an independent and identically distributed manner, then the above protocol will also work with high fidelity  since the probability of two or more phase errors occurring on two or three clocks will be very small.

Finally, what about infinite dimensional covariant codes of finite energy? We leave the motivation to the Outlook (Section \ref{Outlook}) and state our conclusions here. One can embed our finite dimensional clock into an infinite dimensional space, and express $d_\clo$ in terms of the mean energy of the clock, denoted $\langle \hat H_\clo \rangle$. For all the clocks considered here, this corresponds to the mapping 
\be 
d_\clo\mapsto 2 \langle \hat H_\clo \rangle/\omega+1,
\ee (up to additive higher order corrections in $\langle \hat H_\clo \rangle$, for the case of Quasi-Ideal clocks). Our theorems and corollary; when framed in this context, provide bounds as a function of energy for how well the errors on covariant codes can be corrected. In this context, due to the infinite dimensions of the physical space, the Eastin-Knill no-go theorem  does not apply, so in principle one could have perfect covariant error correcting codes. However, as we have seen in Section \ref{sec:perfectcode}, due to the insights of Pauli, covariant error correcting codes based on reference frames can only be decodable without errors (i.e. $f_\wor=1$) iff they use Idealised clocks which necessitate infinite energy. We conjecture that this is true in general. Specifically, that all error correcting codes which are covariant w.r.t. any faithful representation of a non-trivial Lie group, necessitate infinite energy. 
This would be an extension of the Eastin-Knill no-go theorem since all finite dimensional systems have finite energy but not vice-versa.

\section{Outlook}\label{Outlook}





Given any error correcting code, we have seen how to construct simple classes of approximate codes in which compact U(1) groups can be implemented transversally. We study a specific scheme using Quasi-Ideal clocks \cite{woods2016autonomous} that saturates optimal bounds, explore the performance of alternative schemes, and also extend to an error model beyond errors at \emph{known} locations. The present codes are based on the task of reference frame alignment, which has been studied quite extensively \cite{bartlett2007reference} and also formalized within the resource theory of asymmetry \cite{marvian2012symmetry,marvian2014extending}, which we hope will be of further use in the context of error correction \cite{hayden2017error,faist2018prep}. Furthermore, since the clocks need to be measured projectively in our protocols, a preferred basis naturally arises and necessary conditions w.r.t. this basis for a quadratic advantage are identified. 
 
An obvious extension of our results is to groups beyond U(1). This requires the construction of more involved reference frames, such as finite-sized quantum ``gyroscopes" (for the case of SU$(2)$). We hope that our techniques may serve as a starting point for generalising the results found here in this direction | perhaps together with the construction of approximate frames for general groups from \cite{bartlett2009quantum}.


An important question is to understand and characterize the sort of error models and implementations for which the present scheme is adequate. We have mostly explored the case of erasure errors at known locations, and a type of dephasing errors on the clocks at unknown locations. While these are quite natural error models, it may well be that with more involved schemes of codes involving clocks, other types of errors can be dealt with. 

Also, the results of \cite{faist2018prep} show tight bounds that cannot be overcome only for erasure errors at known locations, and it would be interesting to know if their bounds are also tight for other error models. We suspect that the error scaling in Theorem \ref{Thm:3 clock phase error} is optimal, even though it is only as good as that of the $t$-incoherent clock states when the error model was that of erasure at know locations. Intuitively, the limiting factor in this case is that the 3 blocks of clocks are only classically correlated (separable states). While one can easily construct entangled counterparts, it is not clear how this can help when the errors occur over the unknown blocks. 

One key point to be determined is whether the error bounds shown here, even if they are essentially optimal, are not too large to be useful in practice for some type of architecture. In the case of Theorem \ref{re:1clock} and Corollary \ref{re:Nclockoptimal}, explicit bounds on the entanglement fidelity $f_\wor$ for finite $d_\clo$ have been derived in the \app~which can be evaluated numerically for experimentally feasible parameters. Were this the case, we would hope that the reference frames used here might be useful in some near-term applications of quantum technologies in which error correction has started playing a role, such as quantum metrology \cite{demkowicz2017adaptive,zhou2018achieving,layden2018ancilla}. In particular, given Eq. \eqref{eq:up bound general}, the Quasi-Ideal clock states appear to be almost optimally distiguishable along a whole $U(1)$ orbit. This suggests that a setting like that of Theorem \ref{Thm:3 clock phase error} could be of use in the construction of a metrological scheme aiming to determine unknown parameters (such as the time $t$ of the operator $U_C(t)$).

Covariant infinite dimensional error correcting codes are also of interest. Perhaps most prominently is the example of the hypothesized AdS/CFT duality between quantum gravity in asymptotically AdS space (known as the bulk) and a conformal field theory on the boundary, where the theory on the bulk and boundary are related via quantum error correcting codes \cite{almheiri2015bulk}. Specifically, the bulk constitutes the logical space while the boundary is the physical/code space. Global symmetries in the bulk should correspond to symmetries on the boundary which conserve the local structure of the theory | in a similar spirit to how transversal gates act globally in the logical space while locally in the physical space [c.f. Eq. \eqref{eq:cov 2}]. It is argued that variants of the Eastin-Knill no-go theorem have important interpretations in AdS/CFT: all global symmetries in the bulk (both continuous and discrete) are ruled out, since their existence would be in contradiction with the structure of the correspondence\footnote{Local CFT operators acting on a spatial region $R$ in the boundary, can only access information in a limited region in the bulk via \textit{entanglement wedge reconstruction}. This is incompatible with the existence of charged localised objects in the bulk.} \cite{Harlow2018a_short,Harlow2018}. A related issue is time dynamics which is given by a U$(1)$ symmetry with representations $\me^{-\mi t \hat H_\textup{blk}}$ and $\me^{-\mi t \hat H_\textup{bdy}}$ for Hamiltonians $\hat H_\textup{blk}$, $\hat H_\textup{bdy}$ on the bulk and boundary respectively. Often $\hat H_\textup{blk}$, $\hat H_\textup{bdy}$ are quasi-local and thus the corresponding U$(1)$ group action has to also preserve this local structure. An approximate U$(1)$ covariant code with these properties for finite dimensional Hamiltonians has been developed in \cite{TamaToby} using techniques from \cite{Cubitt9497}.
By choosing the number of code blocks to be one i.e. {\footnotesize $K$}~$\,=1$, our protocols allow for such bulk-boundary time dynamics covariance for arbitrary (e.g. quasi-local) finite dimensional Hamiltonians on $\mathcal{H}_\inpp$ and $\mathcal{H}_\code$. An open question is whether further work may allow for clocks with interacting quasi-local Hamiltonians too. 

Since both theories of the duality are infinite-dimensional but of finite energy locally, further quantitative variants of the Eastin-Knill no-go theorem for infinite dimensional physical spaces, seem more appropriate (for instance, by taking into account the ``average energy density" of the code space). Our results suggest that given sufficient energy, the preservation of local symmetries in the boundary is at least approximately possible. 

Finally, it is worth noting that the extension of a physical Hilbert space by including clocks has been proposed in a different context before. The aim of the 
Wheeler-DeWit equation is to describe a time-static theory of quantum gravity in which locally one finds dynamics \cite{PhysRev.160.1113}. This motivated the Page-Wooters mechanism for describing how a quantum clock can allow for time evolution to follow from a static universe using Idealised clocks \cite{PageWooter}. In \app~\ref{Sec:PageWooters} we show that when the logical space has the trivial group representation of the U$(1)$ symmetry, our formalism is an example of an approximate Page-Wooters mechanism where the approximation comes from using non Idealised clocks. 
\\{}\\

\textbf{Note on Related Work}: Around the time this work was being developed, it was realized by the authors that a complementary approach to characterizing how well the Eastin-Knill no-go theorem can be circumvented by allowing for a small error, was being developed independently by the authors of \cite{faist2018prep}. We thank them for their community spirit.

\begin{acknowledgements}
We thank useful conversations with Benjamin Brown, Philippe Faist, Daniel Harlow, Tamara Kohler and Rob Spekkens. This work was initiated at QIP 2018 hosted by QuTech at Delft University of Technology.
M.P.W. acknowledges funding from the Swiss National Science Foundation (AMBIZIONE Fellowship PZ00P2\_179914) in addition to the NCCR QSIT. This research was supported in part by Perimeter Institute for Theoretical Physics. Research at Perimeter Institute is supported by the Government of Canada through the Department of Innovation, Science and Economic Development and by the Province of Ontario through the Ministry of Research, Innovation and Science.
\end{acknowledgements}

\bibliographystyle{unsrtnat}
\bibliography{References}

\input{AppendicesV1.tex}

\end{document}

%% file: MischaCommands.tex
\usepackage{dsfont}
\usepackage{bbm} 
\usepackage[normalem]{ulem} 
\usepackage{color} 
\usepackage{braket} 

\newcommand{\be}{\begin{equation}}
\newcommand{\ee}{\end{equation}}
\newcommand{\nn}{{\mathbbm{N}}}
\newcommand{\rr}{{\mathbbm{R}}}
\newcommand{\cc}{{\mathbbm{C}}}
\newcommand{\zz}{{\mathbbm{Z}}}
\newcommand{\me}{\mathrm{e}}
\newcommand{\mi}{\mathrm{i}}
\newcommand{\bo}{\mathcal{O}}

\newcommand{\map}{\mathcal{E}}

\newcommand{\cl}[1]{\textup{C,#1}}
\newcommand{\clo}{\textup{C}}
\newcommand{\dom}{\textup{Dom}}
\newcommand{\cdom}{\textup{CoDom}}

\newcommand{\inp}{\textup{L}}
\newcommand{\outp}{\textup{P}}

\theoremstyle{definition}

\def\ba#1\ea{\begin{align}#1\end{align}} 

\newcommand{\App}{\text{Supplementary material}}
\newcommand{\app}{\text{supplementary material}}

\newcommand{\mpw}[1]{\textcolor{blue}{MW: #1}}

\newcommand\mpwS[1]{MW:{\let\helpcmd\sout\parhelp#1\par\relax\relax} }
\long\def\parhelp#1\par#2\relax{%
	\helpcmd{#1}\ifx\relax#2\else\par\parhelp#2\relax\fi%
}



%% file: AppendicesV1.tex
\widetext
\appendix

\section{Compatibility with the transversal gates of $\mathcal{E}$}\label{app:tranversal}
In our schemes we start with a code $\mathcal{E}(\cdot)$ which is in principle arbitrary. We can ask what happens if this code already has a group of transversal or covariant gates (possibly of finite order as granted by \cite{eastin2009restrictions}). Does that covariance remain after the reference frame is added? Let $\{V_{\inpp},V_{\code},V_{\F}\}$ be an element of a representation of the group of transversal gates of $\mathcal{E}$ on the logical, physical and reference frame spaces. Thus:
\begin{equation}
\mathcal{E}\left(V_\inpp (\cdot) V_\inpp^\dagger\right)= V_\code^{\otimes \K}\mathcal{E}(\cdot)V_\code^{\dagger \otimes \K},
\end{equation}
where $\otimes \K$ represents the tensor product structure of the physical space of $\mathcal{E}$.
Does the same symmetry hold for $\mathcal{E}_{\textup{cov}}$? Let us write
\begin{equation}\label{eq:code4}
\mathcal{E}_{\textup{cov}}(V_\inpp (\cdot) V_\inpp^\dagger) = \int_G \text{d} g \,  \mathcal{E}_g (V_\inpp (\cdot) V_\inpp^\dagger) \otimes U_\F(g)^{\otimes \M}\,\rho_{\clo }^{(\M)}\,U_\F(g)^{\dagger\otimes \M},
\end{equation}
whereas we have, on the other hand

\begin{align}\label{eq:code3}
& \left(V_\code^{ \otimes \K} \otimes V_\F\right)\mathcal{E}_{\textup{cov}}( \cdot )\left(V_\code^{\dagger \otimes \K}\otimes V_\F^\dag\right) = \int_G \text{d} g \,  V_\code^{\otimes \K}\mathcal{E}_g(\cdot)V_\code^{\dagger \otimes \K} \otimes V_\F(U_\F(g)^{\otimes \M}\,\rho_{\clo }^{(\M)}\,U_\F(g)^{\dagger\otimes \M})V_\F^\dagger,
\end{align}
It is clear that Eqs. \eqref{eq:code4} and \eqref{eq:code3} coincide when $i)$ both commutations $[V_\code,U_\code(g)^{\otimes \K}]=0$ and $[V_\inpp,U_\inpp(g)]=0$ hold for all $g\in \mathcal G$ and $ii)$ $V_\F$ acts trivially on the clocks. Otherwise, the covariance/transversality is lost. While the latter can be obtained by choosing $V_\F$ to act trivially on the clocks, the former seems much more restrictive.

However, there is a sense in which we might still have transversal gates in the code, if we are able to set the representation $U_\code(g)^{\otimes \K}$ on the physical space to the trivial case. After measuring the clocks and obtaining some outcome $g'$, one can apply the gates $V_\outpp^{ \otimes \K} \otimes V_\F$. Then, applying the decoder, the resulting state is 
\begin{equation}
V_\inpp U_\inpp^\dagger (g')\rho_\inpp U_\inpp(g')) V_\inpp^\dagger.
\end{equation}
If $[V_\inpp,U_\inpp(g)]=0$ we may just apply $U_\inpp(g)$ and end up with the desired state to which the gate has been applied transversally. Otherwise, what has happened is that we have applied the gate $V_\inpp U_\inpp^\dagger (g')$. We have knowledge of $g'$, and so to end up with $V_\inpp \rho_\inpp V_\inpp^\dagger$ we may just apply the unitary $V_\inpp U_\inpp(g') V_\inpp^\dagger$. This assumes that one can apply the logical gate both at the logical and the physical level.

This discussion gives a further example, together with the errors of the main results, of why it is advantageous to choose the generator $\hat H_\code$ to be trivial $\hat H_\code \propto \id_\outpp$, if possible.

\section{General encoding-decoding error with finite clocks}\label{app:gen}

Here we explain the form for the encoding-error-decoding protocol for all the schemes of the present work. We refer to this discussion repeatedly in the different proofs. We will here assume that the $M$ clocks are of product form. We will see later that this construction is general enough for our purposes even when the clock are entangled, by considering further divisions of the local sites considered here. 

We define the dimensions of the input and output of the error correcting code $\map$ to be $d_\inp$, $d_\outp$. The dimension of the $i$th clock is $d_i$. If all clocks have the same dimension we use the notation $d_\clo:=d_1=d_2=\ldots$.  If we do not write the range over a summation, it will be over the full range. 


For our case, the unitary representation is $U_\clo(t)=\me^{-\mi t \hat H_\clo}$ where $\hat H_\clo$ is defined in Section \ref{sec:finite}. The integral measure over the group is $\text{d}g=dt/T_0$ on the interval $[0,T_0]$.
 The encoded state in which $M$ clocks in states $\rho_{C,i}$ with $ i \in \{1,..M \}$ are used has the form of Eq. \eqref{eq:code} 
\begin{equation}\label{eq:codefinite}
\mathcal{E}_{\text{cov}}(\rho_\inpp) = \frac{1}{T_0}\int_0^{T_0} \text{d} t \,  \mathcal{E}_t (\rho_\inpp) \bigotimes_{i=1}^M U_\clo(t) \rho_{C,i} U_\clo^\dagger(t).
\end{equation}

Let us first calculate the integral over $t$ by writing the unitary operators $U_\clo(t)$, $U_\inpp(t)$, $U_\code(t)$ in their eigenbasis:
\be 
\map\left( U_\inp^\dag (t) \rho_\inp U_\inp(t) \right)= \sum_{n,n'=0}^{d_\inp-1} \me^{-\mi \omega t (h_{\inp,n}-h_{\inp,n'})} \map \left( {}_\inp\bra{n'}\rho_\inp\ket{n}_\inp \ketbra{n'}{n} \right),
\ee 
and thus,
\be 
\map_{t}\left( \rho_\inp \right)= \sum_{q,q'=0}^{d_\outp-1} \sum_{n,n'=0}^{d_\inp-1} \me^{-\mi \omega t (h_{\code,q}-h_{\code,q'}+h_{\inp,n}-h_{\inp,n'})} \map_{q,q',n,n'}\left( \rho_\inp \right)\ketbra{q}{q'}_\outp,\label{eq:ecoding with t}
\ee 
with
\be \label{eq:map code}
\map_{q,q',n,n'}\left( \rho_\inp \right):={}_\outp\bra{q}\map \left( {}_\inp\bra{n'}\rho_\inp\ket{n}_\inp \ketbra{n'}{n} \right)\ket{q'}_\outp.
\ee 
For simplicity, let us define $Q := h_{\code,q}-h_{\code,q'}+h_{\inp,n}-h_{\inp,n'}$. Therefore
\ba 
\map_{\text{cov}} (\rho_\inp )=&\frac{1}{T_0} \int_{0}^{T_0} dt\, \map_{t}\,(\cdot) \otimes \rho_{\cl1}(t)\otimes\rho_{\cl2}(t)\otimes\ldots\otimes\rho_{\cl M}(t),\\
=& \sum_{q,q'=0}^{d_\outp-1}\sum_{n,n'=0}^{d_\inp-1}\sum_{\vec{r}'_M, \vec{r}_M'} \frac{1}{T_0} \int_0^{T_0} dt\, \me^{-\mi \omega t (Q+r_1-r_1'+\ldots+r_M-r_M')} \map_{q,q',n,n'}\left( \rho_\inp \right) \ketbra{q}{q'} \\
&\otimes \bra{r_1}\rho_{\cl 1}\ket{r_1}\otimes\ldots\otimes \bra{r_M}\rho_{\cl M}\ket{r_M}
\\
=& \sum_{q,q'=0}^{d_\outp-1}\sum_{n,n'=0}^{d_\inp-1}\sum_{\vec{r}_M, \vec{r}'_M} \delta_{Q+r_1-r_1'+\ldots+r_M-r_M',0}\, \map_{q,q',n,n'}\left( \rho_\inp \right) \ketbra{q}{q'} \\
&\otimes \bra{r_1}\rho_{\cl 1}\ket{r_1'}\ketbra{r_1}{r_1'}\otimes\ldots\otimes \bra{r_M}\rho_{\cl M}\ket{r_M'}\ketbra{r_M}{r_M'},\label{eq:covarina map in energy basis before tracing}
\ea 
where $\{\ket{r_i}\}$ is the eigenbasis of the generator $\hat H_{\clo}$ of the $i$-th clock, $\delta_{(\cdot,\cdot)}$ is the Kronecker-Delta function and
\be 
\sum_{\vec{r}_M, \vec{r}_M'} =\sum_{r_1,r_1',=0}^{d_1-1}\ldots\sum_{r_M,r_M',=0}^{d_M-1}.
\ee 
Since by assumption the last $N+1, \ldots, M$ clocks may be lost due to erasure errors, we can trace them out:
\ba 
\tr_{C_{N+1}\ldots C_M}\left[\map_{\text{cov}} (\rho_\inp )\right]= & \sum_{q,q'=0}^{d_\outp-1}\sum_{n,n'=0}^{d_\inp-1}\sum_{\vec{r}_M, \vec{r}'_M} \delta_{Q+r_1-r_1'+\ldots+r_N-r_N',0}\, \map_{q,q',n,n'}\left( \rho_\inp \right) \ketbra{q}{q'} \\
\otimes \bra{r_1}\rho_{\cl 1}\ket{r_1'}&\ketbra{r_1}{r_1'}\otimes\ldots\otimes \bra{r_N}\rho_{\cl N}\ket{r_N'}\ketbra{r_N}{r_N'}\left( \bra{r_{N+1}}\rho_{\cl N+1}\ket{r_{N+1}}\ldots \bra{r_M}\rho_{\cl M}\ket{r_M} \right)\\
=& \sum_{q,q'=0}^{d_\outp-1}\sum_{n,n'=0}^{d_\inp-1}\sum_{\vec{r}_N, \vec{r}'_N} \delta_{Q+r_1-r_1'+\ldots+r_N-r_N',0}\, \map_{q,q',n,n'}\left( \rho_\inp \right) \ketbra{q}{q'} \\
\otimes \bra{r_1}\rho_{\cl 1}\ket{r_1'}&\ketbra{r_1}{r_1'}\otimes\ldots\otimes \bra{r_N}\rho_{\cl N}\ket{r_N'}\ketbra{r_N}{r_N'}.\label{eq:after tracing clocks}
\ea
So by comparing Eqs. \eqref{eq:after tracing clocks}, \eqref{eq:covarina map in energy basis before tracing}, we observe that after tracing out the additional clocks, the resultant channel is of the same form as the original channel, when evaluated for the renaming number of clocks.

We now assume that an error due to the environment occurs. This means that we apply a CPTP error map $\env_j:\mathcal{H}_\outp\rightarrow \mathcal{H}_\outp$ corresponding to an error $j$. We thus denote 
\be \label{eq:map envirom plus code}
\map_{q,q',n,n'}^j \left( \rho_\inp \right):={}_\outp\bra{q}\env_j\left(\map \left( {}_\inp\bra{n'}\rho_\inp\ket{n}_\inp \ketbra{n'}{n} \right)\right)\ket{q'}_\outp.
\ee 
We now measure the remaining clocks, performing projective measurements in the ``time" basis \be 
\left\{ \ket{\theta_k}_m =\frac{1}{\sqrt{d_m}}\sum_{n=0}^{d_m-1} \me^{-\mi 2\pi n k/d_m}\ket{n}_m \right\}_{k=0}^{d_m}
\ee
on the $m$th clock. Let us drop the $m$ subindex for simplicity of notation. The state after the outcome $\vec k:= k_1,k_2,\ldots,k_N$ is: 
\ba 
\rho_\outpp^{\vec{k}}&\otimes \ketbra{\theta_{k_1}}{\theta_{k_1}}\otimes\ldots\otimes \ketbra{\theta_{k_N}}{\theta_{k_N}}:=\\
&\frac{ \ketbra{\theta_{k_1}}{\theta_{k_1}}\otimes\ldots\otimes\ketbra{\theta_{k_N}}{\theta_{k_N}}\,\tr_{C_{N+1}\ldots C_M}\left[\env_j \map_{\text{cov}} (\rho_\inp )\right]  \ketbra{\theta_{k_1}}{\theta_{k_1}}\otimes\ldots\otimes\ketbra{\theta_{k_N}}{\theta_{k_N}}  }{p_{\vec{k}}}\\
=& \frac{1}{ p_{\vec{k}}} \Bigg(\sum_{q,q'=0}^{d_\outp-1}\sum_{n,n'=0}^{d_\inp-1}\sum_{\vec{r}_N, \vec{r}'_N} \delta_{Q+r_1-r_1'+\ldots+r_N-r_N',0}\, \map_{q,q',n,n'}^j\left( \rho_\inp \right) \ketbra{q}{q'} \\
&\bra{r_1}\rho_{\cl 1}\ket{r_1'}\braket{\theta_{k_1}|r_1}\braket{r_1'|\theta_{k_1}}\ldots \bra{r_N}\rho_{\cl N}\ket{r_N'} \braket{\theta_{k_N}|r_N}\braket{r_N'|\theta_{k_N}} \Bigg)\otimes \ketbra{\theta_{k_1}}{\theta_{k_1}}\otimes\ldots\otimes \ketbra{\theta_{k_N}}{\theta_{k_N}}.\label{eq:rho out k vec}
\ea
Thus, after postselecting on a particular outcome represented with the vector $\vec{k}$, we obtain the following state on the Hilbert space of the output of $\env_j \map(\cdot)$
\ba 
\rho_\outpp^{\vec{k}}
=& \frac{1}{ p_{\vec{k}}} \sum_{q,q'=0}^{d_\outp-1}\sum_{n,n'=0}^{d_\inp-1}\sum_{\vec{r}_N, \vec{r}'_N} \delta_{Q+r_1-r_1'+\ldots+r_N-r_N',0}\, \map_{q,q',n,n'}^j\left( \rho_\inp \right) \ketbra{q}{q'} \\
&\bra{r_1}\rho_{\cl 1}\ket{r_1'}\braket{\theta_{k_1}|r_1}\braket{r_1'|\theta_{k_1}}\ldots \bra{r_N}\rho_{\cl N}\ket{r_N'} \braket{\theta_{k_N}|r_N}\braket{r_N'|\theta_{k_N}}.
\ea
An outcome labeled with $\vec k$ happens with a probability given by
\ba
p_{\vec{k}}& = \tr 
\Bigg[ \ketbra{\theta_{k_1}}{\theta_{k_1}}\otimes\ldots\otimes\ketbra{\theta_{k_N}}{\theta_{k_N}}\, \tr_{C_{N+1}\ldots C_M}\left[\map
_j \map_{\text{cov}} (\rho_\textup{in})\right] \ketbra{\theta_{k_1}}{\theta_{k_1}}\otimes\ldots\otimes\ketbra{\theta_{k_N}}{\theta_{k_N}}\,\Bigg] \\
&=\frac{1}{T_0} \int_{t_0}^{T_0+t_0} dt\, \tr\left[\env_j \map_{t}\,(\rho_\inp)\right] \braket{\theta_{k_1}|\rho_{\cl 1}(t)|\theta_{k_1}}\ldots\braket{\theta_{k_N}|\rho_{\cl N}(t)|\theta_{k_N}} \\
&=\frac{1}{T_0} \int_{t_0}^{T_0+t_0} dt\,  \braket{\theta_{k_1}|\rho_{\cl 1}(t)|\theta_{k_1}}\ldots\braket{\theta_{k_N}|\rho_{\cl N}(t)|\theta_{k_N}} \quad \forall \,\,t_0\in\rr.\label{eq:p vec general}
\ea 
Now, let us define the following function. 
\be 
\F_Q(\vec k):=\frac{1}{T_0} \int_{t_0}^{T_0+t_0} dt\, \me^{-\mi \omega Q t} \braket{\theta_{k_1}|\rho_{\cl 1}(t)|\theta_{k_1}}\ldots\braket{\theta_{k_N}|\rho_{\cl N}(t)|\theta_{k_N}}, \quad \forall \,\,t_0\in\rr,\,Q\in\zz,\label{eq: F Q of k Def}
\ee 
where for simplicity we will assume that all the clock dimensions are equal $d_1=d_2=\ldots=d_N=:d_\clo$. 
This definition has the following properties:
\begin{itemize}
	
	\item [1)] Since the integrand has period $T_0$, $F_Q(\vec k)$ is independent of $t_0$ and we can set it to any preferred real number to help perform the calculations.
	\item[ 2)] We can perform the change of variable $t=t'+a$, $a=-l \frac{2\pi}{\omega d_\clo}$, $\l\in\zz$, to show
	
	\ba
	F_Q(\vec k)& =\frac{ \me^{\mi 2\pi l Q /d}}{T_0} \int_{t_0-l \frac{2\pi}{\omega d_\clo}}^{T_0+t_0-l \frac{2\pi}{\omega d_\clo}} dt'\,  \me^{-\mi \omega Q t}  \braket{\theta_{k_1+l}|\rho_{\cl 1}(t)|\theta_{k_1+l}}\ldots\braket{\theta_{k_N+l}|\rho_{\cl N}(t)|\theta_{k_N+l}}\\
	& =\me^{\mi 2\pi l Q /d} F_Q(\vec k+l),\label{eq:F Q shift invarient}
	\ea
	where the $j$th vector component of $\vec k+l$ is defined by $[\vec k+l]_j:= [\vec k]_j +l$ for all vector component $j$. 
	Note how this expression is invariant under $l\mapsto l+ j d_\clo$, $j\in\zz$. So w.l.o.g. we can exchange $\vec{k}+l$ with $[\vec{k}+l]_{\textup{(mod. \textit{d})}}$, where the map acts element-wise on vectors and $\textup{(mod. \textit{d})}$ denotes modular $d$ arithmetic.
	
	\item[3)] By inspection, we see that $F_Q$ is invariant under any pairwise interchanges
	\ba 
	k_q \leftrightarrow k_r \quad \text{and}\quad  \rho_{\cl q}(t) \leftrightarrow \rho_{\cl r}(t), \quad q,r\in 1,\ldots,N,\label{eq:prop 3 interchainge invar}
	\ea 
	where $k_q$ and $k_r$ are the $q$th and $p$th vector elements of $\vec k$.
	\item[4)] By Eq. \eqref{eq:p vec general}, it follows that $F_Q$ encodes the measurement outcome probabilities, 
	\be 
	p_{\vec k}= F_0(\vec k).
	\ee 
	
	This has important consequences when property 2) is taken into account. For example, in the case of one clock, it implies that all measurement outcomes are equally likely, so
	\be 
	p_{\vec k}=\frac{1}{d_\clo}\quad \forall\, k_1=0,\ldots,d_\clo-1.
	\ee 
	
	\item[5)] By making the substitution $U_m(t)=\sum_{r_m=0}^{d_m-1}\me^{-\mi \omega r_m t} \ketbra{r_m}{r_m}$, we find
	\ba
	F_Q(\vec k)&=\sum_{\vec{r}_N, \vec{r}'_N} \delta_{Q+r_1-r_1'+\ldots+r_N-r_N',0}\,\\
	&\bra{r_1}\rho_{\cl 1}\ket{r_1'}\braket{\theta_{k_1}|r_1}\braket{r_1'|\theta_{k_1}}\ldots \bra{r_N}\rho_{\cl N}\ket{r_N'} \braket{\theta_{k_N}|r_N}\braket{r_N'|\theta_{k_N}}.
	\ea
\end{itemize}
Thus using property 5), we can write $\rho_\outpp^{\vec{k}}$ in Eq. \ref{eq:rho out k vec} as

\ba 
\rho_\outpp^{\vec{k}}
=& \frac{1}{ p_{\vec{k}}} \sum_{q,q'=0}^{d_\outp-1}\sum_{n,n'=0}^{d_\inp-1}\sum_{\vec{r}_N, \vec{r}'_N} \delta_{Q+r_1-r_1'+\ldots+r_N-r_N',0}\, \map_{q,q',n,n'}^j\left( \rho_\inp \right) \ketbra{q}{q'} \\
&\bra{r_1}\rho_{\cl 1}\ket{r_1'}\braket{\theta_{k_1}|r_1}\braket{r_1'|\theta_{k_1}}\ldots \bra{r_N}\rho_{\cl N}\ket{r_N'} \braket{\theta_{k_N}|r_N}\braket{r_N'|\theta_{k_N}}.\\
=&  \sum_{q,q'=0}^{d_\outp-1}\sum_{n,n'=0}^{d_\inp-1}  \frac{F_{Q}(\vec k)}{F_{0}(\vec k)}\, \map_{q,q',n,n'}^j\left( \rho_\inp \right) \ketbra{q}{q'}.\label{eq: rho our generix with F}
\ea
It is convenient to introduce an arbitrary phase ${k_\alpha}\in\rr$  which for our purposes has to be defined modulo $d_\clo$. It will depend on $\vec k$, the measurement outcomes. We will choose it later depending on the particular covariant error correcting code set-up and protocol. We can now write 
\ba 
\rho_\outpp^{\vec{k}}
=&  \sum_{q,q'=0}^{d_\outp-1}\sum_{n,n'=0}^{d_\inp-1}  \frac{F_{Q}(\vec k)}{F_{0}(\vec k)}\, \map_{q,q',n,n'}^j\left( \rho_\inp \right) \ketbra{q}{q'}.\\
=&\sum_{q,q'=0}^{d_\outp-1}\sum_{n,n'=0}^{d_\inp-1} \left(\left[ \frac{ F_{Q}(\vec k)}{F_{0}(\vec k)}\,\me^{2\pi \mi {k_\alpha} Q/d_\clo} -1\right] +1\right)\\
& U_\outp^{\otimes \K \dag} (t_\alpha) \map_{q,q',n,n'}^j\left( U_\inp (t_\alpha)\rho_\inp U_\inp^\dag  (t_\alpha)\right) \ketbra{q}{q'} U_\outp^{\otimes \K} (t_\alpha)\\
= & U_\outp^{\otimes \K \dag} (t_\alpha) \left( \hat E'_{\vec k} (\tilde \rho_\inp) + \map^j(\tilde \rho_\inp)  \right) U_\outp^{\otimes \K} (t_\alpha),  \label{eq: rho our generix with F 2}
\ea
where 
\be \label{eq:defU}
\tilde \rho_\inp := U_\inp (t_\alpha) \rho_\inp U_\inp^\dag (t_\alpha),\quad t_\alpha:= {k_\alpha} \,\frac{T_0}{d_\clo},
\ee 
and 
\be 
\hat E'_{\vec k}(\cdot):= \sum_{q,q'=0}^{d_\outp-1}\sum_{n,n'=0}^{d_\inp-1} \left[ \frac{ F_{Q}(\vec k)}{F_{0}(\vec k)}\,\me^{2\pi \mi {k_\alpha} Q/d_\clo} -1\right] \map_{q,q',n,n'}^j (\cdot) \ketbra{q}{q'}\label{eq:error map}
\ee 
In order to proceed with the decoding, let us define the function $p(Q,\vec k)$ as:
\ba\label{eq:ratioF}
p(Q,\vec k)= 1- \frac{ F_{Q}(\vec k)}{F_{0}(\vec k)}\,\me^{2\pi \mi {k_\alpha} Q/d_\clo},
\ea
where $p(Q,\vec k)$ is a complex number for which we expect $\vert p(Q,\vec k) \vert \ll 1 $ (how small will determine the size of the error of the particular sheme). The phase $k_\alpha$ determines the group elements we should apply in the decoding procedure, as per Eq. \eqref{eq:defU}.

Since by assumption the group action only acts non-trivially in the physical space, $U^{\otimes \K}_\outpp(t_\alpha)=U_\code(t_\alpha)\oplus\id_{\outpp/\code}$, the decoder may take the form
\be 
\bar {\mathcal{D}}_{j,t_\alpha}(\cdot)= U_\inpp(t_\alpha) \map^{-1}\left( U_\code (t_\alpha) \map\left( \mathcal{D}_j(\cdot) \right) U_\code(t_\alpha) \right) U_\inpp^\dag(t_\alpha),
\ee 
where $\map^{-1}(\cdot)=\mathcal{D}_j\left(\env_j(\cdot)\right)$ is the inverse channel for the encoder $\map$ and $\mathcal{D}_j$ the decoding map of $\map$. 

 With it we see that if we apply the decoder to the state $\rho_\outpp^{\vec{k}}$, we obtain
\be
\rho_\outpp^{\vec{k}}=\rho_\inp- \hat E_{\vec k}(\rho_\inp),
\ee
where we define 
\ba\label{eq:defE}
\hat E_{\vec k}( \rho_\inp) = \sum_{q,q'=0}^{d_\outp-1}\sum_{n,n'=0}^{d_\inp-1}p(Q,\vec k) \,\bar {\mathcal{D}}_{j,t_\alpha} \left(  \map_{q,q',n,n'}^j\left( \rho_\inp \right) U_\outpp^{\dag\otimes \K}(t_\alpha) \ketbra{q}{q'}U_\outpp^{\otimes \K}(t_\alpha)\right).
\ea
Since each of the outcomes $\vec k$ occurred with probability $F_{0}(\vec k)$, the final decoded state, averaged over all measurement outcomes, is of the form
\begin{equation}\label{eq:finalerror}
\mathcal{K}(\rho_\inp)=\rho_\inp- \sum_{\vec k} F_{0}(\vec k) \hat E_{\vec k}( \rho_\inp).
\end{equation}
In Appendix \ref{app:entfid} we show how a bound on the entanglement fidelity of the code follows from this expression. The last fact that then needs to be shown is the form of the RHS of Eq. \eqref{eq:ratioF} for particular cases of clocks.

\section{Calculation of the entanglement fidelity} \label{app:entfid}
We here give the bounds on the entanglement fidelity which will be used to prove the main results. Again, it is defined as 
\begin{equation}
f_{\wor}(\mathcal{K})=\min_{\phi} \bra{\phi} \mathcal{K} \otimes \mathcal{I} (\ket{\phi}\bra{\phi}) \ket{\phi},
\end{equation}
where the minimization is over all the pure bipartite states $\ket{\phi}$. 

Motivated by the discussion in Appendix \ref{app:gen} leading to Eq. \eqref{eq:finalerror}, let us start with the assumption that after the encoding, error and decoding steps, the map has the following form when applied to a state $\rho$.
\begin{equation}\label{eq:Kmap}
\mathcal{K}(\rho_\inpp) = \rho_\inpp - \hat E(\rho_\inpp),
\end{equation}
A result from \cite{knill1997theory} allows us to lower bound the entanglement fidelity in terms of $\hat E(\rho_\inpp)$. If for all pure states on a single system on $\mathcal{H}_{d_{\text{in}}}$ the following holds
\begin{equation}
\bra{\phi} \mathcal{K} (\ket{\phi}\bra{\phi}) \ket{\phi} \ge 1- \epsilon,
\end{equation}
then we have that $f_{\wor}(\mathcal{K})\ge 1- \frac{3 \epsilon}{2}$. Given that $\bra{\phi} \hat E(\ket{\phi}\bra{\phi}) \ket{\phi} \le \vert\vert \hat E(\ket{\phi}\bra{\phi}) \vert \vert_1$, we can choose $\epsilon =\max_{\ketbra{\phi}{\phi}} \vert\vert \hat E(\ket{\phi}\bra{\phi})  \vert \vert_1$, where the optimization is over all pure states, thus obtaining
\begin{equation}\label{eq:fidEE}
f_{\wor}(\mathcal{K}) \ge 1-\frac{3}{2}\max_{\ketbra{\phi}{\phi}} \vert\vert \hat E(\ket{\phi}\bra{\phi})  \vert \vert_1.
\end{equation}
The next step is to give an upper bound to the $1$-norm of $\hat E$ that holds for any pure state. First, by the triangle inequality
\begin{equation}
\vert \vert \hat E \vert \vert_1= \vert\vert \sum_{\vec k} F_{0}(\vec k) \hat E_{\vec k} \vert\vert_1 \le \sum_{\vec k} F_{0}(\vec k) \vert\vert  \hat E_{\vec k} \vert\vert_1.
\end{equation}
As seen in Appendix \ref{app:gen}, in all of the cases considered here, the operator $ \hat E_{\vec k}$ is defined as $\hat E_{\vec k} = \bar {\mathcal{D}}_{j,t_\alpha} \left( \hat{E}_{\vec k}'  \right) $ where $\bar {\mathcal{D}}_{j,t_\alpha}$ is the decoder,  and 
\begin{equation}
\hat{E}_{\vec k}' = \sum_{q,q'=0}^{d_{\text{out}}-1} \sum_{n,n'=0}^{d_{\text{in}}-1} U_\code^{\dag\otimes \K}(t_\alpha) \ketbra{q}{q'}U_\code^{\otimes \K}(t_\alpha) \bra{q} \env_j\mathcal{E}_0( \ket{n}\bra{n} \bra{n} \rho_\inpp \ket{n'})\ket{q'} p(Q,\vec k),
\end{equation}
where $ p(Q,\vec k)$ is in principle constrained by complete positivity and trace preservation (as defined in Eq. \eqref{eq:defE}).
By contractivity of CPTP maps, $\vert \vert \hat E_{\vec k} \vert\vert_1 \le \vert \vert \hat E_{\vec k}' \vert\vert_1$. Now, we can write
\begin{align}
\vert \vert \hat E_{\vec k}' \vert\vert_1 &= \max_{q'} \sum_q \vert \bra{q} \sum_{n,n'=0}^{d_{\text{in}}-1} \env_j( \ket{n}\bra{n'} \bra{n}\rho_\inpp \ket{n'})\ket{q'} p(Q,\vec k) \vert \\&= \max_{q'} \sum_q \left\vert \tr \left [ \ket{q'}\bra{q} \sum_{n,n'=0}^{d_{\text{in}}-1} \env_j( \ket{n}\bra{n'} \bra{n}\rho_\inpp \ket{n'}) p(Q,\vec k) \right ]\right\vert \\ & \le \max_{q'} \sum_q \vert \vert  \env_j( \sum_{n,n'=0}^{d_{\text{in}}-1}\ket{n}\bra{n'} \bra{n}\rho_\inpp \ket{n'}p(Q,\vec k) )  \vert \vert_1 \\ & \le  \vert \vert \env_j \vert \vert _{1-1} \max_{q'} \sum_q \vert \vert  \sum_{n,n'=0}^{d_{\text{in}}-1}\ket{n}\bra{n'} \bra{n}\rho_\inpp \ket{n'}p(Q,\vec k) \vert \vert_1
\end{align}
where the third line follows from the inequality \cite{wolf2012quantum},
\begin{equation}
\vert \tr[B A^\dagger ] \vert \le \vert \vert A \vert \vert _p \vert \vert B \vert \vert_r,
\end{equation} with $1=1/p+1/r$ and choosing $p=1,r=\infty$, with $B=\ket{q'}\bra{q}$ and $A$ the rest. The fourth line follows from the definition of the $1-1$ norm for CPTP maps, which is $\vert \vert \env_j \vert \vert_{1-1}=\sup_{X} \frac{\vert \vert \env_j (X) \vert \vert_{1}}{\vert \vert X \vert \vert_{1}}$, where the optimization is over operators on the Hilbert space of the input of $ \env_j$ . By contractivity, we have that $\vert \vert \env_j \vert \vert _{1-1} \le 1$. The last step is to bound the left-over $1$-norm. By using $\vert \vert A \vert \vert _1 \le \sqrt{d_\text{in}} \vert\vert A \vert \vert_2$ we can bound the $2$ norm instead, as
\begin{align}
\vert \vert  &\sum_{n,n'=0}^{d_{\text{in}}-1}\ket{n}\bra{n} \bra{n}\rho_\inpp \ket{n'} p(Q,\vec k)  \vert \vert_2^2 = \tr \left[ \left (\sum_{n,n'=0}^{d_{\text{in}}-1}\ket{n}\bra{n} \bra{n}\rho_\inpp \ket{n'}p(Q,\vec k) \right)^2\right] \\&= \sum_{n,n''=0}^{d_{\text{in}}-1} \bra{n}\rho_\inpp \ket{n''} \bra{n''}\rho_\inpp \ket{n} p(h_{\code,q}-h_{\code,q'}+h_{\inpp,n}-h_{\inpp,n''},\vec k) p(h_{\code,q}-h_{\code,q'}+h_{\inpp,n}-h_{\inpp,n''},\vec k) \label{eq:pmax} \\
& \le \max\limits_{Q} \vert p(Q,\vec k)\vert^2 \sum_{n,n''=0}^{d_{\text{in}}-1} \bra{n}\rho_\inpp \ket{n''} \bra{n''}\rho_\inpp \ket{n} = \max\limits_{Q} \vert p(Q,\vec k)\vert^2 \tr[\rho_\inpp^2]= \max\limits_{Q} \vert p(Q,\vec k)\vert^2,\nonumber
\end{align}
since $\rho_\inpp=\ketbra{\phi}{\phi}$. Generically in our examples the optimization is over the range $Q \in \{-(\Delta h_\inp+
\Delta h_\code),\ldots,\Delta h_\inp+
\Delta h_\code\}$ as defined in Sec. \ref{sec:finite}, but if the representation $U^\dag_\code(t)$ is trivial (that is, the generator is the identity) then $Q \in \{-\Delta h_\inp,\Delta h_\inp\}$, which yields the best performance. Finally we have
\begin{align}
\vert \vert \hat E_{\vec k} \vert\vert_1  \le \sqrt{d_{\text{in}}} \max_{q'} \sum_q \max\limits_{Q} \vert p(Q,\vec k)\vert = \sqrt{d_{\text{in}}} d_{\text{out}} \max\limits_{Q} \vert p(Q,\vec k)\vert,
\end{align}
which holds for any $\ketbra{\phi}{\phi}$. Since outcome $\vec k$ occurs with probability $p_{\vec k}=F_0(\vec k)$, we have
\begin{equation}\label{eq:fidpmax}
f_{\wor}(\mathcal{K}) \ge 1-\frac{3}{2}\sqrt{d_{\text{in}}} d_{\text{out}} \sum_{\vec k} F_{0}(\vec k) \max\limits_{Q} \vert p(Q,\vec k)\vert.
\end{equation}
For particular sets of clocks and schemes, we will give bounds for $F_{0}(\vec k) \max\limits_{Q} \vert p(Q,\vec k)\vert$, and then use Eq. \eqref{eq:fidpmax} to estimate the entanglement fidelity.

%

\section{Proof of Theorem \ref{re:1clock}, a bound for a single remaining clock}\label{app:1clockproof}

Following the discussion of appendices \ref{app:gen} and \ref{app:entfid}, here we provide the proof of the bound on the quantity that we use to bound the entanglement fidelity for the case in which, after erasure errors, only a single clock is ensured not to be erased. 
For simplicity of notation, let us use the shorthand $d=d_\clo$.

The first step is to notice that the discussion of Appendices \ref{app:gen} and \ref{app:entfid} shows we only have to compute a bound on $\max\limits_{Q} \vert p(Q,\vec k)\vert$, where
\ba\label{eq:ratioF2}
- p(Q,\vec k)= \frac{ F_{Q}(\vec k)}{F_{0}(\vec k)}\,\me^{2\pi \mi {k_\alpha} Q/d} -1,
\ea
as defined in Eq. \eqref{eq:ratioF}. Then, any upper bound on $\vert p(Q,\vec k) \vert $ $\forall \, Q,\vec k$ will yield a lower bound on the worst-case entanglement fidelity as per Eq. \eqref{eq:fidpmax}. We now compute this ratio explicitly for a single Quasi-Ideal clock, which we do following the notation of Appendix \ref{app:gen}.
Let us recall that
\be 
F_{Q}(\vec k) = \frac{1}{T_0} \int_{t_0}^{T_0+t_0} dt\, \me^{-\mi \omega Q t} \braket{\theta_{k_1}|\rho_{\cl 1}(t)|\theta_{k_1}}, \quad \forall \,\,t_0\in\rr,Q\in\zz.
\ee  
In this case of this proof, we let $k_\alpha = k_1-k_1^0$.
After measuring the single clock, and right before applying the decoding, the state is that of Eq. \eqref{eq: rho our generix with F 2}, that is
\ba 
\rho_\outpp^{\vec{k}}
=&\sum_{q,q'=0}^{d_\outp-1}\sum_{n,n'=0}^{d_\inp-1} \left(\left[ \frac{F_{Q}(\vec k)}{F_{0}(\vec k)}\me^{2\pi \mi {k_\alpha} Q/d} -1\right] +1\right)\\
& U_\code^{\otimes \K \dag} (t_\alpha) \map_{q,q',n,n'}^j\left( U_\inp (t_\alpha)\rho_\inp U_\inp^\dag  (t_\alpha)\right) \ketbra{q}{q'} U_\code^{\otimes \K} (t_\alpha).\label{eq:rho p intermediate}
\ea

Using property 2) of $F_Q(\vec k)$ (see Eq. \eqref{eq:F Q shift invarient}), we observe that the first line of Eq. \eqref{eq:rho p intermediate} is invariant under the change of variable $k_1\mapsto k_1+l$ (Since $F_Q(\vec k)$ maps to $\me^{-\mi 2\pi l Q /d} F_Q(\vec k)$ while $k_\alpha$ to $k_\alpha+l$). Thus the situation is the same for all measurement outcomes $k_1$. As such we will only need to concern ourselves with one of the measurement outcome $k_1$, which we choose for convenience to be $k_1=k_{Ma}:=\max \{ \mathcal{S}_{d_1} (k^0_1) \}$, where the set $\mathcal{S}_d(k^0_1)$ is described in the main text is defined to be
\be\label{eq:set S of k def}
\mathcal{S}_d(k_1^0)= \Big\{ k : k\in\zz, -\frac{d}{2} \leq k_1^0-k< \frac{d}{2}\Big \},
\ee 
for $k_1^0\in\rr.$ Thus $k_{Ma}=k_1^0+d/2$ if $k_1^0+d/2$ is integer (we will assume this is the case here, but one can find analogous results for when it is half integer)\footnote{Recall that $k_1^0$ is the value at which the Gaussian amplitude of the initial Quasi-Ideal clock state is centred. 
} We are now interested in bounding the overlap 
\be 
\braket{\theta_{k_{Ma}} | \me^{-\mi t \hat H_c} |\psi_\textup{nor}(k_1^0)}.\label{eq:the overlap}
\ee
This can be bounded with the results in Theorem 8.1 on page 151 of \cite{woods2016autonomous}. The theorem tells us that for all $t\in\rr$
	\begin{equation}\label{eq:main eq theorem Q continuity}
	\me^{-\mi t  \hat{H}_c} \ket{\Psi_\textup{nor}(k_1^0)} = \ket{\Psi_\textup{nor}(k_1^0+t \frac{d}{T_0})} + \ket{\varepsilon},
	\end{equation}
where
\be 
\ket{\Psi_\textup{nor}(k_1^0)} =\sum_{k\in\mathcal{S}_{d}(k_1^0)} \psi_\textup{nor} (k_1^0;k)  \ket{\theta_k},
\ee
with
	\begin{equation}
	\psi_\textup{nor} (k_1^0;k)=A \me^{- \left(\frac{{\pi}}{\sigma}\right)^2 \left(k-k_1^0\right)^2} \me^{ \mi 2\pi n_1\left(\frac{k-k_1^0}{d}\right)}.
	\end{equation}
The error term $\varepsilon_c:=\vert\vert{ \ket{\varepsilon} }\vert\vert_2$  satisfies 
\begin{align}\label{eq: thorem 1 tot error}
\varepsilon_c(t)&< |t| \frac{d}{T_0} \varepsilon_{total} + \left( |t| \frac{d}{T_0} + 1 \right) \varepsilon_{step} +\varepsilon_\textup{nor},
\end{align}	
where
\begin{align}\label{eq:errtot1}
\varepsilon_{total} =2\pi A d \Bigg( 2 \sigma \left( \frac{\alpha_0}{2} + \frac{1}{2\pi\sigma^2} + \frac{1}{1 - e^{-\pi\sigma^2\alpha_0}} \right) e^{-\frac{\pi\sigma^2}{4}\alpha_0^2} + \left( \frac{1}{1 - e^{-\frac{\pi d}{\sigma^2}}} + \frac{1}{1 - e^{-\frac{\pi d^2}{\sigma^2}}} + \frac{d}{2\sigma^2} + \frac{1}{2\pi d} \right) e^{-\frac{\pi d^2}{4\sigma^2}} \Bigg)
\end{align}
and $A$ is defined in Eq. \eqref{eq:defA2}, and $\alpha_0 \in (0,1]$ 
is defined in \cite{woods2016autonomous} to be:\\
\begin{definition}(Distance of the mean energy from the edge of the spectrum)
	We define the parameter $\alpha_0\in(0,1]$ as a measure of how close $n_0\in(0,d-1)$ is to the edge of the energy spectrum,
	namely
	\begin{align}\label{eq:alpha_0 def}
	\alpha_0&=\left(\frac{2}{d-1}\right) \min\{n_0,(d-1)-n_0\}\\
	&=1-\left|1-n_0\,\left(\frac{2}{d-1}\right)\right|\in(0,1].
	\end{align}
	The maximum value $\alpha_0=1$ is obtained for $n_0=(d-1)/2$ when the mean energy is at the mid point of the energy spectrum, while $\alpha_0\rightarrow 0$ as $n_0$ approaches the edge values $0$ or $d-1$.
\end{definition}
Furthermore,
\begin{align}
\varepsilon_{step} &<
2 A e^{-\frac{ \pi d^2}{4\sigma^2}} 
\\
\varepsilon_\textup{nor}& \leq 
\frac{40\sqrt{2}}{3\sigma}\left(\frac{ \me^{-\frac{\pi d^2}{2\sigma^2}}}{1 - \me^{-\frac{2\pi d}{\sigma^2}}}+\frac{\sigma}{\sqrt{2}}\frac{ \me^{-\frac{\pi \sigma^2}{2}}}{1 - \me^{-\pi \sigma^2}}\right)\label{eq:ep nor simple form above}
\end{align}
where on the r.h.s. of the inequality in Eq. \eqref{eq:ep nor simple form above} we have assumed $\sigma\geq1$ and $d=2,3,4,\ldots$ (tighter bounds can be found in Section E.A.2 of \cite{woods2016autonomous}).

We can now get back to estimating the overlap Eq. \eqref{eq:the overlap}. We find that for all $t\in[0,T_0]$,
\ba
\braket{\theta_{k_{Ma}} | \me^{-\mi t \hat H_c} |\psi_\textup{nor}(k_1^0)}=& \braket{\theta_{k_{Ma}} | \me^{-\mi t \hat H_c} |\psi_\textup{nor}(k_1^0)}+ \braket{\theta_{k_{Ma}} | \varepsilon(t)}\\
&\braket{\theta_{k_{Ma}}| \sum_{k\in\mathcal{S}_{d}(k_1^0)} \psi_\textup{nor} (k_1^0+t{d} /T_0;k)  |\theta_k} + \braket{\theta_{k_{Ma}} | \varepsilon(t)}\\
 =& \psi_\textup{nor} (k_1^0+t{d} /T_0;k_{Ma})  + \braket{\theta_{k_{Ma}} | \varepsilon(t)}.
\ea  

Thus when $\rho_{\cl 1}= \ketbra{\psi_\textup{nor}(k_1^0)}{\psi_\textup{nor}(k_1^0)}$, and $t\in[0,T_0]$,
\ba
\braket{\theta_{k_{Ma}}|\rho_{\cl 1}(t)|\theta_{k_{Ma}}} =& \braket{\theta_{k_{Ma}} | \me^{-\mi t \hat H_c} |\psi_\textup{nor}(k_1^0)}   \braket{\psi_\textup{nor}(k_1^0) | \me^{-\mi t \hat H_c} |\theta_{k_{Ma}} }\\
=&|\psi_\textup{nor} (k_1^0+t{d} /T_0;k_{Ma}) |^2  + |\braket{\theta_{k_{Ma}} | \varepsilon(t)}|^2 \\
&+ 2\, \mathfrak{Re} \left( \psi_\textup{nor} (k_1^0+t{d} /T_0;k_{Ma}) \braket{ \varepsilon(t)| \theta_{k_{Ma}}} \right)\\
=&|\psi_\textup{nor} (k_1^0+td /T_0;k_{Ma}) |^2  + \varepsilon_1(t/T_0)\\
=& | A \me^{-\pi \left(\frac{{d}}{\sigma}\right)^2 \left(\frac{t}{T_0}-\frac{1}{2}\right)^2} \me^{ -\mi 2\pi n_1\left(\frac{t}{T_0}-\frac{1}{2}\right)} |^2 +  \varepsilon_1(t/T_0)\\
=& A^2 \me^{-2\pi \left(\frac{{d}}{\sigma}\right)^2 \left(\frac{t}{T_0}-\frac{1}{2}\right)^2}  +  \varepsilon_1(t/T_0).\label{eq:err1ref}
\ea
Thus using $\omega=2\pi/T_0$ and the change of variable $x=t/T_0-1/2$, and setting $t_0=0$ such that integral is over $t\in[0,T_0]$, we find 

\ba
F_{Q}(\vec k) &=  \frac{1}{T_0} \int_{0}^{T_0} dt\, \me^{-\mi \omega Q t} \braket{\theta_{k_{Ma}}|\rho_{\cl 1}(t)|\theta_{k_{Ma}}}\\
&=  \frac{1}{T_0} \int_{0}^{T_0} dt\, \me^{-\mi \omega Q t} A^{2} \me^{-2\pi \left(\frac{{d}}{\sigma}\right)^2 \left(\frac{t}{T_0}-\frac{1}{2}\right)^2} +  \varepsilon_1(t/T_0)\\
&= \me^{- \pi \mi Q }   \int_{-\frac{1}{2}}^{\frac{1}{2}} dx\, \me^{-\mi 2\pi Q x} A^{2} \me^{-2\pi \left(\frac{{d}}{\sigma}\right)^2 x^2} +  \varepsilon_1(x+1/2)\\
&= \me^{- \pi \mi Q }  \int_{-\infty}^{\infty} dx\, \me^{-\mi 2\pi Q x} A^{2} \me^{-2\pi \left(\frac{{d}}{\sigma}\right)^2 x^2} +  \varepsilon_2\\
&=\me^{- \pi \mi Q }  A^{2} \frac{\sigma}{\sqrt{2} d} \me^{-\frac{\pi}{2}\left(\frac{\sigma}{d}\right)^2 Q^2}  +\varepsilon_2,
\ea  
Recalling that $F_0(\vec k)=1/d$ and that, following Section E.1.1 of \cite{woods2016autonomous} we have 
\begin{equation}\label{eq:defA2}
A^2=\sqrt{2}/\sigma+\varepsilon_A,
\end{equation} 
$\varepsilon_A$ is specified later in Eq. \eqref{eq:ep A bound},  we can write
\begin{align}
\frac{F_{Q}(\vec k)}{F_{0}(\vec k)}\me^{2\pi \mi {k_\alpha} Q/d}&=\left(A^{2} \frac{\sigma}{\sqrt{2} } \me^{-\frac{\pi}{2}\left(\frac{\sigma}{d}\right)^2 Q^2}  +d\varepsilon_2\right) \me^{\pi \mi (2{k_\alpha}-d) Q/d} \\
&=\left((1+ \frac{ \sigma}{\sqrt{2} } \varepsilon_A)\me^{-\frac{\pi}{2}\left(\frac{\sigma}{d}\right)^2 Q^2}+d\varepsilon_2\right) \me^{\pi \mi (2{k_\alpha}-d) Q/d}\\
&=(1+ \frac{ \sigma}{\sqrt{2} } \varepsilon_A)\me^{-\frac{\pi}{2}\left(\frac{\sigma}{d}\right)^2 Q^2}+d\varepsilon_2,
\end{align}
where in the last line we have used $k_\alpha=k_1-k_1^0=d/2$ in order to get rid of the phase.
Finally, we need to find bounds for both $\varepsilon_2$ and $\varepsilon_A$.
Let us start with $\varepsilon_2$, defined as
\begin{equation}
\varepsilon_2= \int_{-\frac{1}{2}}^{\frac{1}{2}} \varepsilon_1(x+1/2) \text{d}x+ A^2 e^{-\frac{\pi}{2}\left(\frac{d^2}{\sigma^2}\right)},\label{eq:ep 2 def}
\end{equation}	
where $\varepsilon_1$ is defined in Eq. \eqref{eq:err1ref}. By inspection, we have 
\begin{equation}
\varepsilon_1(t) \le 2 \varepsilon_c(t) + \varepsilon_c(t)^2
\end{equation}
where $\varepsilon_c$ is reproduced in Eq. \eqref{eq: thorem 1 tot error}. Hence going back to the definition of $\epsilon_2$ in Eq. \eqref{eq:ep 2 def}, we can then conclude that 
\begin{equation}
\vert \varepsilon_2\vert  \le 2 \left( d\varepsilon_{total} + (d+1) \varepsilon_{step} +\varepsilon_\textup{nor}\right)+\left( d\varepsilon_{total} + (d+1) \varepsilon_{step} +\varepsilon_\textup{nor}\right)^2+ A^2 e^{-\frac{\pi}{2}\left(\frac{d^2}{\sigma^2}\right)}
\end{equation}
To bound the error term $\varepsilon_A$, we use the results from Eq. (477-483) from Section E.1.1 on page 84 of \cite{woods2016autonomous}. The bound is
\begin{align}\label{eq:ep A bound}
\vert \varepsilon_A \vert \le \frac{\bar\varepsilon_1+\bar\varepsilon_2}{\frac{\sigma}{\sqrt{2}}\left(\frac{\sigma}{\sqrt{2}}-\bar\varepsilon_1-\bar\varepsilon_2 \right)},
\end{align}
where
\begin{align} \bar\varepsilon_1 :=\frac{2 e^{-\frac{\pi d^2}{2\sigma^2}}}{1 - e^{-\frac{2\pi d}{\sigma^2}}},\label{eq:norm bound ep bar 12} \quad\quad \quad
\bar\varepsilon_2:= \frac{\sigma}{\sqrt{2}}\,\frac{2 e^{-\frac{\pi \sigma^2}{2}}}{1 - e^{-\pi \sigma^2}}.
\end{align}
Thus, the leading contribution to $\varepsilon_A$ comes from $\bar\varepsilon_2$, and the leading contribution to $\varepsilon_2$ comes from the first term of $\varepsilon_{total}$ and the last of $\varepsilon_{\text{nor}}$. It can be seen in Eq. \eqref{eq:errtot1} that the first term of the sum only decays as  $\sqrt{\sigma}d e^{-\frac{\pi\sigma^2}{4}\alpha_0^2}$. Thus, we can write
\begin{align}\label{eq:longerror}
\frac{\tilde F_{Q}(\vec k)}{F_{0}(\vec k)}\me^{2\pi \mi {k_\alpha} Q/d} &=\me^{-\frac{\pi}{2}\left(\frac{\sigma}{d}\right)^2 Q^2}\left(1+\mathcal{O}\left( e^{-\frac{\pi \sigma^2}{2}}\right)+\mathcal{O}(e^{-d^2})\right)+\left(\bo(d^3 e^{-\frac{\pi s\sigma^2}{4}})+\mathcal{O}(e^{-\frac{\pi \sigma^2}{2}})+\mathcal{O}(e^{-d^2})\right).
\end{align}
Going back to the definition of Eq. \eqref{eq:ratioF}, this gives the following bound on $\max\limits_{Q} \vert p(Q,\vec k)\vert$,
\begin{align}
\max\limits_{Q} \vert p(Q,\vec k)\vert &\le 1-\min\limits_{Q} \me^{-\frac{\pi}{2}\left(\frac{\sigma}{d}\right)^2 Q^2}\left(1+\mathcal{O}\left( e^{-\frac{\pi \sigma^2}{2}}\right)+\mathcal{O}(e^{-d^2})\right)+\left(\bo(d^3 e^{-\frac{\pi s\sigma^2}{4}})+\mathcal{O}(e^{-\frac{\pi \sigma^2}{2}})+\mathcal{O}(e^{-d^2})\right),\\
&\le 1- \me^{-\frac{\pi}{2}\left(\frac{\sigma}{d}\right)^2 (\Delta h_\inp+\Delta h_\code)^2}\left(1+\mathcal{O}\left( e^{-\frac{\pi \sigma^2}{2}}\right)+\mathcal{O}(e^{-d^2})\right)+\left(\bo(d^3 e^{-\frac{\pi s\sigma^2}{4}})+\mathcal{O}(e^{-\frac{\pi \sigma^2}{2}})+\mathcal{O}(e^{-d^2})\right),
\end{align}
from which, given the discussion of Appendix \ref{app:entfid}, the bound on the entanglement fidelity follows.

To finalise the proof, we need to pick an optimal value of $\sigma$. Choosing $\sigma=\ln^{3}(d)$, we obtain the best scaling of the leading error term in $d$, which gives
\begin{align}
\max\limits_{Q} \vert p(Q,\vec k)\vert &\le 1-\min\limits_{Q}\me^{-\frac{\pi}{2}\left(\frac{\ln^{3}(d)}{d}\right)^2 Q^2}+\mathcal{O}\left(\frac{\ln^{3/2}(d)}{d^3}\right)
\\&=\frac{\pi}{2}\left(\frac{\ln^{3}(d)}{d}\right)^2 (\Delta h_\inp+\Delta h_\code)^2+\mathcal{O}\left(\frac{\ln^{3/2}(d)}{d^3}\right),
\end{align}
from which, given the discussion of Appendix \ref{app:entfid}, the bound on the entanglement fidelity follows.

The choice $\sigma=\ln^{3}(d)$ corresponds to time squeezed Quasi-Ideal lock state (i.e. $\Delta t < \Delta E$, where $\Delta t$ and  $\Delta E$ are the standard deviation in the time and energy bases as described in the main text).

\section{Proof of bound for clocks diagonal in the time eigenbasis} \label{app:diagonalclockproof}

We here give a proof of Theorem \ref{re:diag}, which is a bound on the entanglement fidelity of clocks which are, at some point in their periodic orbit, diagonal in the basis in which the are measured | the time eigenbasis $\{\ket{\theta_k}\}$, conjugate to the energy eigenbasis. That is, we assume that there exits $t_0\in\rr$ such that the initial clock state $\rho_\textup{inc}$ satisfies 
\be \label{eq:dia after t0 application}
U_\clo(t_0)\rho_\textup{inc} U^\dag_\clo(t_0)=  \sum_{k=0}^{d_\clo-1} |A_k|^2\, \ketbra{\theta_{k}} {\theta_{k}}:=\tilde \rho_{\cl 1},
\ee
for some probability amplitudes $\{\,|A_k|\,\}_{k=0}^{d_\clo-1}$.
Therefore,
\be \label{eq:diagclock1}
\braket{r_1|\tilde\rho_{\cl 1}|r_1'}= \sum_{k=0}^{d_\clo-1} |A_k|^2 \braket{r_1|\theta_k} \braket{\theta_k|r_1}=  \sum_{k=0}^{d_\clo-1} \frac{|A_k|^2}{d_\clo} \me^{-\mi 2\pi k(r_1-r_1')/d_\clo}. 
\ee 
Unlike in the previous case of an optimal clock, the above equation only depends on the difference $r_1-r_1'$ rather than $r_1, r_1'$ individually (c.f. the Quasi-Ideal clock case). 
Before preceding, we need to write the covariant encoding channel in a way which reflects the form of Eq. \eqref{eq:dia after t0 application}. Note that Eq. \eqref{eq:dia clocks thorem} in the theorem has the property
\be
\mathcal{E}_{\text{cov}}(\cdot) 
= \frac{1}{T_0}\int_0^{T_0} dt \,  \mathcal{E}_t (\cdot) \otimes U_\clo(t)\,\rho_\textup{inc}\,U_\clo^\dag(t)=\frac{1}{T_0}\int_{t'}^{T_0+t'} dt \,  \mathcal{E}_t (\cdot) \otimes U_\clo(t)\,\rho_\textup{inc}\,U_\clo^\dag(t),\label{eq:dia clocks thorem 2}
\ee
for all $t'\in\rr$. This is easily provable via a change of variable and is a consequence of the fact that the integrand is periodic and we are integrating over one period. Hence via the change $x=t-t_0$ and setting $t'=t_0$, we find
\be
\mathcal{E}_{\text{cov}}(\rho_\inp) 
= \frac{1}{T_0}\int_0^{T_0} dt \,  \mathcal{E}_t (\cdot) \otimes U_\clo(t)\,\rho_\textup{inc}\,U_\clo^\dag(t)=\frac{1}{T_0}\int_{0}^{T_0} dt \,  \mathcal{E}_{t+t_0} (\tilde\rho_\inp) \otimes U_\clo(t)\,\tilde\rho_\textup{\cl 1}\,U_\clo^\dag(t),\label{eq:dia clocks thorem 3}
\ee
where $\tilde\rho_\inp=U^\dag_\inp(t_0) \rho_\inp U_\inp(t_0)\in\mathcal{H}_\inp$. The above equation shows how to write the covariant encoding channel in terms of the clock state $\tilde \rho_{\cl 1}$ which is diagonal in the time basis. Since our results are stated in terms of the entanglement fidelity which is independent of any particular input, the change in inputs $\rho_\inp$ to $\tilde \rho_\inp$ in the above equation is irrelevant and we will thus hence ignore this difference. The only relevant difference is thus that the encoding map $\mathcal{E}_t$ is shifted to $\mathcal{E}_{t+t_0}$. This small difference can easily be accounted for at the decoding stage without extra complication. 

From Eq. \eqref{eq:ecoding with t}, we see that changing $\map_{t}$ for $\map_{t+t_0}$ is equivalent to changing $\map_{q,q',n,n'}$ to $\me^{-\mi \omega t_0 (h_{\code,q}-h_{\code,q'}+h_{\inp,n}-h_{\inp,n'})}\map_{q,q',n,n'}$ or equivalently, changing $\map^j_{q,q',n,n'}$ to 
\be 
\tilde \map^j_{q,q',n,n'}:=\me^{-\mi \omega t_0 (h_{\code,q}-h_{\code,q'}+h_{\inp,n}-h_{\inp,n'})}\map^j_{q,q',n,n'}
\ee  
Hence we can use Eq. \ref{eq:rho out k vec} by making the replacements $\map_{q,q',n,n'}^j$ (defined in Eq. \eqref{eq:map envirom plus code}) with $\tilde \map_{q,q',n,n'}^j(\cdot)$ and $\rho_{C,1}$ with $\tilde\rho_{C,1}$, followed by plugging in Eq. \eqref{eq:diagclock1}. Hence recalling the short hand notation $Q := h_{\code,q}-h_{\code,q'}+h_{\inp,n}-h_{\inp,n'}$, we find
\ba
\rho_\outpp^{\vec{k}}=& \frac{1}{ p_{\vec{k}}} \sum_{q,q'=0}^{d_\outp-1}\sum_{n,n'=0}^{d_\inp-1}\sum_{{r}_1, {r}'_1=0}^{d_\clo-1} \delta_{Q+r_1-r_1',0}\, \tilde \map_{q,q',n,n'}^j\left( \rho_\inp \right) \ketbra{q}{q'} \\
&\bra{r_1}\tilde\rho_{\cl 1}\ket{r_1'}\braket{\theta_{k_1}|r_1}\braket{r_1'|\theta_{k_1}}\\
=& \frac{1}{ p_{\vec{k}}} \sum_{q,q'=0}^{d_\outp-1}\sum_{n,n'=0}^{d_\inp-1}\sum_{{r}_1, {r}'_1=0}^{d_\clo-1} \delta_{Q+r_1-r_1',0}\, \tilde \map_{q,q',n,n'}^j\left( \rho_\inp \right) \ketbra{q}{q'} \\
& \sum_{k=0}^{d_\clo-1} \frac{|A_k|^2}{d_\clo^2} \me^{-\mi 2\pi (k_1-k)(r_1-r_1')/d_\clo}.\label{eq:intermidiate new}
\ea
Now applying 
\be 
\sum_{r,r'=0}^{d-1} f(r-r')= \sum_{x=-(d-1)}^{d-1} (d-|x|) f(x) \quad\forall f:\rr\to \rr
\ee 
to Eq. \eqref{eq:intermidiate new}, we have
\ba
\rho_\outpp^{\vec{k}}
=& \frac{1}{ p_{\vec{k}}} \sum_{q,q'=0}^{d_\outp-1}\sum_{n,n'=0}^{d_\inp-1}\sum_{x_1=-(d_\clo-1)}^{d_\clo-1}\sum_{p_1=0}^{d_\clo-1-|x_1|} \delta_{Q+x_1,0}\, \tilde \map_{q,q',n,n'}^j\left( \rho_\inp \right) \ketbra{q}{q'} \\
& \sum_{k=0}^{d_\clo-1} \frac{|A_k|^2}{d_\clo^2} \me^{-\mi 2\pi (k_1-k)x_1/d_\clo}\\
=& \frac{1}{ p_{\vec{k}}} \sum_{q,q'=0}^{d_\outp-1}\sum_{n,n'=0}^{d_\inp-1}\sum_{x_1=-(d_\clo-1)}^{d_\clo-1}(d_\clo-|x_1|)\, \delta_{Q+x_1+\ldots+x_N,0}\, \tilde \map_{q,q',n,n'}^j\left( \rho_\inp \right) \ketbra{q}{q'} \\
& \sum_{k=0}^{d_\clo-1} \frac{|A_k|^2}{d_\clo^2} \me^{-\mi 2\pi (k_1-k)x_1/d_\clo}.\\
=& \frac{1}{ p_{\vec{k}}} \sum_{q,q'=0}^{d_\outp-1}\sum_{n,n'=0}^{d_\inp-1}(d_\clo-|Q|)\,  \tilde \map_{q,q',n,n'}^j\left( \rho_\inp \right) \ketbra{q}{q'} \\
& \sum_{k=0}^{d_\clo-1} \frac{|A_k|^2}{d_\clo^2} \me^{-\mi 2\pi (k_1-k)Q/d_\clo}.
\ea
As all the outcomes are equally likely, we have that $p_{\vec{k}}=1/d_\clo$. Hence
\ba \label{eq:incoclock}
\rho_\outpp^{\vec{k}}=&  \sum_{q,q'=0}^{d_\outp-1}\sum_{n,n'=0}^{d_\inp-1}\left(1-\frac{|Q|}{d_\clo}\right)\,  \sum_{k=0}^{d_\clo-1} |A_k|^2\, \hat O_{q,q',n,n'}(t_{k-k_1}),
\ea
where we have defined
\be 
\hat O_{q,q',n,n'}(t_{k-k_1}):=  U_\code^{\otimes \K \dag}(t_{k-k_1}) \left(  \tilde \map_{q,q',n,n'}^j\left( U_\inp(t_{k-k_1})\rho_\inp U_\inp^\dag (t_{k-k_1}) \right) \right) U_\code^{\otimes \K}(t_{k-k_1})
\ee 
with $t_{k-k_1}= \frac{2\pi}{\omega}(k-k_1)$. 
It is convenient to write this as
\be 
\rho_\outpp^{\vec{k}}=   \sum_{k=0}^{d_\clo-1} |A_k|^2 \left(\rho_\outpp(t_{k-k_1})+\frac{1}{d_\clo}\hat \delta (t_{k-k_1})\right),\label{rho out decohered sum form}
\ee 
where
\ba \label{eq:errorinc}
\rho_\outpp(t)&:=      \sum_{q,q'=0}^{d_\outp-1}\sum_{n,n'=0}^{d_\inp-1}  \hat O_{q,q',n,n'}(t)=   U_\code^{\otimes \K \dag}(t+t_0) \left(  \env_j \map \left( U_\inp(t)\rho_\inp U_\inp^\dag (t) \right) \right) U_\code^{\otimes \K}(t+t_0),\\
\hat\delta (t)&:= - \sum_{q,q'=0}^{d_\outp-1}\sum_{n,n'=0}^{d_\inp-1}|Q|\,  \hat O_{q,q',n,n'}(t).
\ea
Observe that if we can apply the map
\be \label{eq:somedecoder}
U_\inp^\dag(t_{k-k_1}) \mathcal{D}_j\left(U_\code^{\otimes \K}(t_{k-k_1}+t_0) (\cdot) U_\code^{\otimes \K \dag}(t_{k-k_1}+t_0)  \right) U_\inp(t_{k-k_1})
\ee 
to $\rho_\outpp(t_{k-{k_1}})$, we have perfect error correction, i.e. 
\be 
U_\inp^\dag(t_{k-k_1}) \mathcal{D}_j\left(U_\code^{\otimes \K}(t_{k-k_1}+t_0) \rho_\outpp(t_{k-k_1}) U_\code^{\otimes \K \dag}(t_{k-k_1}+t_0)  \right) U_\inp(t_{k-k_1})=  \rho_\inp\quad \forall\, t_0,t_{k-k_1} \in\rr.
\ee 
as such we define the error term
\be 
\hat E(t):= U_\inp^\dag(t) \mathcal{D}_j\left(U_\code^{\otimes \K}(t+t_0) \delta(t) U_\code^{\otimes \K \dag}(t+t_0)  \right) U_\inp(t).
\ee 
Let us assume we apply an arbitrary decoder $\mathcal{D}$ to $\rho_\outpp^{\vec{k}}$. The fidelity with an arbitrary initial pure state $\ket{\psi}\bra{\psi}$ is
\begin{align}
\bra{\psi}  \sum_{k=0}^{d_\clo-1} |A_k|^2 \mathcal{D}\left(\rho_\outpp(t_{k-k_1})+\frac{1}{d_\clo}\hat \delta (t_{k-k_1})\right) \ket{\psi} & \le \max_{k}  \bra{\psi}  \mathcal{D}\left(\rho_\outpp(t_{k-k_1})+\frac{1}{d_\clo} \hat \delta (t_{k-k_1})\right) \ket{\psi} \label{eq:ineqper1} \\ &\le 1- \frac{1}{d_\clo} \min_{k}  \bra{\psi}  \mathcal{D}\left(\hat \delta (t_{k-k_1})\right) \ket{\psi} \label{eq:ineqper2}
\end{align} 
which shows that a the Salecker-Wigner-Peres clock  with $|A_k|^2=\delta_{k,k'}$ is the optimal choice. 

Crucially, the term in the optimization $\bra{\psi}  \mathcal{D}\left(\hat \delta (t_{k-k_1})\right) \ket{\psi}$ now only depends on parameters of the  encoding $\mathcal{E}$ and decoding $\mathcal{D}$ independently of the clock and its dimension $d_\clo$. Now, let us assume there exists a sequence of encoding-decoding schemes such that for some fixed $d_\outpp$,
\begin{equation}
\lim_{l\rightarrow \infty} \min_{k}  \bra{\psi}  \mathcal{D}_{l}\left(\hat \delta_l (t_{k-k_1})\right)\ket{\psi}  = 0.
\end{equation}
If this were true, one could achieve an arbitrarily large entanglement fidelity for all pure states without increasing the size of the clock. 
This contradicts the existing no-go results \cite{eastin2009restrictions,faist2018prep}, so there must exist a constant $C$, such that for all $\mathcal{D}$ and $\hat \delta (t_{k-k_1})$ with fixed $d_\outpp$
\begin{equation}
\min_{k}  \bra{\psi}  \mathcal{D}\left(\hat \delta (t_{k-k_1})\right) \ket{\psi} \ge C > 0,
\end{equation}
which can be chosen such that the inequality is saturated for some $\mathcal{D}$ and $\hat \delta (t_{k-k_1})$.
Hence, we have that
\begin{align}
f_{\wor}(\mathcal{K})& \le 1-\frac{C}{d_\clo}\label{eq: error f for classical case}.
\end{align}
On the other hand, since the Salecker-Wigner-Peres clock saturates the inequalities \eqref{eq:ineqper1} and \eqref{eq:ineqper2} we can apply the decoder $U_{\text{\inp}}(t)\mathcal{D}_j U^\dagger_{\text{\inp}}(t)$ in \eqref{eq:somedecoder} to obtain
\begin{equation}
f_{\wor}(\mathcal{K})=1-\frac{C^*}{d_\clo},
\end{equation}
for some $C^*\ge C$, as the only dependence with the dimension of the clock is in the term $\frac{1}{d_\clo}\hat \delta (t_{k-k_1})$.
\section{Proof of Theorem \ref{re:Nclock}}\label{app:Nclock}

We just need to show that with $L$ $d_\clo$-dimensional clocks we can construct a clock state with the same properties as a single clock of dimension $L(d_\clo-1)+1$. Let us take the Hamiltonian of $L$ non-interacting clocks of dimension $d_\clo$,
\begin{equation}
H_{\text{tot}}=\bigoplus_{l=1}^L H_\clo = \sum_{r=0}^{L (d_\clo-1) } \omega r \sum_{\alpha} \ketbra{E_{r,\alpha}}{E_{r,\alpha}}
\end{equation}
which has energy levels $\{0, L (d_\clo-1) \}$, and $\alpha$ is the degeneracy index, as this new Hamiltonian has a very large degeneracy. Let us choose an arbitrary non-degenerate subspace of maximal dimension $d(\Nnew) := L(d_\clo-1)+1$, such that $H_{\text{tot}} = H_{\text{clock}}\, \hat\oplus \,H_{\perp}$.\footnote{Here $\hat\oplus$ denotes the Direct sum.} This can be done for instance by choosing all the eigenvectors with a fixed $\alpha=1$, in which case
\begin{equation}
H_{\text{clock}} = \sum_{r=0}^{d_L-1 } \omega r \ketbra{E_{r,1}}{E_{r,1}}.\label{eq:clock effective Ham}
\end{equation}
This defines a subspace with dimension $d(L)$ and with a Hamiltonian with equally-spaced energy levels with gap $\omega$. Now given any single clock state $\rho_\clo=\sum_{r_1,r_2=0}^{d_\clo-1} A_{r_1,r_2} \ketbra{r_1}{r_2}_\clo\in\mathcal{S}(\mathcal{H}_\clo)$ in a $d_\clo$ dimensional space, we can construct a $d(L)$ dimensional space via the mappings $d_\clo\mapsto d(L)$ and $\rho_\clo\mapsto \rho':=\sum_{r_1,r_2=0}^{d(L)-1} A_{r_1,r_2} \ketbra{E_{r_1,1}}{E_{r_2,1}}\in\mathcal{S}(\mathcal{H}^{\otimes L}_\clo)$. This large dimensional clock state will now be a superposition of product states (the energy eigenstates) and will thus be entangled. Thus, in any scheme that uses a clock with dimension $d_\clo$, such that it achieves a fidelity bounded by $f(d_\clo)$ (either from above or below), one can now use this $d(\Nnew)$ dimensional clock, to achieve a fidelity bounded by $f(d(\Nnew))$ (again, either from above or below). This way, we can use a large amount of clocks to vastly improve the performance of the codes.


\section{The converse bound from \cite{faist2018prep}}\label{app:converse}

Here we show how to adapt the bound from \cite{faist2018prep} to our setting \footnote{We thank Philippe Faist for sharing this argument with us and allowing us to reproduce it here.}. To understand why the bound cannot be straightforwardly applied to our setting, first recall our definition of the covariant code $\map_\textup{cov}(\cdot)$. For all $t \in \mathbb{R}$, and a given encoding $\mathcal{E}(\cdot)$ and group representations $U_\inpp(t),U_\code(t)^{\otimes \K}$, we define
\begin{equation}
\mathcal{E}_t(\cdot) := U_\code(t)^{\otimes \K}\mathcal{E}(U_\inpp^\dagger (t)(\cdot) U_\inpp (t))U_\code(t)^{\dagger \otimes \K}.
\end{equation}
The covariant encoding that we use throughout our work is then the CPTP map:
\begin{equation}\label{eq:code3 2}
\mathcal{E}_{\text{cov}}(\cdot) :=\frac{1}{T_0} \int_0^{T_0} \text{d}t \,  \mathcal{E}_t (\cdot) \otimes U_\clo(t)^{\otimes \M}\,\rho_{\clo }^{(\M)}\,U_\clo(t)^{\dagger\otimes \M}.
\end{equation}
This construction is by definition not an isometry, and in general the output $\mathcal{E}_{\text{cov}}(\cdot)$ is a mixed state even for pure inputs. However, the theorem of \cite{faist2018prep} assumes that the encoding map is an isometry, and as such it does not directly apply to our results. However, since in our setup we assume that the encoding of Eq. \eqref{eq:code3 2} is covariant, it is possible to construct a Stinespring dilation with an isometry to which the bound of \cite{faist2018prep} can be directly applied. 

We can construct the Stinespring dilation with an isometry directly for Eq. \eqref{eq:code3 2}. However, since in our set-up $M-N$ clocks are lost due to erasure errors, and the erased clocks are inaccessible, we can trace out $M-N$ clocks and work with the resultant effective encoding map instead. Since we also assumed no correlations between the erased and non-erased clocks, this leads to the effective channel 
\begin{equation}\label{eq:code3 3}
\tr_{C_{N+1}\ldots C_M}\left[\mathcal{E}_{\text{cov}}(\cdot)\right] =\frac{1}{T_0} \int_0^{T_0} \text{d}t \,  \mathcal{E}_t (\cdot) \otimes U_\clo(t)^{\otimes N}\,\rho_{\clo }^{(N)}\,U_\clo(t)^{\dagger\otimes N}.
\end{equation}
Finally, since in some instances the generator of the unitary group on the clock can be exchanged for an effective generator w.l.o.g. (noticeably Theorem \ref{re:Nclock}), we will assume in Eq. \eqref{eq:code3 3} that the unitary representation of the compact U$(1)$ group on the clock $U_\clo(t)^{\otimes N}$, takes on the generic form $U_\clo(t)^{\otimes N}= \me^{-\mi t \bar H_\clo}$ and specialise it later in this section to specific cases. 

We can now proceed with the isometry. Notice that $\map(\cdot)$ can be dilated to an isometry $V_{\inpp\rightarrow \code \textup{A}}$ defined such that the representation of the symmetry group acts trivially on system $\textup{A}$. Also, the state $\rho_{\clo }^{(N)}$ may not be a pure state, but it can also be dilated to $\ket{\Psi_{\clo\bar{\clo}}}\bra{\Psi_{\clo\bar{\clo}}}^{(N)}$, again such that the symmetry group again acts trivially on system $\bar{\clo}$. We thus define

\begin{equation}\label{eq:codedil}
\bar {\mathcal{E}}_{\text{cov}}(\cdot) :=\frac{1}{T_0} \int_0^{T_0} \text{d}t   \mathcal{V}_t (\cdot) \otimes \left(U_\clo(t)^{\otimes N}\otimes \id_{\bar{\clo}}^{\otimes N}\right)\ketbra{\Psi_{\clo\bar{\clo}}}{\Psi_{\clo\bar{\clo}}}\left(U_\clo(t)^{\dagger\otimes N}\otimes \id_{\bar{\clo}}^{\otimes N}\right),
\end{equation}
where now we have
\begin{equation}
\mathcal{V}_t(\cdot) := \left(U_\code(t)^{\otimes \K}\otimes \id_{\textup{A}}\right)V_{\inpp\rightarrow \code \textup{A}}(U_\inpp^\dagger (t)(\cdot) U_\inpp (t))V^\dagger_{\inpp\rightarrow \code \textup{A}}\left(U_\code(t)^{\dagger \otimes \K}\otimes \id_{\textup{A}}\right).
\end{equation}
The only reason why $\bar {\mathcal{E}}_{\text{cov}}(\cdot)$ is not yet an isometry is the twirling over the group, for which we are also able to define a dilation with the help of the Choi-Jamiolkowski isomorphism. 
First, let $\{\ket{k}_\inpp\}$ be the eigenbasis of $\inpp$ with $\hat H_\inpp=\sum_{k} \omega h_{\inpp, k} \ketbra{k}{k}_\inpp$, and let $\ket{\Phi_{\inpp\bar{\inpp}}}=\sum_{k} \ket{k}_\inpp \otimes \ket{k}_{\bar{\inpp}}$ be an unnormalized maximally entangled state between the logical space $\inpp$ and a copy $\bar{\inpp}$. Moreover, let us define $\hat H_{\bar{\inpp}}=\sum_{k} - \omega h_{\inpp, k} \ketbra{k}{k}_{\bar{\inpp}}$ such that $\left( \hat H_\inpp\otimes \id_{\bar{\inpp}}+\id_{\inpp}\otimes \hat H_{\bar{\inpp}}\right) \ket{\Phi_{\inpp\bar{\inpp}}}=0$.
The Choi-Jamiolkowski representation of channel $\bar {\mathcal{E}}_{\text{cov}}$ is (we now omit writing the trivial representations such as $\id_\textup{A},\id_{\bar{\clo}}^{\otimes N}$ for simplicity of notation)

\begin{align}
\bar {\mathcal{E}}_{\text{cov}}(\ketbra{\Phi_{\inpp\bar{\inpp}}} {\Phi_{\inpp\bar{\inpp}}}) &= \frac{1}{T_0}\int_0^{T_0}\, \text{d}t \,  \mathcal{V}_t (\ketbra{\Phi_{\inpp\bar{\inpp}}}{\Phi_{\inpp\bar{\inpp}}}) \otimes\left( U_\clo(t)^{\otimes N}\,\ketbra{\Psi_{\clo\bar{\clo }}}{\Psi_{\clo\bar{\clo}}}\,U_\clo(t)^{\dagger\otimes N}\right) \\ &=\frac{1}{T_0}\int_0^{T_0} \text{d}t \, U_{\code \clo\bar{\inpp}}(t)\left( V_{\inpp\rightarrow \code \textup{A}}\ketbra{\Phi_{\inpp\bar{\inpp}}}{\Phi_{\inpp\bar{\inpp}}}V_{\inpp\rightarrow \code \textup{A}}^\dagger \otimes \ketbra{\Psi_{\clo\bar{\clo}}}{\Psi_{\clo\bar{\clo}}}\right)U_{\code \clo\bar{\inpp}}^\dagger(t),
\end{align} 
where in the second line we have used the properties of the maximally entangled state, and we define $U_{\code \clo\bar{\inpp}}(t):= U_\code(t)^{\otimes \K}\otimes U_\clo(t)^{\otimes N}  \otimes U_{\bar{\inpp}} (t)=\me^{-\mi t (\hat H_\code \oplus \bar H_\clo \oplus \hat H_{\bar{\inpp}})}$. Thus, we can write $\bar {\mathcal{E}}_{\text{cov}}$ in terms of the projectors onto the degenerate eigenspaces of $\hat H_\code \oplus \bar H_\clo \oplus \hat H_{\bar{\inpp}}$ as
\begin{align}
\bar {\mathcal{E}}_{\text{cov}}(\ketbra{\Phi_{\inpp\bar{\inpp}}} {\Phi_{\inpp\bar{\inpp}}}) &= \sum_x \Pi^x_{\code \clo\bar{\inpp}}\left[V_{\inpp\rightarrow \code \textup{A}}\ketbra{\Phi_{\inpp\bar{\inpp}}}{\Phi_{\inpp\bar{\inpp}}}V_{\inpp\rightarrow \code \textup{A}}^\dagger \otimes \ketbra{\Psi_{\clo\bar{\clo}}}{\Psi_{\clo\bar{\clo}}}\right] \Pi^x_{\code \clo\bar{\inpp}}.
\end{align}
Since the operators $\hat H_\code, \bar H_\clo, \hat H_{\bar\inpp}$ act on different subsystems, we can decompose $\Pi^x_{\code \clo\bar{\inpp}}$ as
\begin{equation}
\Pi^x_{\code \clo\bar{\inpp}}=\sum_{\substack{k,l: \\ \,\,-h_{\inpp, k}+h_{\code\clo, l}=x}} \Pi^l_{\code \clo}\otimes \Pi^k_{\bar{\inpp}},
\end{equation}
where $h_{\code\clo, l}$ are the eigenvalues of $\hat H_\code \oplus \bar H_\clo$ and $h_{\inpp, k}$ are the eigenvalues of $\hat H_{\bar{\inpp}}$. Furthermore we can write $\Pi^k_{\bar{\inpp}}= \ketbra{k}{k}_{\bar{\inpp}}$. If we write the corresponding projector in $\inpp$ as $\Pi^k_{\inpp}= \ketbra{k}{k}_{\inpp}$, using the definition of $\ketbra{\Phi_{\inpp\bar{\inpp}}}{\Phi_{\inpp\bar{\inpp}}}$ we have that

\begin{equation}
\bar {\mathcal{E}}_{\text{cov}}(\ketbra{\Phi_{\inpp\bar{\inpp}}}{\Phi_{\inpp\bar{\inpp}}})=\sum_{\substack{k,l,k',l' \\ -h_{\inpp, k}+h_{\code\clo, l}=-h_{\inpp, k'}+h_{\code\clo, l'}}} \Pi^l_{\code \clo} \left[V_{\inpp\rightarrow \code \textup{A}}\Pi^{k'}_{\inpp} \ketbra{\Phi_{\inpp\bar{\inpp}}}{\Phi_{\inpp\bar{\inpp}}} \Pi^k_{\inpp}V_{\inpp\rightarrow \code \textup{A}}^\dagger\otimes \ketbra{\Psi_{\clo\bar{\clo}}}{\Psi_{\clo\bar{\clo}}}\right] \Pi^{l'}_{\code \clo}.
\end{equation}
Now we can define an extra system $\textup{B}$ on a Hilbert space spanned by  orthonormal basis vectors $\{\ket{x}_B\}$, and such that $\hat H_\textup{B}=\sum_{x \in \{h_{\code\clo, l}-h_{\inpp, k}\}}(x)\ketbra{x}{x}_B$, so that every single energy difference $h_{\code\clo, l}-h_{\inpp, k}$ appears in the sum only once (note that by assumption this set is finite). With this, we can define the following isometry from the logical space labeled by $\inpp$ to $\textup{B\code AC}\bar{\clo}$.
\begin{equation}
W_{\inpp\rightarrow \textup{B\code AC}\bar{\clo}}= \sum_x \ket{x}_B\otimes \sum_{\substack{k,l \\ h_{\code \clo, l}-h_{\inpp, k}=x}}\left[\Pi^l_{\code \clo}(V_{\inpp\rightarrow \code \textup{A}}\Pi_\inpp^{k})\otimes \ket{\Psi_{\clo\bar{\clo}}}\right]. 
\end{equation}
This is indeed an isometry since $W_{\inpp\rightarrow \textup{B\code AC}\bar{\clo}}^\dagger W_{\inpp\rightarrow \textup{B\code AC}\bar{\clo}} =\id_\inpp$. It is covariant, in the sense that by construction
\begin{equation}
W_{\inpp\rightarrow \textup{B\code AC}\bar{\clo}}\, \me^{-\mi t \hat H_\inpp}=\left(\me^{-\mi t(\hat H_\code \oplus \bar H_\clo \oplus \hat H_\textup{B})}\otimes \id_{\textup{A} \bar{\clo}}\right) W_{\inpp\rightarrow \textup{B\code AC}\bar{\clo}}.
\end{equation}
Moreover, it can be easily computed that 
\begin{equation}
\tr_\textup{B}\left[ W_{\inpp\rightarrow \textup{B\code AC}\bar{\clo}}\ketbra{\Phi_{\inpp\bar{\inpp}}} {\Phi_{\inpp\bar{\inpp}}}W_{\inpp\rightarrow \textup{B\code AC}\bar{\clo}}^\dagger\right]=\bar {\mathcal{E}}_{\text{cov}}(\ketbra{\Phi_{\inpp\bar{\inpp}}}{\Phi_{\inpp\bar{\inpp}}})
\end{equation} and subsequently
\begin{equation}
\tr_{\textup{BA}\bar{\clo}}\left[ W_{\inpp\rightarrow \textup{B\code AC}\bar{\clo}}\ketbra{\Phi_{\inpp\bar{\inpp}}}{\Phi_{\inpp\bar{\inpp}}}W_{\inpp\rightarrow \textup{B\code AC}\bar{\clo}}^\dagger\right]=\tr_{C_{N+1}\ldots C_M}\left[\mathcal{E}_{\text{cov}}(\ketbra{\Phi_{\inpp\bar{\inpp}}}{\Phi_{\inpp\bar{\inpp}}})\right], 
\end{equation}
so $W_{\inpp\rightarrow \textup{B\code AC}\bar{\clo}}$ is a covariant Stinespring dilation of channel $\tr_{C_{N+1}\ldots C_M}\left[\mathcal{E}_{\text{cov}}(\cdot)\right]$, which we can see as an encoding isometry for which the error is $i)$ first the loss to the environment of systems $\textup{BA}\bar{\clo}$, and then $ii)$ an erasure error channel for $\mathcal{E}_{\text{cov}}$. The result of \cite{faist2018prep} states that for these isometries, the entanglement fidelity achieved by any decoding scheme is bounded by
\begin{equation}
f_{\wor} \le 1- \frac{\Delta h_\inpp^2}{4 \mathcal{N}^2 \Delta h_{\text{loss}}^2},
\end{equation}
where $\Delta h_\inpp$ is the range of values of the set $\{h_{\inpp, k}\}$ of $\hat H_\inpp$, $\mathcal{N}$ is the number of subsystems in the encoding that can be erased independently by the error model, and where $\Delta h_{\text{loss}}$ is the largest energy difference in the Hamiltonians of all the subsystems that are lost to the environment. We have that $\textup{BA}\bar{\clo}$ are lost and since $\textup{A},\bar{\clo}$ have trivial generators we just have to look at $\hat H_\textup{B}$, so that $\Delta h_{\text{loss}}=\Delta h_\textup{B}$. The eigenvalues of $\hat H_\textup{B}$ are of the form $h_{\code \clo, l}-h_{\inpp, k}$, so that the range of $\hat H_\textup{B}$ is $\Delta h_\textup{B}=\Delta h_\code +\Delta h_\clo$ (the terms with $h_{\inpp, k}$ always contribute negatively and hence $\Delta h_\inpp$ does not enter here). 

We will now specialise the bound to the first case considered in Corollary \ref{re:Nclockoptimal}. Here we have erasure of up to $M-1$ blocks of $L$ entangled clocks so $N=1$. Furthermore, the effective clock generator on the remaining block of $L$ entangled clocks is $\bar H_\clo= H_{\text{clock}}$ (recall Eq. \eqref{eq:clock effective Ham} for an expression for $H_{\text{clock}}$). This effective clock system has $\Delta h_\clo=\inpp d_\clo$. 
Therefore $\Delta h_{\text{loss}}=\Delta h_\code  +\inpp d_\clo$. To compute $\mathcal{N}$, first notice that $\textup{BA}\bar{\clo}$ counts as a single system as it is always lost to the environment. Moreover, as mentioned above, the $M-1$ clocks that are lost to the environment do not appear in the decoding procedure at all, and as detailed above are such that effectively they were not there in the first place. Finally, the error channel on $\outpp $ is later corrected by its own decoding map $\mathcal{D}_j(\cdot)$ and is independent of whether we make the code covariant in the first place, and the block of $L$ entangled clocks that is left (which we have taken to be the whole $\clo$ system) does not get erased, so we can also count both these two as a single subsystem (which gets erased with probability $0$). Hence we have $\mathcal{N}=2$. Putting everything together, we finally obtain
\begin{equation}
f_{\wor} \le 1- \frac{\Delta h_\inpp^2}{16 (\Delta h_\code  +\inpp d_\clo)^2}.\label{eq:f worst up bound}
\end{equation}
The appearance of $\Delta h_\code$ as an additive contribution to the effective clock dimension $L d_\clo$ in the upper bound on the fidelity is to be expected, since in our setup one could have chosen the encoding map $\map$ to be a clock whose decoding map $\mathcal{D}_j$ measures this clock in the same way that the decoding map $\tilde{\mathcal{D}}_{j,q}$ measures the clocks on $\clo$. This scenario would help to make the encoding channel $\map_\textup{cov}$ increase its decoding fidelity, since its effective clock dimension would be $\Delta h_\code  +\inpp d_\clo$. Contrarily, if one considers encoding and decoding channels $\map,\mathcal{D}_j$ which do not help to make the encoding channel $\map_\textup{cov}$ reduce its decoding errors at all, then the value of $\Delta h_\code$ should not be expected to play a role in the decoding fidelity $f_\wor$.

Finally, comparing Eq. \eqref{eq:f worst up bound} with the lower bound of Eq. \eqref{eq:new coherent with L copies}, we see that up to logarithmic factors, the $L$ entangled Quasi-Ideal clocks achieve the optimal error scaling with both their dimension $d_\clo$ and number of clocks $L$. 


\section{Proof of Theorem \ref{Thm:3 clock phase error}}\label{3 clocks proof}
In this section, we will prove Theorem \ref{Thm:3 clock phase error}. We will prove it only for the special case that the three blocks consist of one clock each, i.e. $L=1$. Once Theorem \ref{Thm:3 clock phase error} is proven for this specialised case, the result for larger $L$ is a direct consequence of Theorem \ref{re:Nclock}. Again for simplicity of notation, we will label $d=d_\clo$.

For simplicity, we will start assuming, as for a single clock, that $d$ is even ($d$ odd follows analogously), and that $\sigma$ satisfies
\be 
(0,d)\ni \sigma\rightarrow \infty \text{ and } \frac{d}{\sigma} \rightarrow\infty  \text{ as } d\rightarrow \infty.\label{eq:sigma d limits assumtions}
\ee 

The first goal will be to work out an explicit expression for $F_Q(\vec{k})$ evaluated for the case of three Quasi-Ideal clocks in which a 2-unknown phase error applied to one of them. To indicate this difference (i.e. that un unknown phase has now been applied), we add a tilde to $F$. 
Specifically, we have 
\ba
\tilde F_Q&(k_1,k_2,k_3):=\\
&\frac{1}{T_0} \int_{0}^{T_0} dt\, \me^{-\mi \omega Q t} \braket{\theta_{k_1}|\rho_{\cl 1}(t+\tilde t_{\textup{ph},1})|\theta_{k_1}}\braket{\theta_{k_2}|\rho_{\cl 2}(t+\tilde t_{\textup{ph},2})|\theta_{k_2}}\braket{\theta_{k_3}|\rho_{\cl 3}(t+\tilde t_{\textup{ph},3})|\theta_{k_3}},
\ea
where
\be 
\tilde t_{\textup{ph},q}=
\begin{cases}
	t_{\textup{ph}}  &\mbox{ if } q=r,\\
	0 &\mbox{ otherwise},
\end{cases}
\ee 
and $r=1,2,3$ denotes the clock to which the phase is applied. The variables $r$ and $t_\textup{ph}$ are assumed to be unknown.

Applying Theorem 8.1 (Eq. 71) in \cite{woods2016autonomous} we find 
\ba
\bra{\theta_{k_q}}\rho_{\cl q}&(t+\tilde t_{\textup{ph},q})\ket{\theta_{k_q}} \\ &=\bra{\theta_{k_q}}\me^{-\mi (t+\tilde t_{\textup{ph},q}) \hat H_\clo} \ketbra{\psi_\textup{nor}(k_q^0)}{\psi_\textup{nor}(k_q^0)}_q\, \me^{\mi (t+\tilde t_{\textup{ph},q}) \hat H_\clo}
\ket{\theta_{k_q}}\\
&=\bra{\theta_{k_q}}\me^{\mi t \hat H_\clo}\Big( \ket{\psi_\textup{nor}(\tilde k_q^0)}_{q}+\ket{\varepsilon_c(\tilde t_\textup{ph})}_{q}\Big)\Big(\bra{\varepsilon_c(\tilde t_\textup{ph})}_q+\bra{\psi_\textup{nor}(\tilde k_q^0)}_q\Big) \me^{-\mi t \hat H_\clo}
\ket{\theta_{k_q}}\\
&=\braket{\theta_{k_q}|\rho_{\cl q}(t)|\theta_{k_q}}\bigg{|}_{ k_q^0\mapsto\tilde k_q^0 } + \varepsilon_\textup{I0},\label{eq:adding phase}
\ea
where we have used the Quasi-Ideal clock states $\rho_{\cl q}= \ketbra{\psi_\textup{nor}(k_q^0)}{\psi_\textup{nor}(k_q^0)}_q$ and defined
\be 
\tilde k_q^0=
\begin{cases}
	k_q^0+t_\textup{ph}d/T_0  &\mbox{ if } q=r,\\
	k_q^0 &\mbox{ otherwise}.
\end{cases}
\ee 
In the last line of Eq. \eqref{eq:adding phase} we have defined
\ba 
|\varepsilon_\textup{I0}&|= 
\Big|\bra{\theta_{k_q}}\me^{\mi t \hat H_\clo} \ket{\psi_\textup{nor}(\tilde k_q^0)}_{q}\bra{\varepsilon_c(\tilde t_\textup{ph})}_q \me^{-\mi t \hat H_\clo}
\ket{\theta_{k_q}} \\&+ \bra{\theta_{k_q}}\me^{\mi t \hat H_\clo}\ket{\varepsilon_c(\tilde t_\textup{ph})}_{q}
\bra{\psi_\textup{nor}(\tilde k_q^0)}_q \me^{-\mi t \hat H_\clo}
\ket{\theta_{k_q}}
+        \bra{\varepsilon_c(\tilde t_\textup{ph})}_q \me^{-\mi t \hat H_\clo}
\ket{\theta_{k_q}} \bra{\theta_{k_q}}\me^{\mi t \hat H_\clo}\ket{\varepsilon_c(\tilde t_\textup{ph})}_{q}\Big|\\
&\leq 2\varepsilon_c(\tilde t_\textup{ph},d)+\varepsilon_c^2(\tilde t_\textup{ph},d),
\ea 
where $\varepsilon_c$ is defined in Theorem 8.1 (Eq. 71) in \cite{woods2016autonomous}. Thus from Eqs. \eqref{eq:adding phase} and \eqref{eq: F Q of k Def}, it follows that if the clocks $\rho_\cl{q}$ in question are Quasi-Ideal clocks, then 
\be 
\tilde F_Q(k_1,k_2,k_3)=  F_Q(k_1,k_2,k_3)\Big{|}_{\{k_q^0\mapsto \tilde k_q^0\}_{q=1}^3} + 7\,\varepsilon_\textup{I0},\label{eq:tilde F in terms of F}
\ee  
so we see that a 2-unknown phase error, up to a small error $\varepsilon_c$, simply re-maps the initial time of one of the clocks to an unknown value. 
Given the relation Eq. \eqref{eq:tilde F in terms of F}, our 1st task will be to find an explicit expression for the integral \eqref{eq: F Q of k Def}. We do this in the following section.

\subsection{An Expression for the Integral in $F_Q$}
We start by deriving a general expression for $\braket{\theta_{k}|\rho_{\cl q}(t)|\theta_{k}}$ for $k\in \mathcal{S}_d(k_q^0)$, $q\in\{1,2,3\}$ for Quasi-Ideal clock states. For this we need to consider the overlaps
\ba
\braket{\theta_{k_q}|\psi_\textup{nor}(k_q^0+td/T_0) } &= \bra{\theta_{k_q}} \sum_{k\in \mathcal{S}_d(k_q^0+td/T_0)} \psi_\textup{nor}(k_q^0+td/T_0;k)\ket{\theta_k}\\
&  \bra{\theta_{k_q}} \sum_{k\in \mathcal{S}_d(k_q^0+td/T_0)} \psi_\textup{nor}(k_q^0;k-td/T_0)\ket{\theta_k}.
\ea 
Let $t_q$ be defined via the relation
\be 
k_q-t_q  d/T_0 =\min\{ \mathcal{S}_d(k_q^0)  \}.
\ee 
Using Eq. \eqref{eq:set S of k def} we find $\min\{ \mathcal{S}_d(k_q^0)  \}= \lfloor k_q^0\rfloor-\frac{d}{2}+1$, giving 
\be 
t_q=\left(k_q-\lfloor k_q^0\rfloor+\frac{d}{2}-1\right) \frac{T_0}{d}.
\ee 
Therefore,
\ba 
&\braket{\theta_{k_q}|\psi_\textup{nor}(k_q^0+td/T_0) } = \vspace{2cm} \\
&\;\;\;\;\begin{cases}
	\bra{\theta_{k_q}} \sum_{k\in \mathcal{S}_d(k_q^0)} \psi_\textup{nor}(k_q^0;k-td/T_0)\ket{\theta_k}=\psi_\textup{nor}(k_q^0;{k_q}-td/T_0) &\mbox{ if } t\in[0,t_q]\\
	\bra{\theta_{k_q}} \sum_{k\in \mathcal{S}_d(k_q^0+d)} \psi_\textup{nor}(k_q^0;k-td/T_0)\ket{\theta_k}=\psi_\textup{nor}(k_q^0;{k_q}+d-td/T_0) &\mbox{ if } t\in(t_q,T_0]
\end{cases}&
\ea
Now consider Eq. \eqref{eq: F Q of k Def} with measurement outcomes $k_1 \leq k_2\leq k_3$, with identical clocks other then their starting times denoted $k_1^0,k_2^0,k_3^0$ respectively. We consider this special case w.l.o.g. since the other cases can be reconstructed later using property 3) in Section \ref{app:gen}. Thus using Theorem 9.1 in \cite{woods2016autonomous} we have
\ba
F_Q(k_1,k_2,k_3)=&\frac{1}{T_0} \int_{0}^{T_0} dt\, \me^{-\mi \omega Q t} \braket{\theta_{k_1}|\rho_{\cl 1}(t)|\theta_{k_1}}\braket{\theta_{k_2}|\rho_{\cl 2}(t)|\theta_{k_2}}\braket{\theta_{k_3}|\rho_{\cl 3}(t)|\theta_{k_3}}\\
=&\frac{1}{T_0} \int_{0}^{T_0} dt\, \me^{-\mi \omega Q t} \left| A\psi_\textup{nor}(k_1^0;\hat k_1(t,k_1)-t d/T_0) +\braket{\theta_{k_1}|\varepsilon_c(t)} \right|^2\\
&\quad\quad\quad\quad\quad\quad\quad \left| A\psi_\textup{nor}(k_2^0;\hat k_2(t,k_2)-t d/T_0) +\braket{\theta_{k_2}|\varepsilon_c(t)} \right|^2 \\ &\quad\quad\quad\quad\quad\quad\quad \left| A\psi_\textup{nor}(k_3^0;\hat k_3(t,k_3)-t d/T_0) +\braket{\theta_{k_3}|\varepsilon_c(t)} \right|^2\\
=& \int_{0}^{1} dx\, \me^{-\mi 2\pi x Q } \left| A\psi_\textup{nor}(k_1^0;\hat k_1(x T_0,k_1)-x d) +\braket{\theta_{k_1}|\varepsilon_c(xT_0)} \right|^2\\
&\quad\quad\quad\quad\quad\quad  \left| A\psi_\textup{nor}(k_2^0;\hat k_2(xT_0,k_2)-x d) +\braket{\theta_{k_2}|\varepsilon_c(xT_0)} \right|^2\\
&\quad\quad\quad\quad\quad\quad   \left| A\psi_\textup{nor}(k_3^0;\hat k_3(xT_0,k_3)-x d) +\braket{\theta_{k_3}|\varepsilon_c(xT_0)} \right|^2,\label{eq:FQ inter 1}
\ea
where in the last line we performed the change of variable $x=t/T_0$ and used $\omega=2\pi/T_0$ and defined $ t_m=\left(k_m-\lfloor k_m^0\rfloor+ \frac{d}{2}-1\right) \frac{T_0}{d}$ and for $m=1,\ldots, N$
\be 
\hat k_m(t,k_m)=\begin{cases}
	k_m &\mbox{ if } t\in[0,t_m],\\
	k_m +d &\mbox{ if } t\in(t_m,T_0].
\end{cases}
\ee 
From Eq. \eqref{eq:FQ inter 1} we have
\ba 
F_Q(k_1,k_2,k_3)=  
A^6 \int_{0}^{1} dx\, \me^{-\mi 2\pi x Q } \, &\psi_\textup{nor}^2\left(k_1^0;\hat k_1(x T_0,k_1)-x d\right)\\
&\psi_\textup{nor}^2\left(k_2^0;\hat k_2(x T_0,k_2)-x d\right) \psi_\textup{nor}^2\left(k_3^0;\hat k_3(x T_0,k_3)-x d\right)\\
+\, \varepsilon_\textup{I1}.\label{eq:FQ inter 2}
\ea
Using the triangle inequality and the identity $|R+C|^2=R^2+ \varepsilon$ where $|\varepsilon|\leq |C| (2 R +|C|)$ for $R\geq 0$, $C\in\cc$, we can bound the $\varepsilon_\textup{I1}$ term,
\ba 
|\varepsilon_\textup{I1}| \leq A^6 \int_0^1 dx\, 5 \max_{y\in\rr} \psi_\textup{nor}^2(0;y)\, \varepsilon_c(T_0,d)\leq 5 A^6 \, \varepsilon_c(T_0,d),\label{eq:bound its ep}
\ea 
where $\varepsilon_c(T_0,d)$ is an upper bound to $\| \ket{\varepsilon_c(t)}\|_2$ evaluated at $t=T_0$. This quantity, (which already appeared in Appendix \ref{app:1clockproof}) is given in Theorem 8.1 of \cite{woods2016autonomous} and satisfies 
\ba 
\varepsilon_c(T_0,d)&= \frac{T_0}{T_0} \left(  \bo\left( d^2\sigma^{1/2}  \right)\me^{-\frac{\pi}{4}\sigma^2\alpha_0^2}+\bo\left( 1+\frac{d^3}{\sigma^{3/4}}  \right)\me^{-\frac{\pi}{4}\left(\frac{d}{\sigma}\right)^2}  \right) +\bo\left(\me^{-\frac{\pi}{4}\left(\frac{d}{\sigma}\right)^2}\right)+\bo\left(\me^{-\frac{\pi}{4}\sigma^2}\right)\\
&=   \bo\left( d^2\sigma^{1/2}  \right)\me^{-\frac{\pi}{4}\sigma^2\alpha_0^2}+\bo\left( \frac{d^3}{\sigma^{3/4}}  \right)\me^{-\frac{\pi}{4}\left(\frac{d}{\sigma}\right)^2} \label{eq:ep c def}
\ea
where $\alpha_0\in(0,1]$ is defined in Def. 2 in \cite{woods2016autonomous} and explained in Eq. \eqref{eq:alpha_0 def}.

Using $A^2=\sqrt{2}/\sigma+\varepsilon_A=\bo(1/\sigma)$  from Eq. 483 in \cite{woods2016autonomous}, and Eq. \eqref{eq:bound its ep} we conclude
\ba 
|\varepsilon_\textup{I1}| =&   \bo\left( \frac{d^2}{\sigma^{5/2}}  \right)\me^{-\frac{\pi}{4}\sigma^2\alpha_0^2}+\bo\left( \frac{d^3}{\sigma^{15/4}}  \right)\me^{-\frac{\pi}{4}\left(\frac{d}{\sigma}\right)^2}.\label{eq:ep I1 abs bound}
\ea
Dividing the integral in Eq. \eqref{eq:FQ inter 2} into subintervals and substituting for $\psi_\textup{nor}$ we find
\ba 
F_Q&(k_1,k_2,k_3)=\;  
A^6 \int_0^{t_1/T_0}dx\, \me^{-\mi 2\pi x Q}\, \me^{-\frac{2\pi}{\sigma^2} \left[(k_1^0-k_1+x d)^2+(k_2^0-k_2+x d)^2+(k_3^0-k_3+x d)^2  \right]} \\
&+A^6 \int_{t_1/T_0}^{t_2/T_0}dx\, \me^{-\mi 2\pi x Q}\, \me^{-\frac{2\pi}{\sigma^2} \left[(k_1^0-k_1-d+x d)^2+(k_2^0-k_2+x d)^2+(k_3^0-k_3+x d)^2  \right]} \\
&+A^6 \int_{t_2/T_0}^{t_3/T_0}dx\, \me^{-\mi 2\pi x Q}\, \me^{-\frac{2\pi}{\sigma^2} \left[(k_1^0-k_1-d+x d)^2+(k_2^0-k_2-d+x d)^2+(k_3^0-k_3+x d)^2  \right]} \\
&+A^6 \int_{t_3/T_0}^{1}dx\, \me^{-\mi 2\pi x Q}\, \me^{-\frac{2\pi}{\sigma^2} \left[(k_1^0-k_1-d+x d)^2+(k_2^0-k_2-d+x d)^2+(k_3^0-k_3+x d)^2  \right]}\\
&+\, \varepsilon_\textup{I1}.
\ea
For the rest of the proof, it is convenient to work with $\tilde F_Q$ rather than $F_Q$. Substituting the above into Eq. \eqref{eq:tilde F in terms of F} and performing the mapping $k_q^0\mapsto \tilde k_q^0$, $q=1,2,3,$ we find
\ba
\tilde F_Q&(k_1,k_2,k_3)=\; \\
&\;A^6 \int_0^{\bar k_1/d+ 1/2-1/d}dx\, \me^{-\mi 2\pi x Q}\, \me^{-2\pi \left(\frac{d}{\sigma}\right)^2 \left[ \left(x-\frac{\bar k_1}{d}+\frac{\Delta k_1^0}{d}\right)^2+\left(x-\frac{\bar k_2}{d}+\frac{\Delta k_2^0}{d}\right)^2+\left(x-\frac{{\bar k_3}}{d}+\frac{\Delta k_3^0}{d}\right)^2  \right]} \\
&+A^6 \int_{\bar k_1/d+ 1/2-1/d}^{\bar k_2/d+ 1/2-1/d}dx\, \me^{-\mi 2\pi x Q}\, \me^{-2\pi \left(\frac{d}{\sigma}\right)^2 \left[ \left(x-1-\frac{\bar k_1}{d}+\frac{\Delta k_1^0}{d}\right)^2+\left(x-\frac{\bar k_2}{d}+\frac{\Delta k_2^0}{d}\right)^2+\left(x-\frac{{\bar k_3}}{d}+\frac{\Delta k_3^0}{d}\right)^2  \right]} \\
&+A^6 \int_{\bar k_2/d+ 1/2-1/d}^{ {\bar k_3}/d+ 1/2-1/d}dx\, \me^{-\mi 2\pi x Q}\, \me^{-2\pi \left(\frac{d}{\sigma}\right)^2 \left[ \left(x-1-\frac{\bar k_1}{d}+\frac{\Delta k_1^0}{d}\right)^2+\left(x-1-\frac{\bar k_2}{d}+\frac{\Delta k_2^0}{d}\right)^2+\left(x-\frac{{\bar k_3}}{d}+\frac{\Delta k_3^0}{d}\right)^2  \right]} \\
&+A^6 \int_{{\bar k_3}/d+ 1/2-1/d}^{1}dx\, \me^{-\mi 2\pi x Q}\, \me^{-2\pi \left(\frac{d}{\sigma}\right)^2 \left[ \left(x-1-\frac{\bar k_1}{d}+\frac{\Delta k_1^0}{d}\right)^2+\left(x-1-\frac{\bar k_2}{d}+\frac{\Delta k_2^0}{d}\right)^2+\left(x-1-\frac{{\bar k_3}}{d}+\frac{\Delta k_3^0}{d}\right)^2  \right]} \\
&+\, \varepsilon_\textup{I1}+ 7\,\varepsilon_\textup{I0}.\label{eq:FQ inter 5}
\ea
where we have defined initial clock time independent measurement outcomes,
\be 
\bar k_p:=k_p-\lfloor \tilde k_p^0\rfloor\in\mathcal{S}_d(\tilde k_p^0) - \lfloor \tilde k_p^0 \rfloor= \left\{  -\frac{d}{2}+1,-\frac{d}{2}+2,\ldots, \frac{d}{2}  \right\},\quad p=1,2,3,
\ee 
and $1>\Delta k_p^0:= \tilde k_p^0- \lfloor\tilde k_p^0 \rfloor \geq 0$, $p=1,2,3.$
Before proceeding further, we write explicitly bounds for the $\varepsilon$ terms. Using Eq. \eqref{eq:ep c def} we find
\be 
|\varepsilon_\textup{I0}|  \leq 2\varepsilon_c(\tilde t_\textup{ph},d)+\varepsilon_c^2(\tilde t_\textup{ph},d)= \bo\left( d^2\sigma^{1/2}  \right)\me^{-\frac{\pi}{4}\sigma^2\alpha_0^2}+\bo\left( \frac{d^3}{\sigma^{3/4}}  \right)\me^{-\frac{\pi}{4}\left(\frac{d}{\sigma}\right)^2}.
\ee
Thus taking into account \eqref{eq:ep I1 abs bound}, we conclude 
\be
|\varepsilon_\textup{I1}|+7|\varepsilon_\textup{I0}|= \bo\left( d^2\sigma^{1/2}  \right)\me^{-\frac{\pi}{4}\sigma^2\alpha_0^2}+\bo\left( \frac{d^3}{\sigma^{3/4}}  \right)\me^{-\frac{\pi}{4}\left(\frac{d}{\sigma}\right)^2}.
\label{eq:ep I1 final}
\ee
In order to simplify the equations further, we will set $k_1$ to the smallest measurement outcome possible, i.e. $k_1=\lfloor \tilde k_1^0\rfloor-d/2+1$. We can easily generate the other cases by employing Eq. \eqref{eq:F Q shift invarient}, which we postpone to Section \ref{Final expression for FQ k1 k2 k3}. Substituting into Eq. \eqref{eq:FQ inter 5} gives us
\ba 
\tilde F_Q&(k_1,k_2,k_3)=\\
&\;\;\;\;\, A^6 \int_{0}^{\bar k_2/d+ 1/2-1/d}dx\, \me^{-\mi 2\pi x Q}\, \me^{-2\pi \left(\frac{d}{\sigma}\right)^2 \left[ \left(x-\frac{1}{2}-\frac{1-\Delta k_1^0}{d}\right)^2+\left(x-\frac{\bar k_2}{d}+\frac{\Delta k_2^0}{d}\right)^2+\left(x-\frac{{\bar k_3}}{d}+\frac{\Delta k_3^0}{d}\right)^2  \right]} \label{line:int 1 1}\\
&+A^6 \int_{\bar k_2/d+ 1/2-1/d}^{\bar k_3/d+ 1/2-1/d}dx\, \me^{-\mi 2\pi x Q}\, \me^{-2\pi \left(\frac{d}{\sigma}\right)^2 \left[ \left(x-\frac{1}{2}-\frac{1-\Delta k_1^0}{d}\right)^2+\left(x-1-\frac{\bar k_2}{d}+\frac{\Delta k_2^0}{d}\right)^2+\left(x-\frac{{\bar k_3}}{d}+\frac{\Delta k_3^0}{d}\right)^2  \right]}  \label{line:int 2 1}\\
&+A^6 \int_{\bar k_3/d+ 1/2-1/d}^{1}dx\, \me^{-\mi 2\pi x Q}\, \me^{-2\pi \left(\frac{d}{\sigma}\right)^2 \left[ \left(x-\frac{1}{2}-\frac{1-\Delta k_1^0}{d}\right)^2+\left(x-1-\frac{\bar k_2}{d}+\frac{\Delta k_2^0}{d}\right)^2+\left(x-1-\frac{{\bar k_3}}{d}+\frac{\Delta k_3^0}{d}\right)^2  \right]}\label{line:int 3 1} \\
&+\, \varepsilon_\textup{I1}+ 7\,\varepsilon_\textup{I0}.\label{eq:FQ inter 4}
\ea
The intuition is that each of the three above integrals will be small whenever the the corresponding term in square brackets in the exponent does not pass through zero in the interval over which it is being integrated. We now solve for $\bar k_2$, $\bar k_3$ for when this happens. We start with the 1st integral. From line \eqref{line:int 1 1}, we observe that the term
\be
\left[ \left(x-\frac{1}{2}-\frac{1-\Delta k_1^0}{d}\right)^2+\left(x-\frac{\bar k_2}{d}+\frac{\Delta k_2^0}{d}\right)^2+\left(x-\frac{{\bar k_3}}{d}+\frac{\Delta k_3^0}{d}\right)^2  \right]
\ee
can only be zero iff $x=1/2+1/d-\Delta k_1^0/d$, $\bar k_2/d= 1/2+1/d+\left(\Delta k_2^0-\Delta k_1^0\right)/d$, and $\bar k_3/d= 1/2+1/d+\left(\Delta k_3^0-\Delta k_1^0\right)/d$. In this case, $x\in[0,\bar k_2/d+1/2-1/d]=[0,1+ +\left(\Delta k_2^0-\Delta k_1^0\right)/d ]$, so $x$ passes through $x=1/2+1/d-\Delta k_1^0/d$. Observe that this values of $\bar k_2/d,\bar k_3/d$ are outside the domain of $\bar k_2/d,\bar k_3/d$ by an additive factor of $1/d+\left(\Delta k_2^0-\Delta k_1^0\right)/d$ and $1/d+\left(\Delta k_3^0-\Delta k_1^0\right)/d$ respectively. However, since this quantity tends to zero as $d$ becomes large, the integral on line \eqref{line:int 1 1} will provide a significant contribution. For this case, we will make the parametrization $\bar k_2/d=1+1/d-k/d,$  $\bar k_3/d=1+1/d-l/d$, $k,l=1,\ldots,d$.

Similarly for the integral on line \eqref{line:int 2 1}, we find that
\be 
\left[ \left(x-\frac{1}{2}-\frac{1-\Delta k_1^0}{d}\right)^2+\left(x-1-\frac{\bar k_2}{d}+\frac{\Delta k_2^0}{d}\right)^2+\left(x-\frac{{\bar k_3}}{d}+\frac{\Delta k_3^0}{d}\right)^2  \right]
\ee 
can only be zero iff $x=1/2+1/d-\Delta k_1^0/d$, $\bar k_2/d= -1/2+1/d+\left(\Delta k_2^0-\Delta k_1^0\right)/d$, and $\bar k_3/d= 1/2+1/d+\left(\Delta k_3^0-\Delta k_1^0\right)/d$. In this case, $x\in[0,\bar k_2/d+1/2-1/d]=[0,1+ \left(\Delta k_2^0-\Delta k_1^0\right)/d ]$
For this case, we will make the parametrization $\bar k_2/d=-1/2+1/d+m/d,$  $\bar k_3/d=1+1/d-l/d$, $m=0,\ldots,d-1$, $l=1,\ldots,d$.

Similarly for the integral on line \eqref{line:int 3 1}, we find that
\be 
\left[ \left(x-\frac{1}{2}-\frac{1-\Delta k_1^0}{d}\right)^2+\left(x-1-\frac{\bar k_2}{d}+\frac{\Delta k_2^0}{d}\right)^2+\left(x-1-\frac{{\bar k_3}}{d}+\frac{\Delta k_3^0}{d}\right)^2  \right]
\ee 
can only be zero iff  $x=1/2+1/d-\Delta k_1^0/d$, $\bar k_2/d= -1/2+1/d+\left(\Delta k_2^0-\Delta k_1^0\right)/d$, and $\bar k_3/d= -1/2+1/d+\left(\Delta k_3^0-\Delta k_1^0\right)/d$ with $x\in[\bar k_3/d+1/2-1/d,1]=[ \left(\Delta k_3^0-\Delta k_1^0\right)/d,1]$.

Intuitively, the three cases represent all possibilities (satisfying our assumptions $k_1\leq k_2\leq k_3$ ) in which approximately either all three measured elapsed times (these are all proportional to $k_1-k_1^0$, $k_2-k_2^0$, $k_3-k_3^0$) coincide with the measured elapsed time of the 1st clock (proportional to $k_1-k_1^0$), or the three measured elapsed times correspond to some combination of either the measured elapsed time of the first clock or approximately the time corresponding to one period $T_0$ later. 
In all cases
, these correspond approximately to the initial time, due to the periodic motion of the clock, i.e. the clock resets back to the initial time after one period of its motion. So in all cases, we should expect that to a good approximation, the outcomes which are most likely are those when all the clocks mark (at least approximately) the same elapsed time. 

We will now find a good approximation for the last integral (line \eqref{line:int 3 1}), the others can be approximated in the same way.

\subsection{Approximating the integral in line \eqref{line:int 3 1}.}
We will start by substituting the parametrizations $\bar k_2/d=-1/2+1/d+m/d,$  $\bar k_3/d=-1/2+1/d+p/d$, $m,p=0,\ldots,d-1$ into the integral on line \eqref{line:int 3 1}, obtaining
\ba 
\textup{Int}_3&(m,p):= \int_{\bar k_3/d+ 1/2-1/d}^{1}dx\, \me^{-\mi 2\pi x Q}\, \me^{-2\pi \left(\frac{d}{\sigma}\right)^2 \left[ \left(x-\frac{1}{2}-\frac{1}{d}+\frac{\Delta k_1^0}{d}\right)^2+\left(x-1-\frac{\bar k_2}{d}+\frac{\Delta k_2^0}{d}\right)^2+\left(x-1-\frac{{\bar k_3}}{d}+\frac{\Delta k_3^0}{d}\right)^2  \right]}\\
&= \int_{p/d}^{1}dx\, \me^{-\mi 2\pi x Q}\, \me^{-2\pi \left(\frac{d}{\sigma}\right)^2 \left[ \left(x-\frac{1}{2}-\frac{1}{d}+\frac{\Delta k_1^0}{d}\right)^2+\left(x-\frac{1}{2}-\frac{1}{d}+\frac{\Delta k_2^0}{d}-\frac{m}{d}\right)^2+\left(x-\frac{1}{2}-\frac{1}{d}+\frac{\Delta k_3^0}{d}-\frac{p}{d}\right)^2  \right]}.\label{eq:Int 3 m p 0}
\ea  
We now consider the case in which the integral may contain a significant contribution to Eq. \eqref{eq:FQ inter 4}, and write
\ba
\textup{Int}_3(m,p)=& \int_{-\infty}^{\infty}dx\, \me^{-\mi 2\pi x Q}\, \me^{-2\pi \left(\frac{d}{\sigma}\right)^2 \left[ \left(x-\frac{1}{2}-\frac{1}{d}+\frac{\Delta k_1^0}{d}\right)^2+\left(x-\frac{1}{2}-\frac{1}{d}+\frac{\Delta k_2^0}{d}-\frac{m}{d}\right)^2+\left(x-\frac{1}{2}-\frac{1}{d}+\frac{\Delta k_3^0}{d}-\frac{p}{d}\right)^2  \right]}\\
& + \varepsilon^L_3(m,p) + \varepsilon^R_3(m,p),\label{eq:Int 3 m p}
\ea 
where 
\ba 
\left| \varepsilon^L_3(m,p)\right|&= \left|-   \int_{-\infty}^{p/d}dx\, \me^{-\mi 2\pi x Q}\, \me^{-2\pi \left(\frac{d}{\sigma}\right)^2 \left[ \left(x-\frac{1}{2}-\frac{1}{d}+\frac{\Delta k_1^0}{d}\right)^2+\left(x-\frac{1}{2}-\frac{1}{d}+\frac{\Delta k_2^0}{d}-\frac{m}{d}\right)^2+\left(x-\frac{1}{2}-\frac{1}{d}+\frac{\Delta k_3^0}{d}-\frac{p}{d}\right)^2  \right]}  \right|\\
&\leq  \int_{-\infty}^{p/d}dx\,  \me^{-2\pi \left(\frac{d}{\sigma}\right)^2 \left[ \left(x-\frac{1}{2}-\frac{1}{d}+\frac{\Delta k_1^0}{d}\right)^2+\left(x-\frac{1}{2}-\frac{1}{d}+\frac{\Delta k_2^0}{d}-\frac{m}{d}\right)^2+\left(x-\frac{1}{2}-\frac{1}{d}+\frac{\Delta k_3^0}{d}-\frac{p}{d}\right)^2  \right]}.
\ea 
We now employ the relation
\be 
(x + A)^2 + (x + A - y)^2 + (x + A -  z)^2  =  -\frac{1}{3} (y + z)^2 + y^2 + z^2 + 3 \left(x + A - \frac{1}{3} (y + z)\right)^2   \label{eq:xyzA identity} 
\ee 
for $x,y,z,A\in\rr$ followed by introducing the condition (later we will workout a bound for when this condition is not satisfied)
\be 
\frac{p}{d}-\frac{1}{2}-\frac{1}{d}-\frac{p+m-\Delta k_1^0-\Delta k_2^0-\Delta k_3^0}{3 d} \leq - \varepsilon^L_{3,1} \quad \text{ for some }  \varepsilon^L_{3,1}>0,\label{eq:ep 32L def}
\ee 
we find
\ba
\left| \varepsilon^L_3(m,p)\right|\leq& \me^{-2\pi \left(\frac{d}{\sigma}\right)^2 \left[ -\frac{1}{3}\left(\frac{p+m+2 \Delta k_1^0-\Delta k_2^0-\Delta k_3^0}{d}\right)^2+ \left(\frac{p+\Delta k_1^0-\Delta k_3^0}{d}\right)^2+\left(\frac{m+\Delta k_1^0-\Delta k_2^0}{d}\right)^2 \right]} \\
& \int_{-\infty}^{p/d}dx\,  \me^{-6\pi \left(\frac{d}{\sigma}\right)^2 \left( x-\frac{1}{2}-\frac{1}{d}+\frac{\Delta k_1^0}{d}-\frac{p+m+2\Delta k_1^0-\Delta k_2^0-\Delta k_3^0}{3 d} \right)^2 }\\
\leq& \me^{-2\pi \left(\frac{d}{\sigma}\right)^2 \left[ -\frac{1}{3}\left(\frac{p+m+2 \Delta k_1^0-\Delta k_2^0-\Delta k_3^0}{d}\right)^2+ \left(\frac{p+\Delta k_1^0-\Delta k_3^0}{d}\right)^2+\left(\frac{m+\Delta k_1^0-\Delta k_2^0}{d}\right)^2 \right]} \\ &\int_{-\infty}^{p/d}dx\, \frac{ x-\frac{1}{2}-\frac{1}{d}-\frac{p+m-\Delta k_1^0-\Delta k_2^0-\Delta k_3^0}{3 d} }{ \frac{p}{d}-\frac{1}{2}-\frac{1}{d}-\frac{p+m-\Delta k_1^0-\Delta k_2^0-\Delta k_3^0}{3 d} } \, \me^{-6\pi \left(\frac{d}{\sigma}\right)^2 \left( x-\frac{1}{2}-\frac{1}{d}-\frac{p+m-\Delta k_1^0-\Delta k_2^0-\Delta k_3^0}{3 d} \right)^2 }\\
=& \me^{-2\pi \left(\frac{d}{\sigma}\right)^2 \left[ -\frac{1}{3}\left(\frac{p+m+2 \Delta k_1^0-\Delta k_2^0-\Delta k_3^0}{d}\right)^2+ \left(\frac{p+\Delta k_1^0-\Delta k_3^0}{d}\right)^2+\left(\frac{m+\Delta k_1^0-\Delta k_2^0}{d}\right)^2 \right]} \\
& \frac{-1}{2\pi 6} \left( \frac{\sigma}{d}\right)^2 \frac{\me^{-6\pi \left(\frac{d}{\sigma}\right)^2 \left( \frac{p}{d}-\frac{1}{2}-\frac{1}{d}-\frac{p+m-\Delta k_1^0-\Delta k_2^0-\Delta k_3^0}{3 d} \right)^2}  }{\frac{p}{d}-\frac{1}{2}-\frac{1}{d}-\frac{p+m-\Delta k_1^0-\Delta k_2^0-\Delta k_3^0}{3 d} }\\
\leq &\frac{1}{2\pi 6} \left( \frac{\sigma}{d}\right)^2 \frac{\me^{-2\pi \left(\frac{d}{\sigma}\right)^2 \left[ \left(\frac{p}{d}-\frac{1}{2}-\frac{1}{d}+\frac{\Delta k_1^0}{d}\right)^2+\left(\frac{p}{d}-\frac{1}{2}-\frac{1}{d}+\frac{\Delta k_2^0}{d}-\frac{m}{d}\right)^2+\left(\frac{p}{d}-\frac{1}{2}-\frac{1}{d}+\frac{\Delta k_3^0}{d}-\frac{p}{d}\right)^2  \right]}  }{\varepsilon^L_{3,1}}\\
\leq &\frac{1}{2\pi 6} \left( \frac{\sigma}{d}\right)^2 \frac{\me^{-2\pi \left(\frac{d}{\sigma}\right)^2 \left(\frac{1}{2}+\frac{1-\Delta k_3^0}{d}\right)^2}  }{\varepsilon^L_{3,1}}\\
\leq &\frac{1}{2\pi 6} \left( \frac{\sigma}{d}\right)^2 \frac{\me^{-\frac{\pi}{2} \left(\frac{d}{\sigma}\right)^2}  }{\varepsilon^L_{3,1}}, \quad \forall\, m,p=0,\ldots,d-1.
\ea
where in the penultimate line we have used Eq. \eqref{eq:xyzA identity}  and \eqref{eq:ep 32L def}. Analogously, we can bound $\varepsilon^R_3(m,p)$. We find
\ba 
\left| \varepsilon^R_3(m,p)\right|&= \left|-   \int_1^\infty dx\, \me^{-\mi 2\pi x Q}\, \me^{-2\pi \left(\frac{d}{\sigma}\right)^2 \left[ \left(x-\frac{1}{2}-\frac{1}{d}+\frac{\Delta k_1^0}{d}\right)^2+\left(x-\frac{1}{2}-\frac{1}{d}+\frac{\Delta k_2^0}{d}-\frac{m}{d}\right)^2+\left(x-\frac{1}{2}-\frac{1}{d}+\frac{\Delta k_3^0}{d}-\frac{p}{d}\right)^2  \right]}  \right|\\ 
&\leq \frac{1}{2\pi 6} \left( \frac{\sigma}{d}\right)^2 \frac{\me^{-\frac{\pi}{2} \left(\frac{d}{\sigma}\right)^2 \left(1-\frac{2}{d}\right)^2}  }{\varepsilon^R_{3,1}}, \quad m,p=0,\ldots,d-1.
\ea
where we have introduced the constraint
\be 
1-\frac{1}{2}-\frac{1}{d}-\frac{p+m-\Delta k_1^0-\Delta k_2^0-\Delta k_3^0}{3 d} \geq \varepsilon^R_{3,1} \quad \text{ for some }  \varepsilon^R_{3,1}>0.\label{eq:ep 32R def}
\ee 
Thus computing the integral in Eq. \eqref{eq:Int 3 m p} and combining the epsilons into a single term, we have
\ba
\textup{Int}_3(m,p)= & \frac{1}{\sqrt{6}}\left(\frac{\sigma}{d}\right) \me^{-\mi \pi Q} \me^{\mi\pi\frac{2}{3}\Delta k_\textup{T}^0 Q/d} \me^{-\mi\pi\frac{2}{3}(3+m+p)Q/d}\\
&\me^{-\pi \frac{4}{3}\left(\frac{d}{\sigma}\right)^2 \left(\frac{[\Delta k_1^0-\Delta k_2^0+m]^2+[\Delta k_1^0-\Delta k_3^0+p]^2-[\Delta k_1^0-\Delta k_2^0+m] [\Delta k_1^0-\Delta k_3^0+p]}{d^2}\right)}\me^{-\frac{\pi}{6}\left(\frac{\sigma}{d}\right)^2 Q^2 } + \varepsilon_3(m,p),\\
=& \frac{1}{\sqrt{6}}\left(\frac{\sigma}{d}\right)  \me^{-\mi 2\pi \tilde k Q/d} \me^{-\mi\pi\frac{2}{3}(m+p)Q/d}\\
&\me^{-\pi \frac{4}{3}\left(\frac{d}{\sigma}\right)^2 \left(\frac{[\Delta k_1^0-\Delta k_2^0+m]^2+[\Delta k_1^0-\Delta k_3^0+p]^2-[\Delta k_1^0-\Delta k_2^0+m] [\Delta k_1^0-\Delta k_3^0+p]}{d^2}\right)}\me^{-\frac{\pi}{6}\left(\frac{\sigma}{d}\right)^2 Q^2 } + \varepsilon_3(m,p)
,\label{eq:Int 3 m p final inter}
\ea 
where in the last line we have multiplied by $1=\me^{2\pi \mi Q}$ and defined the quantities
\ba
\tilde k &:= -\frac{1}{2}+\frac{1}{d}-\frac{\Delta k_\textup{T}^0}{3 d},\label{eq: tide k def} \\
\Delta k_\textup{T}^0 &:= \Delta k_1^0+ \Delta k_2^0+ \Delta k_3^0 \label{eq: k tot sum d def}.
\ea 
Note that if inequalities \eqref{eq:ep 32L def} and \eqref{eq:ep 32R def} are both satisfied, then 
\be 
\left|\varepsilon_3(m,p)\right| \leq \frac{1}{2\pi 6} \left( \frac{\sigma}{d}\right)^2 \left(\frac{1 }{\varepsilon^L_{3,1}}+\frac{1 }{\varepsilon^R_{3,1}}\right) \me^{-\frac{\pi}{2} \left(\frac{d}{\sigma}\right)^2 \left(1-\frac{2}{d}\right)^2}, \quad \forall\, m,p=0,\ldots,d-1.\label{eq:rp 3 (m n) up bound}
\ee 
Whenever inequality \eqref{eq:ep 32L def} and/or inequality \eqref{eq:ep 32R def} are/is not satisfied, we can bound the integral by the largest value of the integrand, which is exponentially  small in $(d/\sigma)^2$. In this case since the point at which the Gaussian takes on its maximum value lays outside of the integration region. However, it turns out that $(m^2+p^2-mp)/d^2$ is uniformly bounded away from zero in this case and thus the 1st term in \eqref{eq:Int 3 m p final inter} is exponentially small in $(d/\sigma)^2$. We will prove this result, since it will provide a more useful expression. 

We start by defining three functions for $x\in\rr$
\ba 
C_{3L,\underline{p}} (x) &:= -\underline{p} +\frac{1}{2} +\frac{1}{d}-\frac{\Delta k_3^0}{d} +\frac{1}{3}(\underline{p}+x)-\varepsilon^L_{3,1},\\
C_{3R,\underline{p}} (x) &:= 1 -\frac{1}{2} -\frac{1}{d}+\frac{\Delta k_1^0}{d} -\frac{1}{3}(\underline{p}+x)-\varepsilon^R_{3,1}\\
F_{3,\underline{p}}(x)&:= x^2+\underline{p}^2-x\,\underline{p}.
\ea 
With the identification $\underline{m}=(\Delta k_1^0-\Delta k_2^0+m)/d$ and $\underline{p}=(\Delta k_1^0-\Delta k_3^0+p)/d$, we see that the constraints in Eqs. \eqref{eq:ep 32L def}, \eqref{eq:ep 32R def} are the conditions $C_{3L,\underline{p}} (\underline{m})\leq 0$ and $C_{3R,\underline{p}} (\underline{m})\geq 0$ respectively, while $F_{3,\underline{p}}(\underline{m})$ is the exponent in Eq. \eqref{eq:Int 3 m p final inter} which we aim to prove is uniformly bounded away from zero whenever $C_{3L,\underline{p}} (\underline{m})\leq 0$ and $C_{3R,\underline{p}} (\underline{m})\geq 0$ is \emph{not} satisfied, i.e. when
\be 
C_{3L,\underline{p}} (\underline{m})> 0 \;\text{ and/or }\; C_{3R,\underline{p}} (\underline{m})< 0.\label{eq:C 4 negated contraints}
\ee 
Given the range of $p=0,\ldots,d-1$, it is easy to verify that both Eqs, \eqref{eq:C 4 negated contraints} cannot simultaneously hold since $C_{3L,\underline{p}} (\underline{m})> 0$ implies $C_{3R,\underline{p}} (\underline{m}) \geq 0$ and $C_{3R,\underline{p}} (\underline{m})< 0$ implies $C_{3L,\underline{p}} (\underline{m}) \leq 0$. Therefore, it suffices to consider Eq. \eqref{eq:C 4 negated contraints} with the ``or" case only. It is useful to think of these functions as functions of $\underline{m}$ which are parametrized by $\underline{p}$, as the chosen notation suggests. Furthermore, $\frac{d^2}{dx^2} F_{3,\underline{p}}(x)=2$, and thus $F_{3,\underline{p}}(x)$ is a convex function for all $\underline{p}$. As such, we can lower bound $F_{3,\underline{p}}$ by any tangent straight line. 

First consider the case that $C_{3L,\underline{p}} (\underline{m})> 0$ holds. We will proceed to lower bound $F_{3,\underline{p}}(x)$ by the tangent line $A_{\underline{p}}+B_{\underline{p}} \,x $ which intersects $F_{3,\underline{p}}(x)$ at $x=\underline{m}^\star$, where $\underline{m}^\star$ is defined by $C_{3L,\underline{p}} (\underline{m}^\star) =0$. 

It follows that 
\be 
\underline{m}^\star= 3\,\left(\varepsilon^L_{3,1}+\underline{p}-\frac{1}{2}-\frac{1}{d}+\frac{\Delta k_3^0}{d}\right)-\underline{p},
\ee 
from which we find 
\ba
B_{\underline{p}}&=\frac{d}{dx} F_{3\underline{p}}(x)\Big{|}_{x=\underline{m}^\star}=2\underline{m}^\star-\underline{p}=6\left(\varepsilon^L_{3,1}-\frac{1}{2}-\frac{1}{d}+\frac{\Delta k_3^0}{d}\right)+3 \underline{p} \\
&\leq 6\left(\varepsilon^L_{3,1}-\frac{1}{2}-\frac{1}{d}+\frac{\Delta k_3^0}{d}\right)+3\left(1-\frac{1}{d}+\frac{\Delta k_1^0}{d}-\frac{\Delta k_3^0}{d}\right).
\ea 
It is convenient to demand $B_{\underline{p}}\leq 0 $. We thus set 
\be 
\varepsilon^L_{3,1}:= \frac{3}{2}\, \frac{1}{d}-\frac{1}{2}\frac{\Delta k_1^0+ \Delta k_3^0}{d}>\frac{1}{2 d},\label{eq:ep 3 L}
\ee 
so that $B_{\underline{p}}\leq 0$ for all $p=0,\ldots,d-1$. To find $A_{\underline{p}}$, we need to solve the equation 
\be 
A_{\underline{p}}+B_{\underline{p}} \underline{m}^\star = F_{3\underline{p}}(\underline{m}^\star),
\ee 
giving 
\be 
A_{\underline{p}}=  F_{3\underline{p}}(\underline{m}^\star)-B_{\underline{p}} \underline{m}^\star= - \left(\underline{m}^\star\right)^2+ \underline{p}^2.
\ee 
Therefore, putting it all together we have for all $\underline{m},\underline{p}=0,1/d,\ldots, 1-1/d$, 
\be 
F_{3\underline{p}}(\underline{m}) \geq A_{\underline{p}}+B_{\underline{p}}\, {\underline{m}}= -\left({\underline{m}^\star}\right)^2+ \underline{p}^2 + (2 \underline{m}^\star - \underline{p})\, \underline{m}.
\ee 
Taking into account $B_{\underline{p}}\leq 0$, we have $F_{3\underline{p}}(\underline{m}) \geq A_{\underline{p}}+B_{\underline{p}}\, {\underline{m}}\geq A_{\underline{p}}+B_{\underline{p}}\, {\underline{m}^\star}$ for all $\underline{m}< \underline{m}^\star$. Furthermore, by construction, $C_{3L,\underline{p}} (\underline{m})> 0$ holds iff $\underline{m}< \underline{m}^\star$. Thus simplifying, we find 
\be 
F_{3\underline{p}}(\underline{m}) \geq  \left({\underline{m}^\star}\right)^2+ \underline{p}^2 -\underline{p}\,\underline{m}^\star = \frac{9}{4 d^2} \left(d-1 +\Delta k_1^0-\Delta k_3^0 \right)^2-\frac{9}{2 d} \left(d-1 +\Delta k_1^0-\Delta k_3^0 \right)\underline{p}+3\underline{p}^2,
\ee 
if $C_{3L,\underline{p}} (\underline{m})> 0$ holds. 
The R.H.S. is a convex function in $\underline{p}$, thus calculating its stationary point, it can be lower bounded by 
\be 
F_{3\underline{p}}(\underline{m}) \geq  \frac{9}{16}\left(1-\frac{1}{d}+\frac{\Delta k_3^0-\Delta k_1^0}{d}\right)^2 \geq \frac{9}{16}\left(1-\frac{2}{d}\right)^2 ,
\ee
if $C_{3L,\underline{p}} (\underline{m})> 0$ holds.
Performing the analogous procedure for the case that $C_{3R,\underline{p}} (\underline{m})< 0$ holds, and defining
\be 
\varepsilon^R_{3,1}:= \frac{1}{2d}\left(1+\Delta k_1^0+\Delta k_3^0\right)\geq \frac{1}{2 d},\label{eq:ep 3 R}
\ee 
we find
\be 
F_{3\underline{p}}(\underline{m}) \geq  \frac{9}{16}\left(1-\frac{3}{d}\right)^2.
\ee 
Thus summarising the two cases, we see that 
\be 
F_{3\underline{p}}(\underline{m}) \geq  \frac{9}{16}\left(1-\frac{3}{d}\right)^2.
\ee
if the statement ``Inequalities \eqref{eq:ep 32L def}, \eqref{eq:ep 32R def} are both satisfied" is false. Thus defining $\varepsilon_F$ by the equation
\ba
\varepsilon_F&:= \textup{Int}_3(m,p)-\frac{1}{\sqrt{6}}\left(\frac{\sigma}{d}\right)  \me^{-\mi 2\pi \tilde k Q/d} \me^{-\mi\pi\frac{2}{3}(m+p)Q/d}\\
&\quad \quad \me^{-\pi \frac{4}{3}\left(\frac{d}{\sigma}\right)^2 \left(\frac{[\Delta k_1^0-\Delta k_2^0+m]^2+[\Delta k_1^0-\Delta k_3^0+p]^2-[\Delta k_1^0-\Delta k_2^0+m] [\Delta k_1^0-\Delta k_3^0+p]}{d^2}\right)}\me^{-\frac{\pi}{6}\left(\frac{\sigma}{d}\right)^2 Q^2 } ,\label{eq:Int 3 m p final inter 5}
\ea 
we have that 
\ba
\left|\varepsilon_F \right|\leq &  \Big|\textup{Int}_3(m,p) \Big|\\
&+\, \frac{1}{\sqrt{6}}\left(\frac{\sigma}{d}\right) \me^{-\pi \frac{4}{3}\left(\frac{d}{\sigma}\right)^2 \left(\frac{[\Delta k_1^0-\Delta k_2^0+m]^2+[\Delta k_1^0-\Delta k_3^0+p]^2-[\Delta k_1^0-\Delta k_2^0+m] [\Delta k_1^0-\Delta k_3^0+p]}{d^2}\right)}\me^{-\frac{\pi}{6}\left(\frac{\sigma}{d}\right)^2 Q^2 } \\
&\leq  \Big|\textup{Int}_3(m,p) \Big|+\, \frac{1}{\sqrt{6}}\left(\frac{\sigma}{d}\right) \me^{-\pi \frac{3}{4}\left(\frac{d}{\sigma}\right)^2 \left(1-\frac{3}{d}\right)^2} ,\label{eq:Int 3 m p final inter 6}
\ea 
if ``Inequalities \eqref{eq:ep 32L def}, \eqref{eq:ep 32R def} are both satisfied" is false. We now bound $\big|\textup{Int}_3(m,p) \big|$. From Eq. \eqref{eq:Int 3 m p 0} it follows
\ba 
\Big|\textup{Int}_3(m,p)\Big|& \leq \int_{p/d}^{1}dx\, \me^{-2\pi \left(\frac{d}{\sigma}\right)^2 \left[ \left(x-\frac{1}{2}-\frac{1}{d}+\frac{\Delta k_1^2 }{d}\right)^2+\left(x-\frac{1}{2}-\frac{1}{d}-\frac{m}{d}+\frac{\Delta k_2^0}{d}\right)^2+\left(x-\frac{1}{2}-\frac{1}{d}-\frac{p}{d}+\frac{\Delta k_3^0}{d}\right)^2  \right]}\\
&\leq \me^{-2\pi \left(\frac{d}{\sigma}\right)^2 \left[ -\frac{1}{3}\left(\frac{p+m+2\Delta k_1^0-\Delta k_2^0-\Delta k_3^0}{d}\right)^2+ \left(\frac{p+\Delta k_1^0-\Delta k_3^0}{d}\right)^2+\left(\frac{m+\Delta k_1^0-\Delta k_2^0}{d}\right)^2 \right]}\\
&\;\;\;\,  \int_{p/d}^{1}dx\,  \me^{-6\pi \left(\frac{d}{\sigma}\right)^2 \left( x-\frac{1}{2}-\frac{1}{d}-\frac{p+m-\Delta k_\textup{T}^0}{3 d} \right)^2 } \\
&\leq   \me^{-2\pi \left(\frac{d}{\sigma}\right)^2 \left[ -\frac{1}{3}\left(\frac{p+m+2\Delta k_1^0-\Delta k_2^0-\Delta k_3^0}{d}\right)^2+ \left(\frac{p+\Delta k_1^0-\Delta k_3^0}{d}\right)^2+\left(\frac{m+\Delta k_1^0-\Delta k_2^0}{d}\right)^2 \right]} \\ &\;\;\;\,\left(1-\frac{p}{d}\right)  \me^{-6\pi \left(\frac{d}{\sigma}\right)^2 \min_{x\in[p/d,1]}\left( x-\frac{1}{2}-\frac{1}{d}-\frac{p+m-\Delta k_\textup{T}^0}{3 d} \right)^2 } 
.\label{eq:Int 3 m p 1}
\ea
where we have used Eq. \eqref{eq:xyzA identity}. Since the above minimisation is over a parabola, where the smallest value at $x= \frac{1}{2}+\frac{1}{d}+\frac{p+m-\Delta k_\textup{T}^0}{3 d}$ lays outside the interval $[p/d,1]$ whenever  ``Inequalities \eqref{eq:ep 32L def}, \eqref{eq:ep 32R def} are both satisfied" is false, the solution to the minimization is achieved at one of the boundary points, $x=p/d$ or $x=1$. Specifically we find
\be 
\min_{x\in[p/d,1]}\left( x-\frac{1}{2}-\frac{1}{d}-\frac{p+m-\Delta k_\textup{T}^0}{3 d} \right)^2= \left( \Delta_{3,p}-\frac{1}{2}-\frac{1}{d}-\frac{p+m-\Delta k_\textup{T}^0}{3 d} \right)^2,
\ee 
where
\be 
\Delta_{3,p}= 
\begin{cases}
	\frac{p}{d} \mbox{ if } C_{3L,\underline{p}} (\underline{m})\leq 0 \text{ is false}\\
	1 \mbox{ if } C_{3R,\underline{p}} (\underline{m})\geq 0  \text{ is false}.
\end{cases}
\ee 
Thus from Eq. \eqref{eq:Int 3 m p 1}, it follows
\ba 
\Big|\textup{Int}_3(m,p)\Big|& \leq   \me^{-2\pi \left(\frac{d}{\sigma}\right)^2 \left[ -\frac{1}{3}\left(\frac{p+m+2\Delta k_1^0-\Delta k_2^0-\Delta k_3^0}{d}\right)^2+ \left(\frac{p+\Delta k_1^0-\Delta k_3^0}{d}\right)^2+\left(\frac{m+\Delta k_1^0-\Delta k_2^0}{d}\right)^2 \right]}\\
& \;\;\;\;\;   \me^{-6\pi \left(\frac{d}{\sigma}\right)^2 \left( \Delta_{3,p}-\frac{1}{2}-\frac{1}{d}-\frac{p+m-\Delta k_\textup{T}^0}{3 d} \right)^2 } \\
& \leq \frac{1}{d}\, \me^{-2\pi \left(\frac{d}{\sigma}\right)^2 \left[ \left(\Delta_{3,p}-\frac{1}{2}-\frac{1}{d}+\frac{\Delta k_1^0}{d} \right)^2+ \left(\Delta_{3,p}-\frac{1}{2}-\frac{1}{d}+\frac{\Delta k_3^0}{d}-\frac{p}{d}\right)^2+\left(\Delta_{3,p}-\frac{1}{2}-\frac{1}{d}+\frac{\Delta k_2^0}{d}-\frac{m}{d}\right)^2 \right]} \label{eq:line no d assuption int 4}\\
& \leq \frac{1}{d}\, \me^{-2\pi \left(\frac{d}{\sigma}\right)^2 \left(\frac{1}{2}-\frac{2}{d} \right)^2}= \frac{1}{d}\, \me^{-\frac{\pi}{2} \left(\frac{d}{\sigma}\right)^2 \left(1-\frac{4}{d} \right)^2},\quad d=4,6,8,\ldots,
\label{eq:Int 3 m p 2}
\ea
if  ``Inequalities \eqref{eq:ep 32L def}, \eqref{eq:ep 32R def} are both satisfied" is false.
Hence, finally, taking into account Eq. \eqref{eq:Int 3 m p final inter} and Eqs. \eqref{eq:Int 3 m p final inter 5}, \eqref{eq:Int 3 m p 2}, we find that for all $m,p=0,\ldots,d-1$, such that $m\leq p$,\footnote{This last condition is simply to ensure that $\bar k_2\leq \bar k_3$, as has been assumed from the outset.} 
\ba 
\textup{Int}_3(m,p)=& \frac{1}{\sqrt{6}}\left(\frac{\sigma}{d}\right)  \me^{-\mi 2\pi \tilde k Q/d} \me^{-\mi\pi\frac{2}{3}(m+p)Q/d}\\
&\me^{-\pi \frac{4}{3}\left(\frac{d}{\sigma}\right)^2 \left(\frac{\left[m+\Delta k_1^0-\Delta k_2^0\right]^2+\left[p+\Delta k_1^0-\Delta k_3^0\right]^2-\left[m+\Delta k_1^0-\Delta k_2^0\right] \left[p+\Delta k_1^0-\Delta k_3^0\right]}{d^2}\right)}\me^{-\frac{\pi}{6}\left(\frac{\sigma}{d}\right)^2 Q^2 }  + \varepsilon_{3\textup{T}},\label{eq:Int 3 m p final}
\ea 
where for $d=4,6,8\ldots$
\ba
\left|\varepsilon_{3\textup{T}}\right| &\leq \max\Big\{ \left|\varepsilon_F\right|, \left|\varepsilon_3(m,p)\right|  \Big\}\\
&\leq \max\left\{ \frac{1}{\sqrt{6}}\left(\frac{\sigma}{d}\right)\, \me^{-\pi \frac{3}{4} \left(\frac{d}{\sigma}\right)^2 \left(1-\frac{3}{d}\right)^2} +\frac{1}{d} \me^{-\frac{\pi}{2} \left(\frac{d}{\sigma}\right)^2\left(1-\frac{4}{ d}\right)^2}, \frac{1}{3\pi}\frac{\sigma^2}{d}\, \me^{- \frac{\pi}{2} \left(\frac{d}{\sigma}\right)^2 \left(1-\frac{2}{d}\right)^2} \right\}\\
&= \bo \left( \frac{\sigma^2}{d}\, \me^{- \frac{\pi}{2} \left(\frac{d}{\sigma}\right)^2 \left(1-\frac{4}{d}\right)^2} \right)\quad \text{as }\,\,  d\rightarrow\infty,
\ea
where we have used Eqs. \eqref{eq:Int 3 m p final inter 6}, \eqref{eq:rp 3 (m n) up bound} and the values of $\varepsilon^L_{3,1}$, $\varepsilon^R_{3,1}$ from Eqs. \eqref{eq:ep 3 R}, \eqref{eq:ep 3 L}.

\subsection{Approximating the integrals in lines \eqref{line:int 1 1}, \eqref{line:int 2 1}.}
The steps taken in the previous section for approximating the integral in line \eqref{line:int 3 1}, can be repeated for the integrals in lines \eqref{line:int 1 1}, \eqref{line:int 2 1}. We summarise here the final results, starting with the integrals in line \eqref{line:int 1 1}: For the parametrization $\bar k_2/d=1/2+1/d-k/d,$  $\bar k_3/d=1/2+1/d-l/d$, $k,l=1,\ldots,d$.
\ba
\textup{Int}_1(k,l)&:= \int_{0}^{\bar k_2/d+ 1/2-1/d}dx\, \me^{-\mi 2\pi x Q}\, \me^{-2\pi \left(\frac{d}{\sigma}\right)^2 \left[ \left(x-\frac{1}{2}-\frac{1-\Delta k_1^0}{d}\right)^2+\left(x-\frac{\bar k_2}{d}+\frac{\Delta k_2^0}{d}\right)^2+\left(x-\frac{{\bar k_3}}{d}+\frac{\Delta k_3^0}{d}\right)^2  \right]}\\
&= \frac{1}{\sqrt{6}}\left(\frac{\sigma}{d}\right) \me^{-2\mi \pi \tilde k Q} \me^{\mi\pi\frac{2}{3}(k+l)Q/d}\\
& \;\;\;\;\;   \me^{-\pi \frac{4}{3}\left(\frac{d}{\sigma}\right)^2 \left(\frac{\left[k+\Delta k_2^0-\Delta k_1^0\right]^2+\left[l+\Delta k_3^0-\Delta k_1^0\right]^2-\left[k+\Delta k_2^0-\Delta k_1^0\right]\left[l+\Delta k_3^0-\Delta k_1^0\right]}{d^2}\right)}\me^{-\frac{\pi}{6}\left(\frac{\sigma}{d}\right)^2 Q^2 } + \varepsilon_{1\textup{T}},\label{eq:Int 1 k l final}
\ea 
where
\be 
\left|\varepsilon_{1\textup{T}}\right| = \bo \left( \frac{\sigma^2}{d}\, \me^{- \frac{\pi}{2} \left(\frac{d}{\sigma}\right)^2 \left(1-\frac{4}{d}\right)^2} \right)\quad \text{as }\,\,  d\rightarrow\infty.\label{eq: ep 1T def}
\ee 

In the case of the integral in line Eq. \eqref{line:int 2 1}, for the parametrization $\bar k_2/d=-1/2+1/d+m/d,$  $\bar k_3/d=1/2+1/d-n/d$, $m=0,\ldots,d-1$, $n=1,\ldots,d$; we find

\ba
\textup{Int}_2(m,n)&:=
\int_{\bar k_2/d+ 1/2-1/d}^{\bar k_3/d+ 1/2-1/d}dx\, \me^{-\mi 2\pi x Q}\, \me^{-2\pi \left(\frac{d}{\sigma}\right)^2 \left[ \left(x-\frac{1}{2}-\frac{1-\Delta k_1^0}{d}\right)^2+\left(x-1-\frac{\bar k_2}{d}+\frac{\Delta k_2^0}{d}\right)^2+\left(x-\frac{{\bar k_3}}{d}+\frac{\Delta k_3^0}{d}\right)^2  \right]}   \\
&= \frac{1}{\sqrt{6}}\left(\frac{\sigma}{d}\right) \me^{-\mi 2 \pi \tilde k Q/d} \me^{-\mi\pi\frac{2}{3}(m-n)Q/d}\\
& \;\;\;\;\; \me^{-\pi \frac{4}{3}\left(\frac{d}{\sigma}\right)^2 \left(\frac{\left[m-\Delta k_2^1+\Delta k_1^0\right]^2+\left[n+\Delta k_3^1-\Delta k_1^0\right]^2+\left[n+\Delta k_3^1-\Delta k_1^0 \right]\left[m-\Delta k_2^1+\Delta k_1^0\right]}{d^2}\right)}\me^{-\frac{\pi}{6}\left(\frac{\sigma}{d}\right)^2 Q^2 } + \varepsilon_{2\textup{T}},\label{eq:Int 2 k l final}
\ea 

where
\be 
\left|\varepsilon_{2\textup{T}}\right|=  \bo \left( \frac{\sigma^2}{d}\, \me^{- \frac{\pi}{2} \left(\frac{d}{\sigma}\right)^2 \left(1-\frac{4}{d}\right)^2} \right)\quad \text{as }\,\,  d\rightarrow\infty.\label{eq: ep 2T def}
\ee 

\subsection{Final expression for $\tilde F_Q(k_1,k_2,k_3)$}\label{Final expression for FQ k1 k2 k3}
With the results from the previous section, we can finally formulate a useful approximation for $\tilde F_Q(k_1,k_2,k_3)$. Plugging Eqs. \eqref{eq:Int 1 k l final}, \eqref{eq:Int 2 k l final}, \eqref{eq:Int 3 m p final}, into Eq. \eqref{eq:FQ inter 4} and after some re-parametrization, we find for $\bar k_1/d=-1/2+1/d$, $\bar k_2/d=-1/2+1/d+m/d,$  $\bar k_3/d=-1/2+1/d+p/d$, $m,p=0,\ldots,d-1$, with $\bar k_1\leq \bar k_2\leq \bar k_3$, 

\ba
\tilde F_Q  (k_1,k_2,k_3) = &\frac{A^6}{\sqrt{6}}\left(\frac{\sigma}{d}\right)  \me^{-\mi 2\pi \tilde k Q/d} \me^{-\frac{\pi}{6}\left(\frac{\sigma}{d}\right)^2 Q^2 } G_Q(d;m,p) + \varepsilon_\textup{T},
\label{eq:FQ 3 final fixed k1}
\ea
where
\ba 
G_Q(d;m,p):= &   \me^{-\mi\pi\frac{2}{3}(m+p)Q/d}\, \me^{-\pi \frac{4}{3}\left(\frac{d}{\sigma}\right)^2  H_1(d,m,p
	)}+ \me^{-\mi\pi\frac{2}{3}(m+p-d)Q/d} \me^{-\pi \frac{4}{3}\left(\frac{d}{\sigma}\right)^2  H_2(d,m,p) }\\
&  +\me^{-\mi\pi\frac{2}{3}(m+p-2d)Q/d}\,  \me^{-\pi \frac{4}{3}\left(\frac{d}{\sigma}\right)^2 H_3(d,m,p)},\label{eq: G Q m p def}
\ea
and
\ba 
H_1(d,m,p)
&:= \scriptstyle{\left(\frac{\left[m+\Delta k_1^0-\Delta k_2^0\right]^2+\left[p+\Delta k_1^0-\Delta k_3^0\right]^2-\left[m+\Delta k_1^0-\Delta k_2^0\right] \left[p+\Delta k_1^0-\Delta k_3^0\right]}{d^2}\right)},\\
H_2(d,m,p) &:= \scriptstyle{\left(\frac{\left[m-\Delta k_2^0+\Delta k_1^0\right]^2+\left[-p+d+\Delta k_3^0-\Delta k_1^0\right]^2+\left[-p+d+\Delta k_3^0-\Delta k_1^0 \right]\left[m-\Delta k_2^0+\Delta k_1^0\right]}{d^2}\right)},\\
H_3(d,m,p) &:= \scriptstyle{ \left(\frac{\left[-m+d+\Delta k_2^0-\Delta k_1^0\right]^2+\left[-p+d+\Delta k_3^0-\Delta k_1^0\right]^2-\left[-m+d+\Delta k_2^0-\Delta k_1^0\right]\left[-p+d+\Delta k_3^0-\Delta k_1^0\right]}{d^2}\right)}.
\ea
The small error term $\varepsilon_\textup{T}$ in Eq. \eqref{eq:FQ 3 final fixed k1} is defined via
\be 
\varepsilon_\textup{T} = A^6 \left(\varepsilon_\textup{1T}+\varepsilon_\textup{2T}+\varepsilon_\textup{3T}\right) +\varepsilon_\textup{I1}+7\varepsilon_\textup{I0},
\ee 
and bounded by  
\ba 
\left|  \varepsilon_\textup{T} \right|&=  \left| A^6 \left(\varepsilon_\textup{1T}+\varepsilon_\textup{2T}+\varepsilon_\textup{3T}\right) +\varepsilon_\textup{I1}+7\varepsilon_\textup{I0}\right|\\ 
&\leq \left(\frac{\sqrt{2}}{\sigma}+\left|\varepsilon_A\right|\right)^3 \big( |\varepsilon_\textup{1T}|+|\varepsilon_\textup{2T}|+|\varepsilon_\textup{3T}|\big)+\varepsilon_\textup{I1}+7\varepsilon_\textup{I0}\label{eq:line 298 ep T} 
\\
&= \bo \left( \frac{1}{d\,\sigma}\, \me^{- \frac{\pi}{2} \left(\frac{d}{\sigma}\right)^2 \left(1-\frac{4}{d}\right)^2} \right)+\varepsilon_\textup{I1}+7\varepsilon_\textup{I0}\quad \text{as }\,\,  d\rightarrow\infty,  
\label{eq:up bound to epsilon tot 3 clocks}
\ea
where we have used $A^2=\sqrt{2}/\sigma+\varepsilon_A=\bo(1/\sigma)$ (from page 483 in \cite{woods2016autonomous}) in line \eqref{eq:line 298 ep T}, followed by Eqs. \eqref{eq: ep 1T def}, \eqref{eq: ep 2T def}. Finally, using Eq. \eqref{eq:ep I1 final} we conclude
\be
\left|  \varepsilon_\textup{T} \right|=  \bo \left( \frac{1}{d\,\sigma}\, \me^{- \frac{\pi}{2} \left(\frac{d}{\sigma}\right)^2 \left(1-\frac{4}{d}\right)^2} \right) + \bo\left( \frac{d^2}{\sigma^{5/2}}  \right)\me^{-\frac{\pi}{4}\sigma^2\alpha_0^2}\quad \text{as }\,\,  d\rightarrow\infty.\label{eq: ep T abs final}
\ee

We can now easily generalise this result to any measurement outcomes. Define $ k_1',  k_2',  k_3'\in \mathcal{I}_d=\{0,1,\ldots,d-1\}$, by 
\ba
k_q':= \left[ k_q+l\right]_\textup{(mod. 
	\textit{d})}=\begin{cases}
	\left[ -\frac{d}{2}+1+\lfloor \tilde k_1^0 \rfloor +l\right]_\textup{(mod. \textit{d}
		)} &\mbox{ if } q=1,\\
	\left[ -\frac{d}{2}+1+\lfloor \tilde k_2^0 \rfloor +l+m\right]_\textup{(mod. \textit{d}
		)} &\mbox{ if } q=2,\\
	\left[ -\frac{d}{2}+1+\lfloor \tilde k_3^0 \rfloor +l+p\right]_\textup{(mod. \textit{d}
		)} &\mbox{ if } q=3,
\end{cases} \label{eq:k prime measures def}
\ea
for  $l,m,p=0,\ldots,d-1$. 
Rather than the previous case where the measurement outcomes $k_q$ where chosen $k_q\in\mathcal{S}_d(\tilde k_q^0)$, $q=1,2,3$; in the present case of the measurement outcomes $k_1',  k_2',  k_3'$, for simplicity we will use the convention  $ k_1',  k_2',  k_3'\in \mathcal{I}_d=\{0,1,\ldots, d-1\}$\footnote{Note that this simply corresponds to a shifting the values of all the measurement outcomes by a constant value.}, with outcome $k_q'$ associated with projective measurement $\ketbra{\theta_{k'_q}}{\theta_{k'_q}}_q$, $q=1,2,3$.

Now employing Eqs. \eqref{eq:F Q shift invarient}, \eqref{eq:FQ 3 final fixed k1} we achieve
\ba
\tilde F_Q(k_1',k_2',k_3' )& = \me^{-\mi 2\pi l Q/d}  \tilde F_Q(k_1,k_2, k_3 )\\
&=\frac{A^6}{\sqrt{6}}\left(\frac{\sigma}{d}\right)  \me^{-\mi 2\pi \tilde k_l Q/d} \me^{-\frac{\pi}{6}\left(\frac{\sigma}{d}\right)^2 Q^2 } G_Q(d;m,p) + \varepsilon_\textup{T}\, \me^{-\mi 2\pi l Q/d},\quad \text{if } m\leq p \label{eq:F Q m leq p}
\ea 
Here we have defined the common phase,
\be 
\tilde k_l := -\frac{d}{2}+1-\frac{\Delta k_\textup{T}^0}{3 }+l,\label{eq: tide k l def}
\ee
where recall $\Delta k_\textup{T}^0$ is defined in Eq. \eqref{eq: k tot sum d def}. Finally, we note that not all values of $ k_1',  k_2',  k_3'\in \mathcal{I}_d$ are achievable due to the constraint $m\leq p$. However, we can remedy this by using property 3) [see Eq. \eqref{eq:prop 3 interchainge invar}]. Since in this case the clocks only differ by the values $\tilde k_1^0,\tilde k_2^0,\tilde k_3^0$, interchanging $\rho_{\cl 2}(t)$ with $\rho_{\cl 3}(t)$ is equivalent to interchanging $\tilde k_2^0$ with $\tilde k_3^0$. Furthermore, recalling $k_2=\bar k_2+\lfloor \tilde k_2^0\rfloor= -d/2+1+m +\lfloor \tilde k_2^0\rfloor$, $k_3=\bar k_3+\lfloor \tilde k_3^0\rfloor= -d/2+1+p +\lfloor \tilde k_3^0\rfloor$, we see that interchanging $\tilde k_2^0$ with $\tilde k_3^0$ and $k_2$ with $k_3$ is equivalent to interchanging $\tilde k_2^0$ with $\tilde k_3^0$ and $m$ with $p$.
Finally, by construction $\bar k_2> \bar k_3$ iff $m>p$. Thus using Eq. \eqref{eq:F Q m leq p} we find that if $m>p$, 
\ba
\tilde F_Q(k_1',k_2',k_3' )
&=\Bigg[\frac{A^6}{\sqrt{6}}\left(\frac{\sigma}{d}\right)  \me^{-\mi 2\pi \tilde k_l Q/d} \me^{-\frac{\pi}{6}\left(\frac{\sigma}{d}\right)^2 Q^2 } G_Q(d;m,p) + \varepsilon_\textup{T}\, \me^{-\mi 2\pi l Q/d}\Bigg]_{ \substack{\tilde k_2^0\, \leftrightarrow\, \tilde k_3^0 \\ m \,\leftrightarrow\, p } }\\
&=\frac{A^6}{\sqrt{6}}\left(\frac{\sigma}{d}\right)  \me^{-\mi 2\pi \tilde k_l Q/d} \me^{-\frac{\pi}{6}\left(\frac{\sigma}{d}\right)^2 Q^2 } \Big{[}G_Q(d;p,m)\Big]_{\tilde k_2^0\, \leftrightarrow\, \tilde k_3^0 } + \tilde \varepsilon_\textup{T},\label{eq:F Q p m}
\ea 
where $|\tilde \varepsilon_\textup{T}|$ satisfies the same upper bound in Eq. \eqref{eq:up bound to epsilon tot 3 clocks} that $|\varepsilon_\textup{T}|$ satisfies. Thus, in general, from Eqs. \eqref{eq:F Q m leq p} and \eqref{eq:F Q p m} we conclude
\ba
\tilde F_Q(k_1',k_2',k_3' )
&=\frac{A^6}{\sqrt{6}}\left(\frac{\sigma}{d}\right)  \me^{-\mi 2\pi \tilde k_l Q/d} \me^{-\frac{\pi}{6}\left(\frac{\sigma}{d}\right)^2 Q^2 } \tilde G_Q(d;m,p) + \bar \varepsilon_\textup{T},\label{eq:F Q general}
\ea 
where
\be 
\tilde G_Q(d;m,p):=
\begin{cases}
	G_Q(d;m,p) & \mbox{if } m\leq p \\
	\Big{[}G_Q(d;p,m)\Big]_{\tilde k_2^0\, \leftrightarrow\, \tilde k_3^0 } & \mbox{if } p< m 
\end{cases}
\ee
and $|\bar \varepsilon_\textup{T}|$ is upper bounded by the r.h.s. of Eq. \eqref{eq:up bound to epsilon tot 3 clocks}.

\subsection{Bounding $\|\hat E\|_1$}\label{Bounding hat E in the case no random phase error}
Now that we have from the previous section an expression for $\tilde F_Q$ for all measurement outcomes of the 3 clocks, the next task is to bound $\|\hat E\|_1$ defined in  \ref{eq: rho our generix with F 2} to be
\ba
\vert \vert \hat E \vert \vert_1\le& d_\inp \sqrt{d_\outp}  \sum_{k_1',k_2',k_3' } \tilde F_{0}(k_1',k_2',k_3' ) \max\limits_{Q} \vert p(Q,\vec k)\vert^2\\
= & d_\inp \sqrt{d_\outp} \sum_{k_1',k_2',k_3' } \tilde F_{0}(k_1',k_2',k_3' ) \max_{Q} \left| \frac{ \tilde F_{Q}(k_1',k_2',k_3' )}{\tilde F_{0}(k_1',k_2',k_3' )}\,\me^{2\pi \mi {k_\alpha} Q/d} -1\right|\\
= & d_\inp \sqrt{d_\outp}  \max_{Q} \sum_{k_1',k_2',k_3' } \left|  \tilde F_{Q}(k_1',k_2',k_3' )\,\me^{2\pi \mi {k_\alpha} Q/d} -\tilde F_{0}(k_1',k_2',k_3' )\right|\\
= & d_\inp \sqrt{d_\outp} \frac{A^6}{\sqrt{6}}\left(\frac{\sigma}{d}\right)   \max_{Q}\\ &\sum_{l, m,p=0 }^{d-1} \left| \me^{-\frac{\pi}{6}\left(\frac{\sigma}{d}\right)^2 Q^2 } \tilde G_Q(d;m,p) \me^{2\pi \mi (k_\alpha-\tilde k_l) Q/d} -  \tilde G_0(d;m,p) + \tilde\varepsilon_\textup{T}+\bar\varepsilon_\textup{T}\right|.\label{eq: hat E norm 3}
\ea
In all cases, the optimization is over $Q\in\{-(\Delta h_\code+\Delta h_\inp),\ldots,(\Delta h_\code+\Delta h_\inp)\}$. 

\begin{figure}[]
	\includegraphics[width=0.8\linewidth]{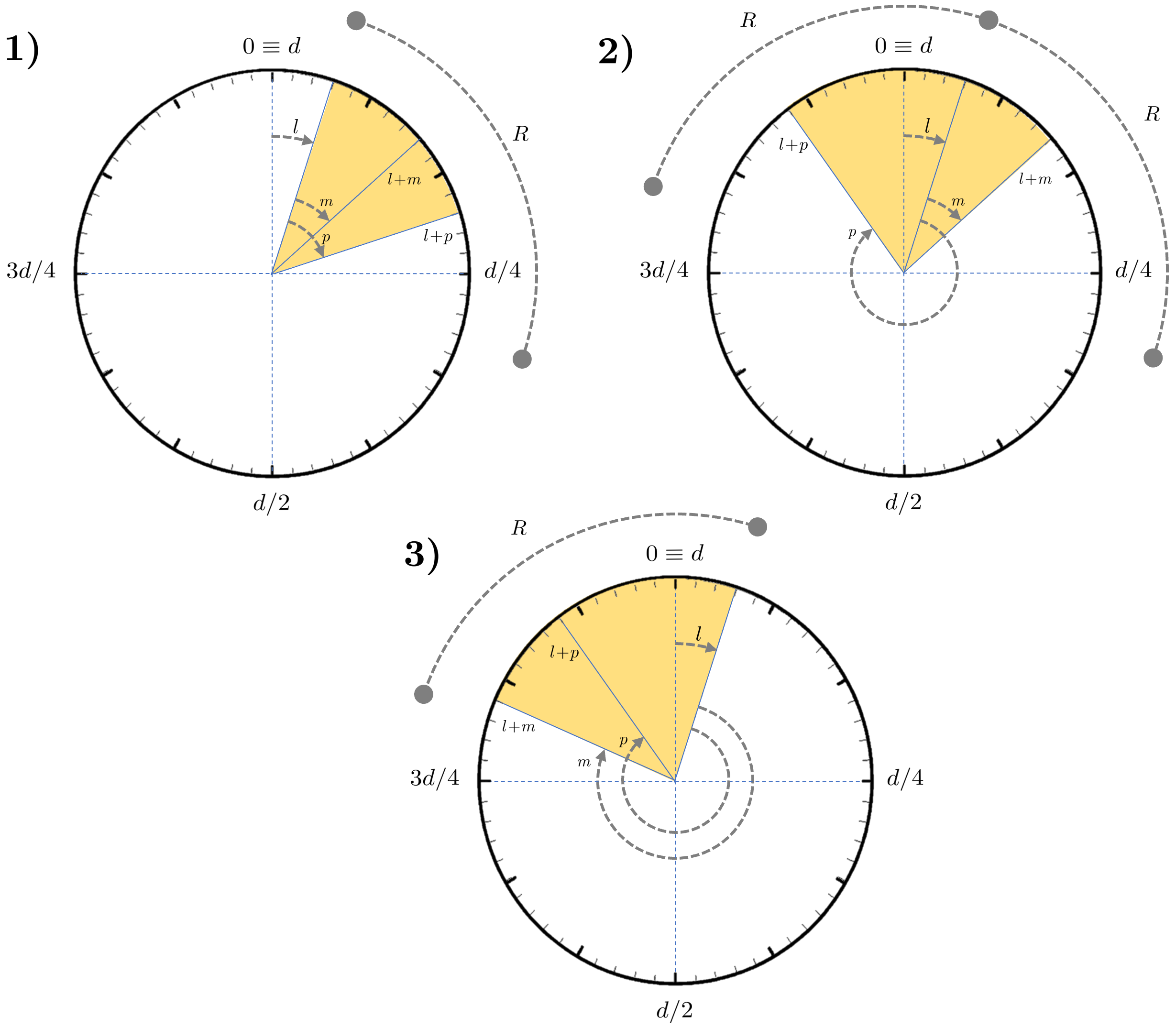}
	\caption{\textbf{ $\dom(\gamma_{l,m,p})$ for $(m,p)$ in different intervals and $l=0,1,\ldots,d-1$.}  \newline
		\textbf{Case 1):} Yellow highlighted interval represents $\dom(\gamma_{l,m,p})$ for $(m,p)\in \mathcal{S}_{11}(R).$ \newline
		\textbf{Case 2):} Yellow highlighted interval represents $\dom(\gamma_{l,m,p})$ for $(m,p)\in \mathcal{S}_{31}(R).$	\newline
		\textbf{Case 3):} Yellow highlighted interval represents $\dom(\gamma_{l,m,p})$ for $(m,p)\in \mathcal{S}_{33}(R).$\newline
		The case $\dom(\gamma_{l,m,p})$ for $(m,p)\in \mathcal{S}_{13}(R)$ is the same as case 2) above under the interchange $m \leftrightarrow p.$ In all four cases, $\cdom(2\pi \gamma_{l,m,p}/d)$ is ``any angle between angles $l$, $ m$, $ p$."}\label{fig 4 plots}
\end{figure}

In order to proceed further, we need to specify $k_\alpha$ for this protocol. If there were no phase errors, it turns out that defining it equal to either of the three quantities $\bar k_1':=-d/2+1+l$, $\bar k_2':=-d/2+1+l+m$, $\bar k_3':=-d/2+1+l+p$,  would suffice. Note how doing so would only require information from one clock measurement | the other two clock results would be redundant information. Intuitively, this is because with high probability, all three clocks will have measurement outcomes corresponding to approximately the same elapsed time (if there were no phase errors). For the case of no phase error, it is convenient to prove that the protocol works up to the specified error when $k_\alpha$  is chosen to be ``any angle \footnote{Throughout this proof, we call ``angle" to real numbers which only need to be specified up to modulo $d$. These angles can be converted to radian by multiplying them by $2\pi/d$.} in-between the three angles $\bar k_1'$, $\bar k_2'$, $\bar k_3'$" when $\bar k_1'$, $\bar k_2'$, $\bar k_3'$ belong to a to-be specified domain, and ``any angle otherwise". Later in section \ref{sec:3 clock protocol} we will show how such a result is sufficient in the case of the unknown phase error.  Specifically, we define 
\be 
k_\alpha=-\frac{d}{2}+1+\gamma_{l,m,p}(\alpha)\label{eq: def of k alpha 3 clocks}
\ee
where $\gamma_{l,m,p}$ are functions indexed by $l,m,p=0,\ldots,d-1$ of the form 
\be 
\gamma_{l,m,p}(\alpha)=\alpha.
\ee 
Due to the modular arithmetic, depending on which sector $l,m,p$ belong to, the domain of $\gamma_{l,m,p}$ changes.
Specifically, for $R\in[1,d/4]$  we define the sets
\ba 
\mathcal{S}_{11}(R) &:=  \left\{(m,p)\in\zz^2\;\; |\;\; 0\leq m,p\leq R-1\right\},\label{es:set 11 def}\\
\mathcal{S}_{31}(R) &:=  \{(m,p)\in\zz^2\;\; |\;\; 0\leq m \leq R-1 \,\,\,\&\,\,\, d-R \leq p\leq d-1\},\\
\mathcal{S}_{13}(R) &:=  \{(m,p)\in\zz^2\;\; |\;\; d-R \leq m\leq d-1 \,\,\,\&\,\,\,  0\leq p \leq R-1 \},\\
\mathcal{S}_{33}(R) &:=  \{(m,p)\in\zz^2\;\; |\;\; d-R \leq m,p\leq d-1 \}.\label{es:set 33 def}
\ea 
Now we can define $\dom(\gamma_{l,m,p})$:
\be 
\dom(\gamma_{l,m,p}):=
\begin{cases}
	l,l+1,\ldots,l+\max\{m,p\}  &\mbox{ if } (m,p)\in \mathcal{S}_{11}(R)\\
	l+p,l+p+1,\ldots,l+m+d  &\mbox{ if } (m,p)\in \mathcal{S}_{31}(R)\\
	l+m,l+m+1,\ldots,l+p+d  &\mbox{ if } (m,p)\in \mathcal{S}_{13}(R)\\
	l+\min\{m,p\},l+\min\{m,p\}+1,\ldots,l+d  &\mbox{ if } (m,p)\in \mathcal{S}_{33}(R)\\
	l,l+1,\ldots, l+d-1  &\mbox{ otherwise }
\end{cases}
\ee
Observer that $l,(l+m)_{\textup{(mod. \textit{d})}},(l+p)_{\textup{(mod. \textit{d})}}\in\dom(\gamma_{l,m,p})_{\textup{(mod. \textit{d})}}$ always holds. In particular, the function $\gamma_{m,p}(\alpha)$ covers up to (but not more than), all angles $0,1,\ldots, d-1$ which are between the three angles $l,l+ m,l+p$ whenever $(m,p)\in \mathcal{S}_{11}(R)\cup\mathcal{S}_{31}(R)\cup\mathcal{S}_{13}(R)\cup\mathcal{S}_{33}(R)$ and $l\in\mathcal{I}_d$. See Fig. \ref{fig 4 plots}.

Writing Eq. \eqref{eq: hat E norm 3} in terms of this new notation, we have 
\ba
\vert \vert \hat E \vert \vert_1\le& 
d_\inp \sqrt{d_\outp} \frac{A^6 }{\sqrt{6}}  \frac{\sigma}{d} \max_{Q\in\{-(\Delta h_\code+\Delta h_\inp),\ldots,(\Delta h_\code+\Delta h_\inp)\}}\\ &\left( \sum_{ l,m,p=0 }^{d-1}\left|  \me^{-\frac{\pi}{6}\left(\frac{\sigma}{d}\right)^2 Q^2 } \tilde G_Q(d;m,p) \me^{2\pi \mi \left(\gamma_{m,p}(\alpha)+\frac{\Delta k_\textup{T}}{3}\right) Q/d} -  \tilde G_0(d;m,p)\right| + 2 d\, |\hat  \varepsilon_\textup{T}|\right)\\
= & d_\inp \sqrt{d_\outp} \frac{A^6 }{\sqrt{6}}\frac{\sigma}{d}   \max_{Q\in\{-(\Delta h_\code+\Delta h_\inp),\ldots,(\Delta h_\code+\Delta h_\inp)\}}\left( \sum_{ l,m,p=0 }^{d-1}\Delta F(d;l,m,p)  + 2 d\, |\hat  \varepsilon_\textup{T}|\right)
,\label{eq: hat E 1 norm bound 5}
\ea
where $|\hat \varepsilon_\textup{T}|= \max_{k_1'} | \tilde\varepsilon_\textup{T}+\bar\varepsilon_\textup{T}|/2$ is upper bounded by the r.h.s. of Eq. \eqref{eq:up bound to epsilon tot 3 clocks}. It is now convenient to bound the summand $\Delta F(d;l,m,p)$. Using Eqs. \eqref{eq: G Q m p def} and \eqref{eq:F Q general}, we find
\ba
\Delta F(d;l,m,p) :=&\Big|\me^{-\frac{\pi}{6}\left(\frac{\sigma}{d}\right)^2 Q^2 } \tilde G_Q(d;m,p)  \me^{2\pi \mi \left(\gamma_{m,p}(\alpha)-l+\frac{\Delta k_\textup{T}}{3}\right) Q/d} -  \tilde G_0(d;m,p)\Big|\label{eq: hat E 1 norm bound 7}\\
\leq &  \left| \me^{-\frac{\pi}{6}\left(\frac{\sigma}{d}\right)^2 Q^2}  \me^{-\mi\pi\frac{2}{3}(m+p)Q/d} \me^{2\pi \mi \left(\gamma_{m,p}(\alpha)-l+\frac{\Delta k_\textup{T}}{3}\right) Q/d} - 1\right| \me^{-\pi \frac{4}{3}\left(\frac{d}{\sigma}\right)^2 \tilde H_1(d,m,p)}\label{line:Delta F H1}\\
& + \left| \me^{-\frac{\pi}{6}\left(\frac{\sigma}{d}\right)^2 Q^2}  \me^{-\mi\pi\frac{2}{3}(m+p-d)Q/d} \me^{2\pi \mi \left(\gamma_{m,p}(\alpha)-l+\frac{\Delta k_\textup{T}}{3}\right) Q/d} - 1 \right| \me^{-\pi \frac{4}{3}\left(\frac{d}{\sigma}\right)^2 \tilde H_2(d,m,p)}\label{line:Delta F H2}\\
& + \left| \me^{-\frac{\pi}{6}\left(\frac{\sigma}{d}\right)^2 Q^2} 
\me^{-\mi\pi\frac{2}{3}(m+p-2d)Q/d}\me^{2\pi \mi \left(\gamma_{m,p}(\alpha)-l+\frac{\Delta k_\textup{T}}{3}\right) Q/d} - 1 \right|\me^{-\pi \frac{4}{3}\left(\frac{d}{\sigma}\right)^2 \tilde H_3(d,m,p)},\label{line:Delta F H3}
\ea
where for $q=1,2,3,$ we have defined
\ba
\tilde H_q(d,m,p) := 
\begin{cases} 
	H_q(d,m,p) &\mbox{ if } m\leq p\\
	\Big{[}H_q(d,p,m)\Big]_{k_2^0\, \leftrightarrow\, k_3^0 } &\mbox{ if } p<m
\end{cases}
\ea

\begin{figure}[]
	\includegraphics[scale=0.6]{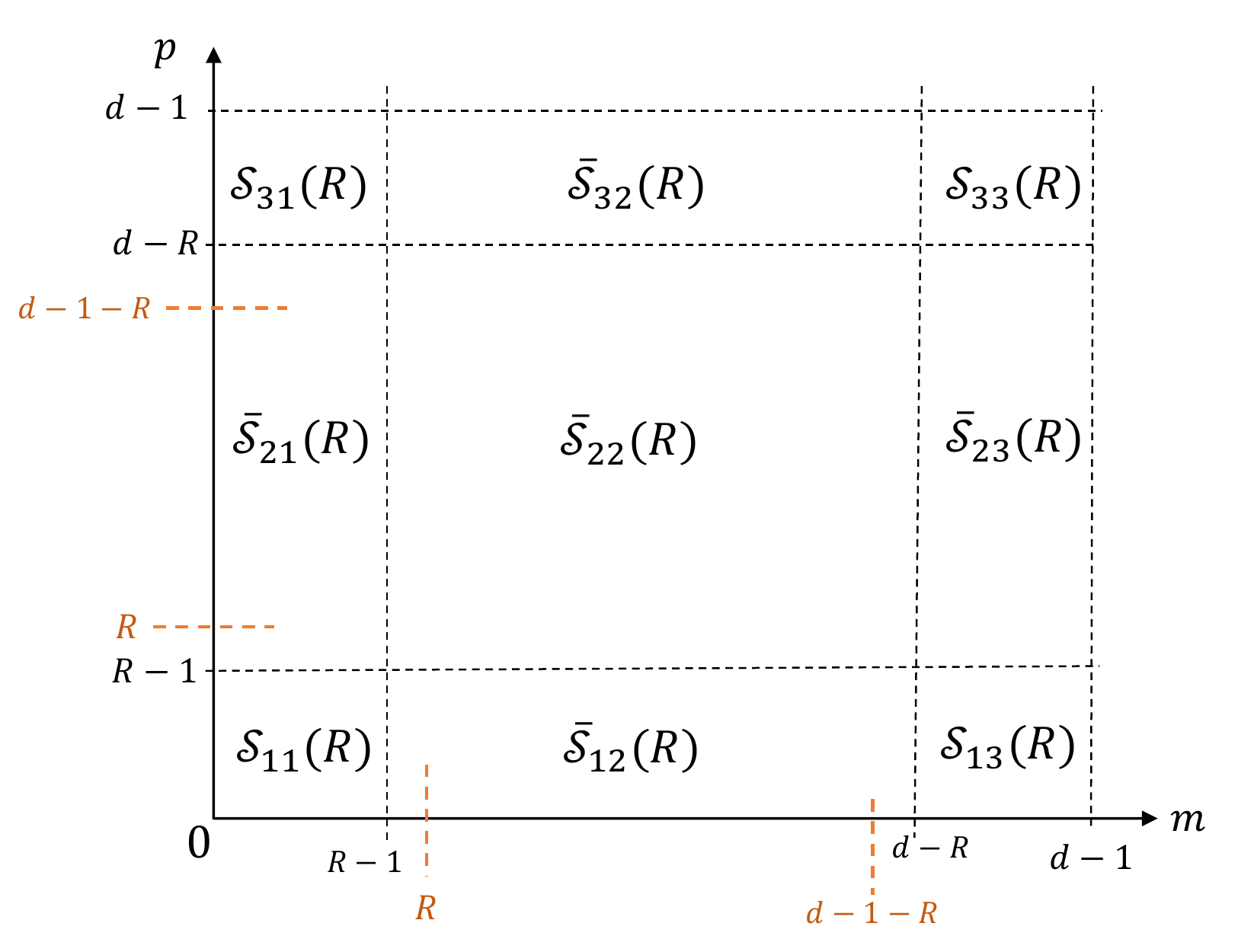}
	\caption{Diagrammatic view of the nine distinct sets used in the proof. Note that all pairs have the empty set as their intersection and the union of all sets is the set $\mathcal{S}_\textup{tot}$ given by Eq. \eqref{eq:tot set}.}  \label{fig:square}
\end{figure}

The evaluation of the summations over $m,p$ for $ \Delta F(d;l,m,p)$ is more complicated than the summation over $l$ and it is convenient to perform the summation separately over subsets of the domain of $m,p$. With this in mind, we define for $R\in[1,d/4]$ the following sets.
\be 
\mathcal{S}_\textup{tot}= \mathcal{S}_{11}(R) \,\cup\, \mathcal{\bar S}_{12}(R)\,\cup\, \mathcal{ S}_{13}(R) \,\cup\, \mathcal{\bar S}_{21}(R)\,\cup\, \mathcal{\bar S}_{22}(R)\,\cup\, \mathcal{\bar S}_{23}(R)\,\cup\, \mathcal{ S}_{31}(R) \,\cup\, \mathcal{\bar S}_{32}(R)\,\cup\, \mathcal{ S}_{33}(R).\label{eq:tot set}
\ee 
where $\mathcal{S}_\textup{tot}=\{(m,p)\in\zz^2\;\; |\;\; 0\leq m,p\leq d-1\}$ is the set $m$ and $p$ run over in the summation in Eq. \eqref{eq: hat E 1 norm bound 5} and $\mathcal{S}_{11}(R)$, $\mathcal{S}_{31}(R)$, $\mathcal{S}_{11}(R)$, $\mathcal{S}_{33}(R)$, are the sets which when $R$ is chosen to scale with $d$ appropriately, will include all measurement outcomes $k_2',k_3'$ which, as a whole, occur with high probability. On the other hand, the sets
\ba 
\mathcal{\bar S}_{12}(R) &:=   \{(m,p)\in\zz^2\;\; |\;\; R \leq m\leq d-1-R \,\,\,\&\,\,\,  0\leq p \leq R-1 \}\label{eq:set 12 def}\\
\mathcal{\bar S}_{21}(R) &:=   \{(m,p)\in\zz^2\;\; |\;\; 0\leq m\leq R-1 \,\,\,\&\,\,\,  R \leq p \leq d-1-R \}\\
\mathcal{\bar S}_{22}(R) &:=   \{(m,p)\in\zz^2\;\; |\;\; R\leq m,p\leq d-1-R  \}\\
\mathcal{\bar S}_{23}(R) &:=   \{(m,p)\in\zz^2\;\; |\;\; d-R\leq m\leq d-1 \,\,\,\&\,\,\,  R \leq p \leq d-1-R \}\\
\mathcal{\bar S}_{32}(R) &:=   \{(m,p)\in\zz^2\;\; |\;\; R\leq m\leq d-1-R \,\,\,\&\,\,\,  d-R \leq p \leq d-1 \},\label{eq:set 32 def}
\ea
correspond to all measurement outcomes $k_2',k_3'$ which, as a whole, occur with very low probability. See graphical depiction Fig. \ref{fig:square}.

In addition to the limits Eqs. \eqref{eq:sigma d limits assumtions} assumed thought, we now demand that $R$ satisfies the following limits. The exact choice of $R$ will be determined later. 
%
\ba\label{eq:R limits def}
\lim_{d\rightarrow +\infty}\frac{R}{\sigma}=+\infty, \quad & \quad \lim_{d\rightarrow +\infty}\frac{R}{d}  Q_{\max} =0,
\ea
\footnote{Note that $Q_{\max}\geq 1$ uniformly in $d$ and thus it follows from Eq. \eqref{eq:R limits def} that $\lim_{d\rightarrow \infty}R/d=0$.}
where $Q_{\max}:= \max_{Q\in\{-(\Delta h_\outp+\Delta h_\inp),\ldots,(\Delta h_\outp+\Delta h_\inp)\}} |Q|=\Delta h_\inp+
\Delta h_\outp$.
We now want to bound the $\Delta F(d,l,m,p)$ in Eq. \eqref{eq: hat E 1 norm bound 7} for measurement outcomes in sets Eqs. \eqref{es:set 11 def} to \eqref{es:set 33 def}. We start with the set $\mathcal{S}_{12}$. For all $(m,p)\in\mathcal{ S}_{12}(R)$, we have that $\lim_{d\rightarrow \infty} \tilde H_2(d,m,p)=\lim_{d\rightarrow \infty} \tilde H_3(d,m,p) = 1$, so the terms in lines Eqs. \eqref{line:Delta F H2},\eqref{line:Delta F H3} are exponentially small in $(d/\sigma)^2$. On the other hand, $\lim_{d\rightarrow \infty} \tilde H_2(d,m,p)$ converges to zero and the common factor in line \eqref{line:Delta F H1} is significant.  For the term in line Eq. \eqref{line:Delta F H1}, it is the terms within the absolute value which are small. Specifically, for all $\alpha_{l,m,p}\in\dom(\gamma_{l,m,p})$ and for all $(m,p)\in\mathcal{ S}_{12}(R)$, $l\in\mathcal{I}_d$,
\ba
\Bigg|\me^{-\frac{\pi}{6}\left(\frac{\sigma}{d}\right)^2 Q^2}&  \me^{-\mi\pi\frac{2}{3}(m+p)Q/d} \me^{2\pi \mi \left(\gamma_{l,m,p}(\alpha_{l,m,p})-l+\frac{\Delta k_\textup{T}}{3}\right) Q/d}-1\Bigg|\\
=&\left|\left(-\frac{\pi}{6}\left(\frac{\sigma}{d}\right)^2 Q^2 -\mi\pi\frac{2}{3}(m+p)Q/d +2\pi \mi \left(\gamma_{l,m,p}(\alpha_{l,m,p})-l+\frac{\Delta k_\textup{T}}{3}\right)Q/d \right)\right|\\
&+\bo\left(-\frac{\pi}{6}\left(\frac{\sigma}{d}\right)^2 Q^2 -\mi\pi\frac{2}{3}(m+p)Q/d +2\pi \mi \left(\gamma_{l,m,p}(\alpha_{l,m,p})-l+\frac{\Delta k_\textup{T}}{3}\right)Q/d \right)^2\\
\leq & \left|\pi\frac{4}{3}  \frac{R}{d}Q_{\max} +2\pi \frac{R}{d}Q_{\max} \right|+\bo\left(\pi\frac{4}{3}  \frac{R}{d}Q_{\max} +2\pi \frac{R}{d}Q_{\max} \right)^2\\
=&\pi \frac{10}{3} \frac{R}{d} Q_{\max}+ \bo\left(\pi \frac{10}{3} \frac{R}{d} Q_{\max}\right)^2,
\ea
where we have used that for $(m,p)\in\mathcal{ S}_{11}(R)$, $\dom(\gamma_{l,m,p})=l,\ldots,l+\max\{m,p\}$.
Thus we conclude that for all $\alpha_{l,m,p}\in\dom(\gamma_{l,m,p})$ and for all $(m,p)\in\mathcal{ S}_{12}(R)$, $l\in\mathcal{I}_d$
\be 
\Delta F(d;l,m,p) \leq \pi \frac{10}{3}  \frac{R}{d} Q_{\max}+ \bo\left(\pi \frac{10}{3}  \frac{R}{d} Q_{\max}\right)^2 + \bo \left(  2 \me^{-\pi \frac{4}{3}\left(\frac{d}{\sigma}\right)^2}  \right).
\ee 
Now consider the set $\mathcal{S}_{33}$. For all $(m,p)\in\mathcal{ S}_{33}(R)$, we have that $\lim_{d\rightarrow \infty} \tilde H_1(d,m,p)=\lim_{d\rightarrow \infty} \tilde H_2(d,m,p) = 1$, so the terms in lines Eqs. \eqref{line:Delta F H1},\eqref{line:Delta F H2} are exponentially small in $(d/\sigma)^2$. For the other term, i.e. line Eq. \eqref{line:Delta F H3}, we proceed by bounding the part within absolute values. Specifically, for all $\alpha_{l,m,p}\in\dom(\gamma_{l,m,p})$ and for all $(m,p)\in\mathcal{ S}_{33}(R)$, $l\in\mathcal{I}_d$,
\ba
\bigg|\me^{-\frac{\pi}{6}\left(\frac{\sigma}{d}\right)^2 Q^2} &
\me^{-\mi\pi\frac{2}{3}(m+p-2d)Q/d}\me^{2\pi \mi \left(\gamma_{l,m,p}(\alpha_{l,m,p})-l+\frac{\Delta k_\textup{T}}{3}\right) Q/d} - 1 \bigg| \\
=&  \bigg|\me^{-\frac{\pi}{6}\left(\frac{\sigma}{d}\right)^2 Q^2} 
\me^{-\mi\pi\frac{2}{3}(m+p-2d)Q/d}\me^{2\pi \mi \left(\gamma_{l,m,p}(\alpha_{l,m,p})-l-d+\frac{\Delta k_\textup{T}}{3}\right) Q/d} - 1 \bigg|\\
\leq &  \left|\pi\frac{4}{3}  \frac{R}{d}Q_{\max} +2\pi \frac{R}{d}Q_{\max} \right|+\bo\left(\pi\frac{4}{3}  \frac{R}{d}Q_{\max} +2\pi \frac{R}{d}Q_{\max} \right)^2\\
=&\pi \frac{10}{3}  \frac{R}{d} Q_{\max}+ \bo\left(\pi \frac{10}{3} \frac{R}{d} Q_{\max}\right)^2,\label{eq:exp phase exmaple}
\ea 
where we have used that for $(m,p)\in\mathcal{ S}_{33}(R)$, $\cdom(\gamma_{l,m,p}(\alpha_{l,m,p})-l-d)=\min\{m,p\}-d,\dots,0$ and $- R\leq  \min\{m,p\}-d$, giving us $|\gamma_{l,m,p}(\alpha_{l,m,p})-l-d|\leq R$.  Thus we conclude that for all $\alpha_{l,m,p}\in\dom(\gamma_{l,m,p})$ and for all $(m,p)\in\mathcal{ S}_{33}(R)$, $l\in\mathcal{I}_d$
\be 
\Delta F(d;l,m,p) \leq \pi \frac{10}{3}  \frac{R}{d} Q_{\max}+ \bo\left(\pi \frac{10}{3}  \frac{R}{d} Q_{\max}\right)^2 + \bo \left(  2 \me^{-\pi \frac{4}{3}\left(\frac{d}{\sigma}\right)^2}  \right).
\ee 
In general, for every set $\mathcal{ S}_{xy}(R)$, $x,y\in\{1,3\}$, two out of the three terms $\tilde H_1(d;m,p)$, $\tilde H_2(d;m,p)$, $\tilde H_3(d;m,p)$ converge to unity as $d$ tends to infinity, resulting in two of the lines Eqs. \eqref{line:Delta F H1}, \eqref{line:Delta F H2}, \eqref{line:Delta F H3} being exponentially small in $(d/\sigma)^2$. The terms in the  remaining line can be expended in a power expansion similarly to Eq. \eqref{eq:exp phase exmaple}, resulting in an order $Q_{\max}R/d$ contribution.

We therefore conclude that the following is true for all $x,y\in\{1,3\}$: For all $\alpha_{l,m,p}\in\dom(\gamma_{l,m,p})$ and for all $(m,p)\in\mathcal{ S}_{xy}(R)$, $l\in\mathcal{I}_d$,
\be 
\Delta F(d;l,m,p) \leq \pi \frac{10}{3} \frac{R}{d} Q_{\max}+ \bo\left(\pi \frac{10}{3} \frac{R}{d} Q_{\max}\right)^2 + \bo \left(  2 \me^{-\pi \frac{4}{3}\left(\frac{d}{\sigma}\right)^2}  \right).\label{eq:Delta F for S}
\ee 
From the definition of the sets $x,y\in\{1,3\}$ with $m,p\in\mathcal{S}_{xy}(R)$ we see that they all have cardinality $R^2$. Thus from Eq. \eqref{eq:Delta F for S}, it follows that for all $\alpha_{l,m,p}\in\dom(\gamma_{l,m,p})$ and for all $(m,p)\in$ $ \mathcal{S}_{11}(R)\cup\mathcal{S}_{13}(R)\cup\mathcal{S}_{31}(R)\cup\mathcal{S}_{33}(R)$, $l\in\mathcal{I}_d$,
\ba
\sum_{l\in\mathcal{I}_d} \,\,\sum_{(m,p)\in \mathcal{S}_{11}(R)\cup\mathcal{S}_{13}(R)\cup\mathcal{S}_{31}(R)\cup\mathcal{S}_{33}(R)}& \Delta F(d;l,m,p) \\
&\leq d R^2\left(  \pi \frac{10}{3} \frac{R}{d} Q_{\max}+ \bo\left(\pi \frac{10}{3} \frac{R}{d} Q_{\max}\right)^2 + \bo \left(  2 \me^{-\pi \frac{4}{3}\left(\frac{d}{\sigma}\right)^2}  \right) \right).\label{eq:sum over S sets}
\ea 
We will now bound $\Delta F(d;l,m,p)$ for $(m,p)$ belonging to the sets Eqs. \eqref{eq:set 12 def} to \eqref{eq:set 12 def}. We start with the set $\mathcal{\bar S}_{12}$. For all $(m,p)\in\mathcal{\bar S}_{12}(R)$, we have for sufficiently large $d$
\ba
 & \left(\frac{d}{\sigma}\right)^2\tilde H_1(d,m,p)\geq  \left( \frac{R}{\sigma}\right)^2,\\
 & \left(\frac{d}{\sigma}\right)^2\tilde H_2(d,m,p)\geq  \left( \frac{R}{\sigma}\right)^2,\\
 & \left(\frac{d}{\sigma}\right)^2\tilde H_3(d,m,p)\geq  \left(\frac{d}{\sigma}\right)^2.
\ea
Thus from Eq. \eqref{eq: hat E 1 norm bound 7} we conclude
that for all $\alpha_{l,m,p}\in\dom(\gamma_{l,m,p})$ and for all $(m,p)\in\mathcal{\bar S}_{12}(R)$, $l\in\mathcal{I}_d$,
\be 
\Delta F(d;l,m,p) \leq \bo \left(2\,  \me^{-\pi \frac{4}{3}\left(\frac{d}{\sigma}\right)^2} \right) + \bo \left( 4\, \me^{-\pi \frac{4}{3}\left(\frac{R}{\sigma}\right)^2} \right).\label{eq:Delta F 2 4 d sigma R sigma}
\ee 
Similarly, one finds that Eq. \eqref{eq:Delta F 2 4 d sigma R sigma} holds for all $\alpha_{l,m,p}\in\dom(\gamma_{l,m,p})$ and for all $(m,p)\in\mathcal{\bar S}_{xy}(R)$, $l\in\mathcal{I}_d$, 
for $xy=21,23,32$. In the case of the set $\mathcal{\bar S}_{22}(R)$ we find that for all $\alpha_{l,m,p}\in\dom(\gamma_{l,m,p})$ and for all $(m,p)\in\mathcal{\bar S}_{22}(R)$, $l\in\mathcal{I}_d$,
\be 
\Delta F(d;l,m,p) \leq \bo \left( 6\, \me^{-\pi \frac{4}{3}\left(\frac{R}{\sigma}\right)^2} \right).
\ee 
Finally, since the cardinality of the sets $\mathcal{\bar S}_{12}(R)$, $\mathcal{\bar S}_{21}(R)$, $\mathcal{\bar S}_{22}(R)$, $\mathcal{\bar S}_{23}(R)$, $\mathcal{\bar S}_{32}(R)$ are all upper bounded by $d^2$, we have that for all $\alpha_{l,m,p}\in\dom(\gamma_{l,m,p})$, and for all $(m,p)\in \mathcal{\bar S}_{12}(R)\cup \mathcal{\bar S}_{21}(R)\cup \mathcal{\bar S}_{22}(R)\cup \mathcal{\bar S}_{23}(R)\cup \mathcal{\bar S}_{32}(R)$, $l\in\mathcal{I}_d$,
\ba
\sum_{l\in\mathcal{I}_d}\,\,\sum_{(m,p)\in \mathcal{\bar S}_{12}(R)\cup \mathcal{\bar S}_{21}(R)\cup \mathcal{\bar S}_{22}(R)\cup \mathcal{\bar S}_{23}(R)\cup \mathcal{\bar S}_{32}(R)}& \Delta F(d;l,m,p) \\
&\leq  \bo \left(  d^3 \, \me^{-\pi \frac{4}{3}\left(\frac{d}{\sigma}\right)^2}\right) + \bo \left(   d^3 \,\me^{-\pi \frac{4}{3}\left(\frac{R}{\sigma}\right)^2}  \right).\label{eq:bar S sets sum}
\ea 
Hence, combining Eqs. \eqref{eq:sum over S sets} and \eqref{eq:bar S sets sum}, and plugging into Eq. \eqref{eq: hat E 1 norm bound 5} we find for all $\alpha_{l,m,p}\in\dom(\gamma_{l,m,p})$ and for all $(m,p)\in\mathcal{S}_\textup{tot}$, $l\in\mathcal{I}_d$
\ba
\vert \vert \hat E \vert \vert_1\le& 
\sqrt{d_\inp} d_\outp \frac{A^6\sigma }{\sqrt{6}}   \Bigg[ R^2\left(  \pi \frac{10}{3} \frac{R}{d} Q_{\max}+ \bo\left( \frac{R}{d} Q_{\max}\right)^2 + \bo \left(   \me^{-\pi \frac{4}{3}\left(\frac{d}{\sigma}\right)^2}  \right) \right)\\
&+  \bo \left(  d^2 \, \me^{-\pi \frac{4}{3}\left(\frac{d}{\sigma}\right)^2}\right) + \bo \left(   d^2 \,\me^{-\pi \frac{4}{3}\left(\frac{R}{\sigma}\right)^2}  \right) + 2 d^2\, |\hat  \varepsilon_\textup{T}|\Bigg]\\
= & 
\sqrt{d_\inp} d_\outp \frac{A^6\sigma }{\sqrt{6}}   \Bigg[ R^2\left(  \pi \frac{10}{3} \frac{R}{d} Q_{\max}+ \bo\left( \frac{R}{d} Q_{\max}\right)^2  \right)\\
& + \bo \left(   d^2 \,\me^{-\pi \frac{4}{3}\left(\frac{R}{\sigma}\right)^2}  \right) + 2 d^2\, |\hat  \varepsilon_\textup{T}|\Bigg],
\ea
where, to find the higher order terms, we have used in the last equality
\be 
\lim_{d\rightarrow \infty}\frac{R/\sigma}{d/\sigma} =0,
\ee 
which follows from Eq. \eqref{eq:R limits def}. Finally, using $A^2=\sqrt{2}/\sigma+\varepsilon_A=\bo(1/\sigma)$  from page 483 in \cite{woods2016autonomous} and the upper bound on in Eq. \eqref{eq: ep T abs final}, we find for all $\alpha_{l,m,p}\in\dom(\gamma_{l,m,p})$ and for all $(m,p)\in\mathcal{S}_\textup{tot}$, $l\in\mathcal{I}_d$,
\ba
\vert \vert \hat E \vert \vert_1
\leq & 
\sqrt{d_\inp} d_\outp \frac{2 }{\sqrt{3}}   \Bigg[ \left(\frac{R}{\sigma}\right)^2\left(  \pi \frac{10}{3} \frac{R}{d} Q_{\max}+ \bo\left( \frac{R}{d} Q_{\max}\right)^2  \right)\\
& + \bo \left(   \left(\frac{d}{\sigma}\right)^2 \,\me^{-\pi \frac{4}{3}\left(\frac{R}{\sigma}\right)^2}  \right) +\bo\left( \frac{d^4}{\sigma^{9/2}}  \right)\me^{-\frac{\pi}{4}\sigma^2\alpha_0^2}\Bigg],\quad \text{as }\,\,  d\rightarrow\infty. \label{up bound hat E 1 norm for 3 clocks}
\ea

We have the error term $\vert \vert \hat E \vert \vert_1$ in terms of $R$, and $\sigma$. These are free parameters that we can choose. The optimal scaling is given by
\begin{equation}
\sigma = \ln^{5/2}{d} \,\,\,\quad R=\ln^4 d,
\end{equation}
as the errors become
\ba
\vert \vert \hat E \vert \vert_1
\leq  
\sqrt{d_\inp} d_\outp \frac{2 }{\sqrt{3}}   \Bigg[&\frac{10 \pi \ln^7(d) Q_{\max}}{3 d} +\bo\left( \frac{\ln^{11}(d)}{d^2} Q^2_{\max}\right) \\ &+\bo \left(\frac{1}{d \ln^{5/2} (d)} \right) +\bo \left(\frac{1}{d \ln^{45/4} (d)} \right)\Bigg],\quad \text{as }\,\,  d\rightarrow\infty, \label{up bound hat E 1 norm for 3 clocks2}
\ea
for all $\alpha_{l,m,p}\in\dom(\gamma_{l,m,p})$ and for all $(m,p)\in\mathcal{S}_\textup{tot}$, $l\in\mathcal{I}_d$. 
Asymptotically, the first two error terms will be the dominant ones, and thus we conclude that for all $\alpha_{l,m,p}\in\dom(\gamma_{l,m,p})$ and for all $(m,p)\in\mathcal{S}_\textup{tot}$, $l\in\mathcal{I}_d$,
\ba
\vert \vert \hat E \vert \vert_1
\leq  
\sqrt{d_\inp} d_\outp \frac{2 }{\sqrt{3}}   \Bigg[&\frac{10 \pi \ln^7(d) Q_{\max}}{3 d} +\bo\left( \frac{\ln^{11}(d)}{d^2} Q^2_{\max}\right)\Bigg],\quad \text{as }\,\,  d\rightarrow\infty, \label{up bound hat E 1 norm for 3 clocks3}
\ea
making the error effectively scale as $\bo \left(\frac{d_\inp \sqrt{d_\outp} \ln^7(d) Q_{\max}}{d} \right)$. Finally using Eq. \eqref{eq:fidEE}, we achieve for all $\alpha_{l,m,p}\in\dom(\gamma_{l,m,p})$ and for all $(m,p)\in\mathcal{S}_\textup{tot}$, $l\in\mathcal{I}_d$. 
\be 
f_{worst}^2\geq 1-\sqrt{d_\inp} d_\outp   \Bigg[\frac{10 \pi \ln^7(d) Q_{\max}}{\sqrt{3} d} +\bo\left( \frac{\ln^{11}(d)}{d^2} Q^2_{\max}\right)\Bigg],\quad \text{as }\,\,  d\rightarrow\infty,\label{eq:f worst 3 clcoks in proof}
\ee
where $Q_{\max}=\Delta h_\inp+\Delta h_\code$. 

\subsection{The Protocol}\label{sec:3 clock protocol}
Here we provide the protocol (based on the results from the previous subsection) which will achieve the statement of Theorem \ref{Thm:3 clock phase error} . Recall that the general protocol for all set-ups in this paper is described in Appendix \ref{app:gen}, thus here we only need to specificity how to choose the parameter $k_\alpha$ given the clock measurement outcomes.

After obtaining the measurement outcomes $k_1',k_2',k_3'\in\{0,1,\ldots,d-1\}$, the second task will be to work out the quantities $\tilde l,$ $\tilde l+ \tilde m$, $\tilde l+ \tilde p$ which are defined via the relations
\ba
k_q'=	\left[ -\frac{d}{2}+1+\lfloor  k_q^0 \rfloor +\tilde \Gamma_q\right]_\textup{(mod.  \textit{d})
},\quad\quad
\tilde \Gamma_q=
\begin{cases}
	\tilde l &\mbox{ if } q=1,\\
	\tilde l+\tilde m  &\mbox{ if } q=2,\\
	\tilde l+\tilde p &\mbox{ if } q=3,\label{eq:tilde l m p def}
\end{cases}
\ea
where $\tilde l, \tilde m, \tilde p\in\{0,1,\ldots,d-1\}$ and $\mathcal{I}_d=\{0,1,\ldots, d-1\}$. Importantly $\tilde \Gamma_1, \tilde \Gamma_2, \tilde \Gamma_3$ are always calculable since $d$, $k_1^0$, $k_2^0$, $k_3^0$ are known.

Next, out of the three angles %
$ \tilde \Gamma_1$, $ [\tilde \Gamma_2]_{\textup{(mod. } d)}$, $ [\tilde \Gamma_3]_{\textup{(mod. } d)} \in[0,d-1]$, find which one of these is the \emph{middle angle}; denoted $\alpha_\textup{middle}$. Specifically, this angle can be calculated as follows:
\begin{itemize}
	\item [1.] First, for $q_1<q_2$ with  $q_1,q_2\in\{1,2,3\}$, calculate the quantities 
	\be 
	\Delta_{q_1, q_2}:=\begin{cases}
		\Big{|}[\tilde \Gamma_{q_1}]_{\textup{(mod. \textit{d})}}-[\tilde \Gamma_{q_2}]_{\textup{(mod. \textit{d}) }}\Big{|} &\mbox{if  }\quad 	\Big{|}[\tilde \Gamma_{q_1}]_{\textup{(mod. \textit{d})}}-[\tilde \Gamma_{q_2}]_{\textup{(mod. \textit{d})}}\Big{|} \leq \frac{d}{2}-1 \vspace{0.31cm}\\
		d-	\Big{|}[\tilde \Gamma_{q_1}]_{\textup{(mod. \textit{d})}}-[\tilde \Gamma_{q_2}]_{\textup{(mod. \textit{d})}}\Big{|} &\mbox{if  }\quad 	\Big{|}[\tilde \Gamma_{q_1}]_{\textup{(mod. \textit{d})}}-[\tilde \Gamma_{q_2}]_{\textup{(mod. \textit{d})}}\Big{|} \geq \frac{d}{2}
	\end{cases}
	\ee 
	\item [2.] Second, find the set containing the smallest two out of the values $\Delta_{1,2}$, $\Delta_{1,3}$, $\Delta_{2,3}$:
	\ba 
	\text{Case 2.1) }&\quad\quad \{\Delta_{1,3}, \Delta_{2,3} \} \text{ if } \Delta_{1,2} \geq \Delta_{1,3},\, \Delta_{1,2} \geq \Delta_{2,3}\\
	\text{Case 2.2) }&\quad\quad  \{\Delta_{1,3}, \Delta_{1,2} \} \text{ if } \Delta_{2,3} \geq \Delta_{1,2},\, \Delta_{2,3} \geq \Delta_{1,3} \\
	\text{Case 2.3) }&\quad\quad  \{\Delta_{1,2}, \Delta_{2,3} \} \text{ if } \Delta_{1,3} \geq \Delta_{1,2},\, \Delta_{1,3} \geq \Delta_{2,3}. 
	\ea
	\item[3.] Third, find the common index $w$ of the smallest two:
	\ba 
	w:= \begin{cases}
		3 &\mbox{ in case 2.1)}\\
		1 &\mbox{ in case 2.2)}\\
		2 &\mbox{ in case 2.3)}
	\end{cases}
	\ea
	\item[4.] Fourth, the middle angle is
	\be 
	\alpha_\textup{middle}=\tilde \Gamma_w.\label{eq:middle anlge def}
	\ee  
\end{itemize}

We call $w\in{1,2,3}$ in Eq. \eqref{eq:middle anlge def} corresponding to the middle angle, the \textit{location of the middle angle}.

Set $\alpha$ in Eq. \eqref{eq: def of k alpha 3 clocks} to the middle angle, $\alpha=\alpha_\textup{middle}$,
\be 
k_\alpha=-\frac{d}{2}+1+\gamma_{l,m,p}(\alpha_\textup{middle})=-\frac{d}{2}+1+\alpha_\textup{middle}.\label{eq:k alpha in middle}
\ee
Since Eq. \eqref{eq:f worst 3 clcoks in proof} holds for all $\alpha_{l,m,p}\in\dom(\gamma_{l,m,p})$ and for all $(m,p)\in\mathcal{S}_\textup{tot}$, $l\in\mathcal{I}_d$, the protocol would be complete, since calculation of the r.h.s. of Eq. \eqref{eq:k alpha in middle} only requires the knowledge of known parameters $d, k_1', k_2', k_3',k_1^0, k_2^0, k_3^0$.  However, we are not quite done, since what is left to prove is that $\alpha_\textup{middle}\in\dom(\gamma_{l,m,p})$ indeed holds true for all $(m,p)\in\mathcal{S}_\textup{tot}$, $l\in\mathcal{I}_d$, since this has been assumed in Eq. \eqref{eq:k alpha in middle} without justification. To do so, we have to analyse two possible scenarios. First, however, recall Eq. \eqref{eq:k prime measures def}; reproduced here for convenience:
\ba
k_q':= \left[ k_q+l\right]_\textup{(mod. \textit{d})
}=\left[ -\frac{d}{2}+1+\lfloor \tilde k_1^0 \rfloor +\Gamma_q\right]_\textup{(mod. \textit{d})
},\quad\quad
\Gamma_q=
\begin{cases}
	l &\mbox{ if } q=1,\\
	l+m &\mbox{ if } q=2,\\
	l+p &\mbox{ if } q=3,\label{eq:k prime def again}
\end{cases}
\ea
where $\mathcal{I}_d=\{0,1,\ldots, d-1\}$ and 
\be 
\tilde k_q^0=
\begin{cases}
	k_q^0+t_\textup{ph}d/T_0  &\mbox{ if } q=r,\\
	k_q^0 &\mbox{ otherwise},
\end{cases}
\ee 
with $r\in{1,2,3}$ the unknown location of the unknown phase $t_\textup{ph}\in\rr$. The two scenarios are:\\

\textbf{Case 1)} The unknown location $r\in{1,2,3}$ does not coincide with the location of the middle angle ($w\neq r$): In this case, by comparing Eqs. \eqref{eq:tilde l m p def}, \eqref{eq:k prime def again}, we see that the angle $\alpha_\textup{middle}$ corresponds to one of the angles $\Gamma_1$, $\Gamma_2$, $\Gamma_3$. By definition, all three of these angles belong to the domain of $\gamma_{m,p}$ for all $(m,p)\in\mathcal{S}_\textup{tot}$ and for all $l\in\mathcal{I}_d$.

\textbf{Case 2)} The unknown location $r\in{1,2,3}$ coincides with the location of the middle angle ($w = r$): In this case, it follows that the two angles out of the three angles $\Gamma_1$, $\Gamma_2$, $\Gamma_3$ which do \emph{not} correspond to the middle angle $\alpha_\textup{middle}$, (namely angles $\Gamma_h,\Gamma_k$, where $h\neq k\neq r$)  have \emph{not} had an unknown phase applied to them (since by definition, there is only one unknown phase). Hence from Eqs. \eqref{eq:tilde l m p def}, \eqref{eq:k prime def again} we have $\Gamma_h=\tilde \Gamma_h,\Gamma_k=\tilde \Gamma_k$. Hence since the middle angle is always between the other two angles $\Gamma_h,\Gamma_k$, and $\dom(\gamma_{l,m,p})$ includes all angles which are between all three angles $\Gamma_1, \Gamma_2,\Gamma_3$ (See Fig. \ref{fig 4 plots}), it follows that the middle angle $\alpha_\textup{middle}=\tilde \Gamma_r\in\dom(\gamma_{l,m,p})$ for all $(m,p)\in\mathcal{S}_\textup{tot}$, $l\in\mathcal{I}_d$.\\

Hence cases 1) and 2) together prove the assumption made in Eq. \eqref{eq:k alpha in middle}, namely that $\alpha_\textup{middle}\in\dom(\gamma_{l,m,p})$ indeed holds true for all $(m,p)\in\mathcal{S}_\textup{tot}$, $l\in\mathcal{I}_d$. This completes the proof.


\section{Quantum communication without a shared reference frame}\label{app:quantumcomm}

Throughout the paper we have shown that the Quasi-Ideal clock states $\ket{\psi_\textup{nor}(k_1^0)}$ achieve the best bounds for the task of maximizing the entanglement fidelity of a covariant code. Here, following \cite{bartlett2009quantum}, we also show how they can improve on a pre-existing scheme for another related task: the transmission of quantum states between parties which do not share a common reference frame. 

The setting is as follows: Alice and Bob are connected by a noiseless quantum channel, but separated by an indeterminate amount of ``time". This means that the channel connecting then takes the form
\begin{equation}
\mathcal{E}(\cdot)=\mathcal{G}(\cdot):= \lim_{T\rightarrow \infty} \int_0^T \frac{\text{d}t}{T} U(t) (\cdot) U^\dagger(t),
\end{equation}
where $ U(t)$ is the time translation of the systems that are sent. This is a decoherence map in the basis of the generator of $U(t)$. In order to damp the loss of information from the channel, what Alice can do is to send the message $\rho_M$ together with a reference frame $\F$, as $\mathcal{G}(\rho_M \otimes \ket{\psi_\F}\bra{\psi_\F})$ which Bob then uses to ``decode" the message. To decode, Bob measures the clock to learn the time that separates them, which can be used to retrieve the message up to some error (we label this decoding procedure with $\mathcal{D}$). In \cite{bartlett2009quantum} it was shown that for a qubit, the resultant channel is given by
\begin{equation}
\mathcal{D}\circ \mathcal{E}(\rho) =\lim_{T\rightarrow \infty} \int_0^T \frac{\text{d}t}{T} \vert \bra{\psi}U_
\F(t)\ket{\psi}\vert^2  U_M(t) \rho U_M^\dagger(t)=(1-p)\mathcal{I}+p \mathcal{G}(\rho),
\end{equation}
where the probability of fully decohering $p$ is given by
\begin{align}
p=\frac{A_1-A_2}{A_1},
\end{align}
where
\begin{align}
A_1 = \lim_{T\rightarrow \infty} \int_0^T \frac{\text{d}t}{T} \vert \bra{\psi}U_
\F(t)\ket{\psi}\vert^2 ,\\
A_2 = \lim_{T\rightarrow \infty} \int_0^T \frac{\text{d}t}{T} \vert \bra{\psi}U_
\F(t)\ket{\psi}\vert^2 e^{-i \frac{2\pi t}{T}}.
\end{align}
In \cite{bartlett2009quantum}, it was found that for the Salecker-Wigner-Peres clock, $p_{PSW}=\frac{1}{d_\clo+1}$ with $d$ the dimension of the clock. We now show how this changes when one uses the Quasi-Ideal clock $\ket{\psi(0)}$ as defined in the main text. In that case, it can be shown that
\begin{equation}\label{eq:charac}
\vert \bra{\psi}U_\F(t)\ket{\psi}\vert^2 \propto \left(e^{-\frac{\pi}{2 \sigma^2}\left(\frac{2 \pi d_\clo t}{\omega}\right)^2}+e^{-\frac{\pi}{2 \sigma^2}\left(d_\clo-\frac{2 \pi d_\clo t}{\omega}\right)^2}\right)^2+\epsilon_1 ,
\end{equation}
where $\omega$ is the frequency at which the qubit oscillates, and
\begin{equation}
\vert \epsilon_1\vert \le \mathcal{O}(e^{-c_1 \sigma^2})+\mathcal{O}(e^{-c_2 \frac{d_\clo^2}{\sigma^2}}),
\end{equation}
where $c_1,c_2$ are positive constants independent of $d_\clo$ and $\sigma$.
Let us now sketch how to derive Eq. \eqref{eq:charac}. First one writes out the characteristic function $\vert \bra{\psi}U_\F(t)\ket{\psi}\vert$, as a finite sum of terms corresponding to the elements of the eigenbasis $\{\ket{\phi_k}\}$. This is 
\begin{align}
\vert \bra{\psi}U_\F(t)\ket{\psi}\vert&=A^2  \sum_{k\in \{-\frac{d_\clo}{2}+1,-\frac{d_\clo}{2}+\lfloor \frac{2 \pi d_\clo t}{\omega}\rfloor\}} e^{-\frac{\pi}{\sigma^2}[k^2+(k-\frac{2 \pi d_\clo t}{\omega})^2]}+A^2\sum_{k\in \{-\frac{d_\clo}{2}+\lfloor \frac{2 \pi d_\clo t}{\omega}\rfloor+1,\frac{d_\clo}{2}\}} e^{-\frac{\pi}{\sigma^2}[k^2+(k+d_\clo-\frac{2 \pi d_\clo t}{\omega})^2]} \\&=A^2  \sum_{k\in \mathbb{Z}} e^{-\frac{\pi}{\sigma^2}[k^2+(k-\frac{2 \pi d_\clo t}{\omega})^2]}+ e^{-\frac{\pi}{\sigma^2}[k^2+(k+d_\clo-\frac{2 \pi d_\clo t}{\omega})^2]}+\mathcal{O}(e^{-c_2 \frac{d_\clo^2}{\sigma^2}})\\
&=A^2  \sum_{m\in \mathbb{Z}}\frac{\sigma}{\sqrt{2d_\clo}} e^{-\frac{\pi}{2}\left[\left(\frac{2 \pi d_\clo t}{\omega} \right)^2 +m^2 \sigma^2 -2i d_\clo^2 m \frac{2 \pi d_\clo t}{\omega})\sigma^2\right]}\\&\quad \quad \quad \quad + \frac{\sigma}{\sqrt{2d_\clo}}e^{-\frac{\pi}{2}\left[\left(d_\clo-\frac{2 \pi d_\clo t}{\omega} \right)^2 +m^2 \sigma^2 -2i d_\clo^2 m (d_\clo-\frac{2 \pi d_\clo t}{\omega})\sigma^2\right]}+\mathcal{O}(e^{-c_2 \frac{d_\clo^2}{\sigma^2}})\\
&=A^2 \frac{\sigma}{\sqrt{2d_\clo}}\left( e^{-\frac{\pi}{2}\left(\frac{2 \pi d_\clo t}{\omega} \right)^2} + e^{-\frac{\pi}{2}\left(d_\clo-\frac{2 \pi d_\clo t}{\omega} \right)^2}+\mathcal{O}(e^{-c_1 \sigma^2})\right)+\mathcal{O}(e^{-c_2 \frac{d_\clo^2}{\sigma^2}}),
\end{align} 
where in going from the second to the third line we have used the Poisson summation formula (see for instance Corollary C.0.2 from \cite{woods2016autonomous}), and in the third to the fourth we have kept the $m=0$ term and bounded the size of the others.

With that, we obtain
\begin{align}
&A_1/C = \frac{\sigma}{d_\clo}+\mathcal{O}\left(\left(\frac{\sigma}{d_\clo} \right)^4 \right)+\mathcal{O}(e^{-c_1 \sigma^2})+\mathcal{O}(e^{-c_2 \frac{d_\clo^2}{\sigma^2}}) \\
&A_2/C = \frac{\sigma}{d_\clo}-\pi \frac{\sigma^3}{d_\clo^3}+\mathcal{O}\left(\left(\frac{\sigma}{d_\clo} \right)^4 \right)+\mathcal{O}(e^{-c_1 \sigma^2})+\mathcal{O}(e^{-c_2 \frac{d_\clo^2}{\sigma^2}}),
\end{align}
with $C>0$ independent of $d_\clo$ and $\sigma$.
Thus, to leading order, it gives a probability of decoherence of
\begin{equation}
p_{id}=\pi \frac{\sigma^2}{d_\clo^2}+\mathcal{O}\left(\left(\frac{\sigma}{d_\clo} \right)^3 \right)+\mathcal{O}(e^{-c_1 \sigma^2})+\mathcal{O}(e^{-c_2 \frac{d_\clo^2}{\sigma^2}}).
\end{equation}
The best choice for the width of the clock is $\sigma= (\ln{d_\clo})^{\frac{3}{2}}$, which gives a scaling of
\begin{equation}
p_{id}=\pi \left(\frac{(\ln{d_\clo})^{\frac{3}{2}}}{d_\clo}\right)^2+\mathcal{O}\left(\left(\frac{(\ln{d_\clo})^{\frac{3}{2}}}{d_\clo} \right)^3\right),
\end{equation}
which (up to the logarithmic factor) is quadratically smaller than $p_{PSW}$. This is essentially the same advantage as the one obtained in covariant error correction as shown in the main text.

\section{Evolution without evolution: connection with the Page-Wootters mechanism}\label{Sec:PageWooters}
In this Section we briefly re-cap the Page-Wooters mechanism (Section III. ``Evolution in a Stationary Universe" in \cite{PageWooter}) and show how our formalism and the results in this paper fit into that picture. These authors consider a bipartite setup consisting of a clock and a system $S$ of interest. The system and clock are considered as a closed system evolving under a Hamiltonian of the form
\be 
\hat H= \hat H_S\otimes\id_\textup{clock} + \id_S\otimes \hat H_\textup{clock}.\label{eq:Ham PW ham}
\ee 
Their idea is to show that there is a global state $\rho_{S\textup{clock}}$ which does not evolve in time, i.e. $[\rho_{S\textup{clock}},\hat H]=0$ yet the dynamics of the system, after conditioning on the clock being in a state  
\be 
\ket{\psi(\tau)}_\textup{clock}=\me^{-\tau\mi \hat H_\textup{clock}} \ket{\psi(0)}_\textup{clock}, \label{eq:PWcondition exact}
\ee 
at time $\tau$, is given by the free evolution of the system $S$ according to its own Hamiltonian $\hat H_S$. Namely,
\be 
 \tr_\textup{clock}[\hat P_\textup{clock}(\tau)  \rho_{S\textup{clock}} \hat P_\textup{clock}(\tau) ]= \me^{-\mi\tau \hat H_S} \tr_\textup{clock}[ \rho_{S\textup{clock}}] \me^{\mi\tau \hat H_S},\label{eq:PWcondition}
\ee 
 where $\hat P_\textup{clock}(\tau) := \id_S\otimes\ketbra{\psi(\tau)}{\psi(\tau)}_\textup{clock}$ is the projector onto the clock state at time $\tau$. So in the above sense, even though the global state $\rho_{S\textup{clock}}$ is stationary, the state of the system conditioned on the clock being at a particular time, is evolving according to the Schr\"odinger equation for its own Hamiltonian $\hat H_S$.\\
 
 Before seeing how this is connected to our work, a few remarks are pertinent. Firstly, time is assumed to be continuous, i.e. if one wishes Eq. \eqref{eq:PWcondition} to hold for all $\tau$ in some finite subinterval of the real line, then one needs to use an Idealised clock (these are discussed in Section \ref{Introduction and Overview of Results} in the main text). These, however, are unphysical in the sense of requiring infinite energy. To the best knowledge of the authors, while many aspects of the Page-Wooters mechanism have been explored (see \cite{PhysRevD.92.045033,PhysRevD.79.041501,PhysRevD.95.043510,PhysRevA.89.052122} and references therein), it remains an open question to how well finite dimensional clocks (or infinite dimensional clocks with finite energy) can realize this. It is also noteworthy that in our setup, the clock and system are only classically correlated, and thus our results demonstrate that quantum entanglement between the clock and system $S$ is not necessary to fulfil the Page-Wooters mechanism. We now find the Hamiltonian and time invariant state which realise the Page-Wooters mechanism up to a quantifiable error incurred due to using physical clocks.
 
 Firstly, we can identify the physical space with the system in the Page-Wooters model and the clocks in our setup with those in Page-Wooters model. The total Hamiltonian on the system and the clock will be the Kronecker sum of the generators of the physical space Lie group and those of the clock's Lie group, namely (c.f. Eq. \eqref{eq:Ham PW ham})
 \be 
 \hat H=\hat H_{\outpp}\otimes\id_\clo^{\otimes \M}+ \id_\outpp\otimes\bigoplus_{i=1}^{\M} \hat H_\clo. \label{eq: Ham PW mech}
 \ee 
 Secondly, we use our covariant encoding channel $\mathcal{E}_{\text{cov}}(\cdot)$ in Eq. \eqref{eq:code} to realise the stationary state $\rho_{S\textup{clock}}$ of the Page-Wooters mechanism
 \be 
 \rho_{S\textup{clock}}=\map_{\textup{cov}}(\rho_\inpp),
 \ee 
 for any $\rho_\inpp\in\mathcal{S}\left(\mathcal{H}_\inpp\right)$ and for the choice of trivial representation on the Logical space, namely $U_\inpp(t)=\id_\inpp$ for all $t\in\rr$. To see how this works, note that 1) the unitary group representation on the physical and clock space has the Hamiltonian in Eq. \eqref{eq: Ham PW mech} as its generator, and 2) that the encoding map is covariant. Specifically 
 \be 
 \me^{-\mi t\hat H} \map_{\textup{cov}}(\rho_\inpp)  \me^{\mi t\hat H}= U_\code(t)\otimes U_\clo^{\otimes \M}(t)\,\map_{\textup{cov}}(\rho_\inpp)\, U^\dag_\code(t)\otimes U_\clo^{\otimes \M \dag}(t)= \map_{\textup{cov}}(U_\inpp(t)\rho_\inpp U_\inpp^\dag(t))= \map_{\textup{cov}}(\rho_\inpp)\quad \forall\, t\in\rr,\label{eq:con state PW}
 \ee 
 where in the last line, we have used $U_\inpp(t)=\id_\inpp$ for all $t\in\rr$. Hence differentiating w.r.t. $t$ on both sides on Eq. \eqref{eq:con state PW}, we achieve the Page-Wooters condition, 
 \be 
 [\hat H, \map_{\textup{cov}}(\rho_\inpp)]=0\quad \forall\, t\in\rr.
 \ee 
Thirdly, our clock state must, at least approximately, satisfy Eq. \eqref{eq:PWcondition}. We have already calculated the state of $ \map_{\textup{cov}}(\rho_\inpp)$ when the clocks were projected onto the time basis $\{ \ketbra{\theta_k}{\theta_k} \}_{k=0}^{d_\clo-1}$ in Section \ref{app:gen}. Now, we want to perform a similar calculation but for when the clock is projected onto the state of the clock at time $\tau$. In the case of one Quasi-Ideal clock, this corresponds to covariant positive-operator valued measure, which up to a vanishing error in the large $d_\clo/\sigma$ and $\sigma$ limit, are $\{\hat P_\textup{QI}(\tau):= U_\clo(\tau)\ketbra{\psi_\textup{nor}(k_1^0)}{\psi_\textup{nor}(k_1^0)} U_\clo(\tau)^\dag \}_{\tau\in[0,T_0]}$. Proceeding analogously to Section \ref{app:gen}, but this time not applying the error map $\env_j$, we find

\ba 
\rho_\outpp(\tau):= \tr_\clo\left[\hat P_\textup{QI}(\tau)  \map_\textup{cov}(\rho_L)  \hat P_\textup{QI}(\tau)  \right]
= & U_\outp^{\otimes \K \dag} (\tau) \left( \hat E'_{\tau} ( \rho_\inp) + \map( \rho_\inp)  \right) U_\outp^{\otimes \K} (\tau),  \label{eq: rho our generix with F 2 ver2}
\ea
where 
\be 
\hat E'_{\tau}(\cdot):= \sum_{q,q'=0}^{d_\outp-1}\sum_{n,n'=0}^{d_\inp-1} \left[ \frac{ F_{Q}(\tau)}{F_{0}(\tau)}\,\me^{2\pi \mi \tau\, Q/{T_0}} -1\right] \map_{q,q',n,n'} (\cdot) \ketbra{q}{q'},\label{eq:error map ver2}
\ee
with $\map_{q,q',n,n'}$ give by Eq. \eqref{eq:map code} and $F_{Q}(\tau)$ given by
\ba
F_{Q}(\tau)&:= \frac{1}{T_0} \int_0^{T_0} dt \me^{-\mi \omega Q t} \bra{\psi_\textup{nor}(k_0)} U_\clo^\dag(\tau) \rho_{\cl 1}(t) U_\clo(\tau) \ket{\psi_\textup{nor}(k_0)}\\
&= \frac{1}{T_0} \int_0^{T_0} dt \me^{-\mi 2\pi Q t/{T_0}} \left| \bra{\psi_\textup{nor}(k_0)}U_\clo(t-\tau)\ket{\psi_\textup{nor}(k_0)} \right|^2,\label{eq:new F for PW}
\ea 
where $\ket{\psi_\textup{nor}(k_0)}$ is the Quasi-Ideal clock described previously in the manuscript.
Note that Eq. \eqref{eq: rho our generix with F 2 ver2} is analogous to Eq. \eqref{eq: rho our generix with F 2} but with a trivial error map $\env_j=\mathcal{I}$ (no clock errors), $\tilde \rho_\inp = U_\inp (t_\alpha) \rho_\inp U_\inp^\dag(t_\alpha)= \rho_\inp$ (trivial logical group representation), and a time $\tau$ rather than the discrete times $t_\alpha:= {k_\alpha} \,\frac{T_0}{d_\clo}$ (since we are not projecting onto the time basis any-more). One can bound Eqs. \eqref{eq:new F for PW} and \eqref{eq:error map ver2}, analogously to the calculations in Sections \ref{app:1clockproof} \ref{app:entfid}. We will not perform such calculations here for brevity, but one finds that $F_{Q}(\tau)/F_{0}(\tau)\approx \me^{-2\pi \mi \tau Q/{T_0}}$ where the approximation becomes exact in the large $d_\clo$ limit. So one can see that the quantity in square brackets in Eq. \eqref{eq:error map ver2} vanishes in said limit. Hence for large $d_\clo$, from Eq. \eqref{eq: rho our generix with F 2 ver2}
\be 
 \tr_\clo\left[\hat P_\textup{QI}(\tau)  \map_\textup{cov}(\rho_L)  \hat P_\textup{QI}(\tau)  \right] \approx U_\code^{\otimes \K \dag}  \map( \rho_\inp) U_\code^{\otimes \K} (\tau)= \me^{-\mi \tau \hat H_\code} \map( \rho_\inp) \me^{\mi \tau \hat H_\code} \quad\forall \,\tau\in\rr, 
\ee  
where the approximate equality becomes exact in the large $d_\clo$ limit. This is an approximate Page-Wooters condition (Eq. \eqref{eq:PWcondition exact}).

%% file: Cov-CodeV20-Quantumv3.bbl
\begin{thebibliography}{51}
\providecommand{\natexlab}[1]{#1}
\providecommand{\url}[1]{\texttt{#1}}
\expandafter\ifx\csname urlstyle\endcsname\relax
  \providecommand{\doi}[1]{doi: #1}\else
  \providecommand{\doi}{doi: \begingroup \urlstyle{rm}\Url}\fi

\bibitem[Hayden et~al.(2017)Hayden, Nezami, Popescu, and
  Salton]{hayden2017error}
Patrick Hayden, Sepehr Nezami, Sandu Popescu, and Grant Salton.
\newblock Error correction of quantum reference frame information.
\newblock \emph{ArXiv:1709.04471}, 2017.
\newblock URL \url{https://arxiv.org/abs/1709.04471}.

\bibitem[Zeng et~al.(2011)Zeng, Cross, and Chuang]{zeng2011transversality}
Bei Zeng, Andrew Cross, and Isaac~L Chuang.
\newblock Transversality versus universality for additive quantum codes.
\newblock \emph{IEEE Trans. Inf. Theory}, 57\penalty0 (9):\penalty0 6272--6284,
  2011.
\newblock \doi{10.1109/TIT.2011.2161917}.

\bibitem[Bravyi and K{\"o}nig(2013)]{bravyi2013classification}
Sergey Bravyi and Robert K{\"o}nig.
\newblock Classification of topologically protected gates for local stabilizer
  codes.
\newblock \emph{Phys. Rev. Lett.}, 110\penalty0 (17):\penalty0 170503, 2013.
\newblock \doi{10.1103/PhysRevLett.110.170503}.

\bibitem[Pastawski and Yoshida(2015)]{pastawski2015fault}
Fernando Pastawski and Beni Yoshida.
\newblock Fault-tolerant logical gates in quantum error-correcting codes.
\newblock \emph{Phys. Rev. A}, 91\penalty0 (1):\penalty0 012305, 2015.
\newblock \doi{10.1103/PhysRevA.91.012305}.

\bibitem[Jochym-O'Connor et~al.(2018)Jochym-O'Connor, Kubica, and
  Yoder]{jochym2018disjointness}
Tomas Jochym-O'Connor, Aleksander Kubica, and Theodore~J Yoder.
\newblock Disjointness of stabilizer codes and limitations on fault-tolerant
  logical gates.
\newblock \emph{Phys. Rev. X}, 8\penalty0 (2):\penalty0 021047, 2018.
\newblock \doi{10.1103/PhysRevX.8.021047}.

\bibitem[Brown et~al.(2016)Brown, Loss, Pachos, Self, and
  Wootton]{brown2016quantum}
Benjamin~J Brown, Daniel Loss, Jiannis~K Pachos, Chris~N Self, and James~R
  Wootton.
\newblock Quantum memories at finite temperature.
\newblock \emph{Rev. Mod. Phys.}, 88\penalty0 (4):\penalty0 045005, 2016.
\newblock \doi{10.1103/RevModPhys.88.045005}.

\bibitem[Eastin and Knill(2009)]{eastin2009restrictions}
Bryan Eastin and Emanuel Knill.
\newblock Restrictions on transversal encoded quantum gate sets.
\newblock \emph{Phys. Rev. Lett.}, 102\penalty0 (11):\penalty0 110502, 2009.
\newblock \doi{10.1103/PhysRevLett.102.110502}.

\bibitem[Bravyi and Kitaev(2005)]{bravyi2005universal}
Sergey Bravyi and Alexei Kitaev.
\newblock Universal quantum computation with ideal clifford gates and noisy
  ancillas.
\newblock \emph{Phys. Rev. A}, 71\penalty0 (2):\penalty0 022316, 2005.
\newblock \doi{10.1103/PhysRevA.71.022316}.

\bibitem[Knill et~al.(1996)Knill, Laflamme, and Zurek]{knill1996threshold}
Emanuel Knill, Raymond Laflamme, and W~Zurek.
\newblock Threshold accuracy for quantum computation.
\newblock \emph{ArXiv:quant-ph/9610011}, 1996.
\newblock URL \url{https://arxiv.org/abs/quant-ph/9610011}.

\bibitem[Bomb{\'\i}n and Mart{\'\i}n-Delgado(2007)]{bombin2007topological}
H{\'e}ctor Bomb{\'\i}n and Miguel~\'Angel Mart{\'\i}n-Delgado.
\newblock Topological computation without braiding.
\newblock \emph{Physical review letters}, 98\penalty0 (16):\penalty0 160502,
  2007.
\newblock \doi{10.1103/PhysRevLett.98.160502}.

\bibitem[Paetznick and Reichardt(2013)]{paetznick2013universal}
Adam Paetznick and Ben~W Reichardt.
\newblock Universal fault-tolerant quantum computation with only transversal
  gates and error correction.
\newblock \emph{Phys. Rev. Lett.}, 111\penalty0 (9):\penalty0 090505, 2013.
\newblock \doi{10.1103/PhysRevLett.111.090505}.

\bibitem[Jochym-O'Connor and Laflamme(2014)]{jochym2014using}
Tomas Jochym-O'Connor and Raymond Laflamme.
\newblock Using concatenated quantum codes for universal fault-tolerant quantum
  gates.
\newblock \emph{Phys. Rev. Lett.}, 112\penalty0 (1):\penalty0 010505, 2014.
\newblock \doi{10.1103/PhysRevLett.112.010505}.

\bibitem[Bomb{\'\i}n(2015)]{bombin2015gauge}
H{\'e}ctor Bomb{\'\i}n.
\newblock Gauge color codes: optimal transversal gates and gauge fixing in
  topological stabilizer codes.
\newblock \emph{New J. Phys.}, 17\penalty0 (8):\penalty0 083002, 2015.
\newblock \doi{10.1088/1367-2630/17/8/083002}.

\bibitem[Yoder et~al.(2016)Yoder, Takagi, and Chuang]{yoder2016universal}
Theodore~J Yoder, Ryuji Takagi, and Isaac~L Chuang.
\newblock Universal fault-tolerant gates on concatenated stabilizer codes.
\newblock \emph{Phys. Rev. X}, 6\penalty0 (3):\penalty0 031039, 2016.
\newblock \doi{10.1103/PhysRevX.6.031039}.

\bibitem[Bartlett et~al.(2007)Bartlett, Rudolph, and
  Spekkens]{bartlett2007reference}
Stephen~D Bartlett, Terry Rudolph, and Robert~W Spekkens.
\newblock Reference frames, superselection rules, and quantum information.
\newblock \emph{Rev. Mod. Phys.}, 79\penalty0 (2):\penalty0 555, 2007.
\newblock \doi{10.1103/RevModPhys.79.555}.

\bibitem[Pauli(1933)]{pauli1}
Wolfgang Pauli.
\newblock {Handbuch der Physik}.
\newblock \emph{Springer, Berlin}, 24:\penalty0 83 -- 272, 1933.
\newblock \doi{10.1007/978-3-642-52619-0_2}.

\bibitem[Pauli(1958)]{pauli2}
Wolfgang Pauli.
\newblock {Encyclopedia of Physics}.
\newblock \emph{Springer, Berlin}, 1:\penalty0 60, 1958.

\bibitem[Garrison and Wong(1970)]{Garrison1970}
John~C. Garrison and Jack Wong.
\newblock {Canonically conjugate pairs, uncertainty relations, and phase
  operators}.
\newblock \emph{J. Math. Phys.}, 11\penalty0 (8):\penalty0 2242--2249, Aug
  1970.
\newblock \doi{10.1063/1.1665388}.

\bibitem[Salecker and Wigner(1958)]{salecker1958quantum}
Helmut Salecker and EP~Wigner.
\newblock Quantum limitations of the measurement of space-time distances.
\newblock \emph{Phys. Rev.}, 109\penalty0 (2):\penalty0 571, 1958.
\newblock \doi{10.1103/PhysRev.109.571}.

\bibitem[Peres(1980)]{peres1980measurement}
Asher Peres.
\newblock Measurement of time by quantum clocks.
\newblock \emph{American Journal of Physics}, 48\penalty0 (7):\penalty0
  552--557, 1980.
\newblock \doi{10.1119/1.12061}.

\bibitem[Woods et~al.(2018{\natexlab{a}})Woods, Silva, and
  Oppenheim]{woods2016autonomous}
Mischa~P. Woods, Ralph Silva, and Jonathan Oppenheim.
\newblock {Autonomous Quantum Machines and Finite-Sized Clocks}.
\newblock \emph{Annales Henri Poincar{\'e}}, Oct 2018{\natexlab{a}}.
\newblock ISSN 1424-0661.
\newblock \doi{10.1007/s00023-018-0736-9}.

\bibitem[Bu\ifmmode~\check{z}\else \v{z}\fi{}ek
  et~al.(1999)Bu\ifmmode~\check{z}\else \v{z}\fi{}ek, Derka, and
  Massar]{PhysRevLett.82.2207}
Vladimir Bu\ifmmode~\check{z}\else \v{z}\fi{}ek, Radoslav Derka, and Serge
  Massar.
\newblock Optimal quantum clocks.
\newblock \emph{Phys. Rev. Lett.}, 82:\penalty0 2207--2210, Mar 1999.
\newblock \doi{10.1103/PhysRevLett.82.2207}.

\bibitem[Erker et~al.(2017)Erker, Mitchison, Silva, Woods, Brunner, and
  Huber]{Pauletal2017}
Paul Erker, Mark~T. Mitchison, Ralph Silva, Mischa~P. Woods, Nicolas Brunner,
  and Marcus Huber.
\newblock Autonomous quantum clocks: Does thermodynamics limit our ability to
  measure time?
\newblock \emph{Phys. Rev. X}, 7:\penalty0 031022, Aug 2017.
\newblock \doi{10.1103/PhysRevX.7.031022}.
\newblock URL \url{https://link.aps.org/doi/10.1103/PhysRevX.7.031022}.

\bibitem[Rankovi{\'{c}} et~al.(2015)Rankovi{\'{c}}, Liang, and
  Renner]{RaLiRe15}
Sandra Rankovi{\'{c}}, Yeong-Cherng Liang, and Renato Renner.
\newblock Quantum clocks and their synchronisation - the alternate ticks game.
\newblock \emph{ArXiv:1506.01373v1}, 2015.
\newblock URL \url{https://arxiv.org/abs/1506.01373}.

\bibitem[Woods et~al.(2018{\natexlab{b}})Woods, Silva, P{\"u}tz, Stupar, and
  Renner]{woods2018quantum}
Mischa~P Woods, Ralph Silva, Gilles P{\"u}tz, Sandra Stupar, and Renato Renner.
\newblock Quantum clocks are more accurate than classical ones.
\newblock \emph{ArXiv:1806.00491}, 2018{\natexlab{b}}.
\newblock URL \url{https://arxiv.org/abs/1806.00491}.

\bibitem[Faist et~al.(2019)Faist, Nezami, V.~Albert, Salton, Pastawski, Hayden,
  and Preskill]{faist2018prep}
Philippe Faist, Sepehr Nezami, Victor V.~Albert, Grant Salton, Fernando
  Pastawski, Patrick Hayden, and John Preskill.
\newblock Continuous symmetries and approximate quantum error correction.
\newblock \emph{ArXiv:1902.07714}, 2019.
\newblock URL \url{https://arxiv.org/abs/1902.07714}.

\bibitem[Lunardi et~al.(2011)Lunardi, Manzoni, and Nystrom]{LUNARDI2011415}
Jos\'e~T. Lunardi, Luiz~A. Manzoni, and Andrew~T. Nystrom.
\newblock {Salecker Wigner Peres clock and average tunneling times}.
\newblock \emph{Phys. Lett. A}, 375\penalty0 (3):\penalty0 415 -- 421, 2011.
\newblock ISSN 0375-9601.
\newblock \doi{10.1016/j.physleta.2010.11.055}.

\bibitem[Sokolovski(2017)]{PhysRevA.96.022120}
Dmitri Sokolovski.
\newblock Salecker-wigner-peres clock, feynman paths, and a tunneling time that
  should not exist.
\newblock \emph{Phys. Rev. A}, 96:\penalty0 022120, Aug 2017.
\newblock \doi{10.1103/PhysRevA.96.022120}.

\bibitem[Cal\ifmmode~\mbox{\c{c}}\else \c{c}\fi{}ada
  et~al.(2009)Cal\ifmmode~\mbox{\c{c}}\else \c{c}\fi{}ada, Lunardi, and
  Manzoni]{PhysRevA.79.012110}
Marcos Cal\ifmmode~\mbox{\c{c}}\else \c{c}\fi{}ada, Jos\'e~T. Lunardi, and
  Luiz~A. Manzoni.
\newblock {Salecker-Wigner-Peres clock and double-barrier tunneling}.
\newblock \emph{Phys. Rev. A}, 79:\penalty0 012110, Jan 2009.
\newblock \doi{10.1103/PhysRevA.79.012110}.

\bibitem[Teeny et~al.(2016)Teeny, Keitel, and Bauke]{Teeny2016}
Nicolas Teeny, Christoph~H. Keitel, and Heiko Bauke.
\newblock {Salecker-Wigner-Peres quantum clock applied to strong-field tunnel
  ionization}.
\newblock \emph{ArXiv:1608.02854}, Aug 2016.
\newblock URL \url{http://arxiv.org/abs/1608.02854}.

\bibitem[Bartlett et~al.(2009)Bartlett, Rudolph, Spekkens, and
  Turner]{bartlett2009quantum}
Stephen~D Bartlett, Terry Rudolph, Robert~W Spekkens, and Peter~S Turner.
\newblock Quantum communication using a bounded-size quantum reference frame.
\newblock \emph{New J. Phys.}, 11\penalty0 (6):\penalty0 063013, 2009.
\newblock \doi{10.1088/1367-2630/11/6/063013}.

\bibitem[Knill and Laflamme(1997)]{knill1997theory}
Emanuel Knill and Raymond Laflamme.
\newblock Theory of quantum error-correcting codes.
\newblock \emph{Phys. Rev. A}, 55\penalty0 (2):\penalty0 900, 1997.
\newblock \doi{10.1103/PhysRevA.55.900}.

\bibitem[Nielsen(1996)]{nielsen1996entanglement}
Michael~A Nielsen.
\newblock The entanglement fidelity and quantum error correction.
\newblock \emph{ArXiv: quant-ph/9606012}, 1996.
\newblock URL \url{https://arxiv.org/abs/quant-ph/9606012}.

\bibitem[Giovannetti et~al.(2011)Giovannetti, Lloyd, and Maccone]{StephLody}
Vittorio Giovannetti, Seth Lloyd, and Lorenzo Maccone.
\newblock Advances in quantum metrology.
\newblock \emph{Nat. Photonics}, 5\penalty0 (222), Mar 2011.
\newblock \doi{10.1038/nphoton.2011.35}.

\bibitem[Marvian~Mashhad(2012)]{marvian2012symmetry}
Iman Marvian~Mashhad.
\newblock Symmetry, asymmetry and quantum information.
\newblock \emph{PhD thesis, University of Waterloo}, 2012.
\newblock URL \url{http://hdl.handle.net/10012/7088}.

\bibitem[Marvian and Spekkens(2014)]{marvian2014extending}
Iman Marvian and Robert~W Spekkens.
\newblock Extending noether's theorem by quantifying the asymmetry of quantum
  states.
\newblock \emph{Nat. Commun.}, 5, 2014.
\newblock \doi{10.1038/ncomms4821}.

\bibitem[Demkowicz-Dobrza{\'n}ski et~al.(2017)Demkowicz-Dobrza{\'n}ski,
  Czajkowski, and Sekatski]{demkowicz2017adaptive}
Rafa{\l} Demkowicz-Dobrza{\'n}ski, Jan Czajkowski, and Pavel Sekatski.
\newblock Adaptive quantum metrology under general markovian noise.
\newblock \emph{Phys. Rev. X}, 7\penalty0 (4):\penalty0 041009, 2017.
\newblock \doi{10.1103/PhysRevX.7.041009}.

\bibitem[Zhou et~al.(2018)Zhou, Zhang, Preskill, and Jiang]{zhou2018achieving}
Sisi Zhou, Mengzhen Zhang, John Preskill, and Liang Jiang.
\newblock Achieving the heisenberg limit in quantum metrology using quantum
  error correction.
\newblock \emph{Nat. Commun.}, 9\penalty0 (1):\penalty0 78, 2018.
\newblock \doi{10.1038/s41467-017-02510-3}.

\bibitem[Layden et~al.(2019)Layden, Zhou, Cappellaro, and
  Jiang]{layden2018ancilla}
David Layden, Sisi Zhou, Paola Cappellaro, and Liang Jiang.
\newblock Ancilla-free quantum error correction codes for quantum metrology.
\newblock \emph{Physical review letters}, 122\penalty0 (4):\penalty0 040502,
  2019.
\newblock \doi{10.1103/PhysRevLett.122.040502}.

\bibitem[Almheiri et~al.(2015)Almheiri, Dong, and Harlow]{almheiri2015bulk}
Ahmed Almheiri, Xi~Dong, and Daniel Harlow.
\newblock {Bulk locality and quantum error correction in AdS/CFT}.
\newblock \emph{J. High Energy Phys.}, 2015\penalty0 (4):\penalty0 163, Apr
  2015.
\newblock ISSN 1029-8479.
\newblock \doi{10.1007/JHEP04(2015)163}.

\bibitem[Harlow and Ooguri(2019)]{Harlow2018a_short}
Daniel Harlow and Hirosi Ooguri.
\newblock Constraints on symmetries from holography.
\newblock \emph{Physical review letters}, 122\penalty0 (19):\penalty0 191601,
  2019.
\newblock \doi{10.1103/PhysRevLett.122.191601}.

\bibitem[Harlow and Ooguri(2018)]{Harlow2018}
Daniel Harlow and Hirosi Ooguri.
\newblock {Symmetries in quantum field theory and quantum gravity}.
\newblock \emph{ArXiv:1810.05338}, Oct 2018.
\newblock URL \url{http://arxiv.org/abs/1810.05338}.

\bibitem[Kohler and Cubitt(2019)]{TamaToby}
Tamara Kohler and Toby Cubitt.
\newblock Toy models of holographic duality between local hamiltonians.
\newblock \emph{Journal of High Energy Physics}, 2019\penalty0 (8):\penalty0
  17, Aug 2019.
\newblock ISSN 1029-8479.
\newblock \doi{10.1007/JHEP08(2019)017}.

\bibitem[Cubitt et~al.(2018)Cubitt, Montanaro, and Piddock]{Cubitt9497}
Toby~S. Cubitt, Ashley Montanaro, and Stephen Piddock.
\newblock {Universal quantum Hamiltonians}.
\newblock \emph{Proc. Natl. Acad. Sci. U.S.A}, 115\penalty0 (38):\penalty0
  9497--9502, 2018.
\newblock ISSN 0027-8424.
\newblock \doi{10.1073/pnas.1804949115}.

\bibitem[DeWitt(1967)]{PhysRev.160.1113}
Bryce~S. DeWitt.
\newblock Quantum theory of gravity. i. the canonical theory.
\newblock \emph{Phys. Rev.}, 160:\penalty0 1113--1148, Aug 1967.
\newblock \doi{10.1103/PhysRev.160.1113}.

\bibitem[Page and Wootters(1983)]{PageWooter}
Don~N. Page and William~K. Wootters.
\newblock {Evolution without evolution: Dynamics described by stationary
  observables}.
\newblock \emph{Phys. Rev. D}, 27:\penalty0 2885--2892, Jun 1983.
\newblock \doi{10.1103/PhysRevD.27.2885}.

\bibitem[Wolf(Lecture notes, July 2012)]{wolf2012quantum}
Michael~M Wolf.
\newblock \emph{Quantum channels \& operations: A Guided tour}.
\newblock Lecture notes, July 2012.
\newblock URL
  \url{https://www-m5.ma.tum.de/foswiki/pub/M5/Allgemeines/MichaelWolf/QChannelLecture.pdf}.

\bibitem[Giovannetti et~al.(2015)Giovannetti, Lloyd, and
  Maccone]{PhysRevD.92.045033}
Vittorio Giovannetti, Seth Lloyd, and Lorenzo Maccone.
\newblock Quantum time.
\newblock \emph{Phys. Rev. D}, 92:\penalty0 045033, Aug 2015.
\newblock \doi{10.1103/PhysRevD.92.045033}.

\bibitem[Gambini et~al.(2009)Gambini, Porto, Pullin, and
  Torterolo]{PhysRevD.79.041501}
Rodolfo Gambini, Rafael~A. Porto, Jorge Pullin, and Sebasti\'an Torterolo.
\newblock Conditional probabilities with dirac observables and the problem of
  time in quantum gravity.
\newblock \emph{Phys. Rev. D}, 79:\penalty0 041501, Feb 2009.
\newblock \doi{10.1103/PhysRevD.79.041501}.

\bibitem[Marletto and Vedral(2017)]{PhysRevD.95.043510}
Chiara Marletto and Vlatko Vedral.
\newblock Evolution without evolution and without ambiguities.
\newblock \emph{Phys. Rev. D}, 95:\penalty0 043510, Feb 2017.
\newblock \doi{10.1103/PhysRevD.95.043510}.

\bibitem[Moreva et~al.(2014)Moreva, Brida, Gramegna, Giovannetti, Maccone, and
  Genovese]{PhysRevA.89.052122}
Ekaterina Moreva, Giorgio Brida, Marco Gramegna, Vittorio Giovannetti, Lorenzo
  Maccone, and Marco Genovese.
\newblock Time from quantum entanglement: An experimental illustration.
\newblock \emph{Phys. Rev. A}, 89:\penalty0 052122, May 2014.
\newblock \doi{10.1103/PhysRevA.89.052122}.

\end{thebibliography}
